\documentclass[twocolumn,astrosymb,tighten,twocolappendix,longbib,trackchanges]{aastex7} 
\usepackage{mathrsfs}

\usepackage{nccmath,gensymb} 
\usepackage{siunitx,xspace,tabularx,multirow}
\usepackage{xcolor,pifont,url}
\usepackage{hyperref}

\definecolor{tabgreen}{RGB}{27,133,62}
\definecolor{tabred}{RGB}{192,28,40}

\shorttitle{DESI DGL Maps}
\shortauthors{Saydjari et al.}

\defcitealias{Brandt_2012_ApJ}{BD12}
\defcitealias{Chellew_2022_ApJ}{C22}
\defcitealias{Bertoldi_1996_ApJ} {BD96}
\defcitealias{Draine_1996_ApJ}{DB96}

\graphicspath{{./}{figures/}}

\begin{document}

\title{Spatially-Resolved Spectra of Diffuse Galactic Light using 10.8 M DESI Sky Fibers}

\suppressAffiliations
\author[orcid=0000-0002-6561-9002,gname=Andrew,sname=Saydjari]{Andrew~K.~Saydjari}
\altaffiliation{Hubble Fellow}
\affiliation{Department of Astrophysical Sciences, Princeton University,
Princeton, NJ 08544 USA}
\affiliation{Department of Astronomy, New Mexico State University,
Las Cruces, NM 88001 USA}
\email[show]{aksaydjari@gmail.com}
\correspondingauthor{Andrew~K.~Saydjari}

\author[0000-0002-0846-936X,gname=Bruce,sname=Draine]{Bruce~T.~Draine}
\affiliation{Department of Astrophysical Sciences, Princeton University,
Princeton, NJ 08544 USA}
\email{draine@astro.princeton.edu}

\author[0000-0003-2630-8073,gname=Timothy,sname=Brandt]{Timothy~D.~Brandt}
\affiliation{Space Telescope Science Institute, 3700 San Martin Dr., Baltimore, MD 21218, USA}
\email{tbrandt@stsci.edu}

\author[0000-0002-3569-7421,gname=Edward,sname=Schlafly]{E.~F.~Schlafly}
\affiliation{Space Telescope Science Institute, 3700 San Martin Drive, Baltimore, MD 21218, USA}
\email{eschlafly@stsci.edu}

\author[orcid=0000-0002-4928-4003,gname=Arjun,sname=Dey]{Arjun~Dey}
\affiliation{NSF NOIRLab, 950 North Cherry Avenue, Tucson, AZ 85719, USA}
\email{arjun.dey@noirlab.edu}

\author[orcid=0000-0003-2808-275X,gname=Douglas,sname=Finkbeiner]{D.~Finkbeiner}
\affiliation{Center for Astrophysics $|$ Harvard \& Smithsonian, 60 Garden St., Cambridge, MA, 02138, USA}
\affiliation{Department of Physics, Harvard University, 17 Oxford St., Cambridge, MA 02138, USA}
\email{dfinkbeiner@cfa.harvard.edu}

\author[gname={Jessica Nicole},sname=Aguilar]{J.~Aguilar}
\affiliation{Lawrence Berkeley National Laboratory, 1 Cyclotron Road, Berkeley, CA 94720, USA}
\email{jaguilar@lbl.gov}

\author[orcid=0000-0001-6098-7247,gname=Steven,sname=Ahlen]{S.~Ahlen}
\affiliation{Department of Physics, Boston University, 590 Commonwealth Avenue, Boston, MA 02215 USA}
\email{ahlen@bu.edu}

\author[orcid=0000-0002-0084-572X,gname=Carlos,sname={Allende Prieto}]{C.~Allende~Prieto}
\affiliation{Departamento de Astrof\'{\i}sica, Universidad de La Laguna (ULL), E-38206, La Laguna, Tenerife, Spain}
\affiliation{Instituto de Astrof\'{\i}sica de Canarias, C/ V\'{\i}a L\'{a}ctea, s/n, E-38205 La Laguna, Tenerife, Spain}
\email{carlos.allende.prieto@iac.es}

\author[orcid=0000-0003-2923-1585,gname=Abhijeet,sname=Anand]{A.~Anand}
\affiliation{Inter-University Centre for Astronomy and Astrophysics, Post Bag 4, Ganeshkhind, Pune 411 007, India}
\email{abhijeetanand2011@gmail.com}

\author[orcid=0000-0003-0467-5438,gname=Florian,sname=Beutler]{F.~Beutler}
\affiliation{Institute for Astronomy, University of Edinburgh, Royal Observatory, Blackford Hill, Edinburgh EH9 3HJ, UK}
\email{florian.beutler@ed.ac.uk}

\author[orcid=0000-0001-9712-0006,gname=Davide,sname=Bianchi]{D.~Bianchi}
\affiliation{Dipartimento di Fisica ``Aldo Pontremoli'', Universit\`a degli Studi di Milano, Via Celoria 16, I-20133 Milano, Italy}
\affiliation{INAF-Osservatorio Astronomico di Brera, Via Brera 28, 20122 Milano, Italy}
\email{davide.bianchi1@unimi.it}

\author[gname=David,sname=Brooks]{D.~Brooks}
\affiliation{Department of Physics \& Astronomy, University College London, Gower Street, London, WC1E 6BT, UK}
\email{david.brooks@ucl.ac.uk}

\author[orcid=0000-0003-3044-5150,gname=Aurelio,sname={Carnero Rosell}]{A.~Carnero Rosell}
\affiliation{Departamento de Astrof\'{\i}sica, Universidad de La Laguna (ULL), E-38206, La Laguna, Tenerife, Spain}
\affiliation{Instituto de Astrof\'{\i}sica de Canarias, C/ V\'{\i}a L\'{a}ctea, s/n, E-38205 La Laguna, Tenerife, Spain}
\email{acarnero@iac.es}

\author[gname=Todd,sname=Claybaugh]{T.~Claybaugh}
\affiliation{Lawrence Berkeley National Laboratory, 1 Cyclotron Road, Berkeley, CA 94720, USA}
\email{tmclaybaugh@lbl.gov}

\author[orcid=0000-0002-1769-1640,gname=Axel,sname={de la Macorra}]{A.~de la Macorra}
\affiliation{Instituto de F\'{\i}sica, Universidad Nacional Aut\'{o}noma de M\'{e}xico,  Circuito de la Investigaci\'{o}n Cient\'{\i}fica, Ciudad Universitaria, Cd. de M\'{e}xico  C.~P.~04510,  M\'{e}xico}
\email{macorra@fisica.unam.mx}

\author[gname=Peter,sname=Doel]{P.~Doel}
\affiliation{Department of Physics \& Astronomy, University College London, Gower Street, London, WC1E 6BT, UK}
\email{apd@star.ucl.ac.uk}

\author[orcid=0000-0002-3033-7312,gname=Andreu,sname=Font-Ribera]{A.~Font-Ribera}
\affiliation{Instituci\'{o} Catalana de Recerca i Estudis Avan\c{c}ats, Passeig de Llu\'{\i}s Companys, 23, 08010 Barcelona, Spain}
\affiliation{Institut de F\'{i}sica d'Altes Energies (IFAE), The Barcelona Institute of Science and Technology, Edifici Cn, Campus UAB, 08193, Bellaterra (Barcelona), Spain}
\email{afont@ifae.es}

\author[orcid=0000-0002-2890-3725,gname={Jaime E.},sname=Forero-Romero]{J.~E.~Forero-Romero}
\affiliation{Departamento de F\'isica, Universidad de los Andes, Cra. 1 No. 18A-10, Edificio Ip, CP 111711, Bogot\'a, Colombia}
\affiliation{Observatorio Astron\'omico, Universidad de los Andes, Cra. 1 No. 18A-10, Edificio H, CP 111711 Bogot\'a, Colombia}
\email{je.forero@uniandes.edu.co}

\author[orcid=0000-0001-9632-0815,gname=Enrique,sname={Gaztañaga}]{E.~Gaztañaga}
\affiliation{Institut d'Estudis Espacials de Catalunya (IEEC), c/ Esteve Terradas 1, Edifici RDIT, Campus PMT-UPC, 08860 Castelldefels, Spain}
\affiliation{Institute of Cosmology and Gravitation, University of Portsmouth, Dennis Sciama Building, Portsmouth, PO1 3FX, UK}
\affiliation{Institute of Space Sciences, ICE-CSIC, Campus UAB, Carrer de Can Magrans s/n, 08913 Bellaterra, Barcelona, Spain}
\email{gaztanaga@gmail.com}

\author[orcid=0000-0003-3142-233X,gname=Satya,sname={Gontcho A Gontcho}]{Satya~{Gontcho A Gontcho}}
\affiliation{University of Virginia, Department of Astronomy, Charlottesville, VA 22904, USA}
\email{satya@virginia.edu}

\author[gname=Gaston,sname=Gutierrez]{G.~Gutierrez}
\affiliation{Fermi National Accelerator Laboratory, PO Box 500, Batavia, IL 60510, USA}
\email{gaston@fnal.gov}

\author[orcid=0000-0001-9822-6793,gname=Julien,sname=Guy]{J.~Guy}
\affiliation{Lawrence Berkeley National Laboratory, 1 Cyclotron Road, Berkeley, CA 94720, USA}
\email{jguy@lbl.gov}

\author[orcid=0000-0002-6550-2023,gname=Klaus,sname=Honscheid]{K.~Honscheid}
\affiliation{Center for Cosmology and AstroParticle Physics, The Ohio State University, 191 West Woodruff Avenue, Columbus, OH 43210, USA}
\affiliation{Department of Physics, The Ohio State University, 191 West Woodruff Avenue, Columbus, OH 43210, USA}
\affiliation{The Ohio State University, Columbus, 43210 OH, USA}
\email{kh@physics.osu.edu}

\author[orcid=0000-0002-5652-8870,gname=Tanveer,sname=Karim]{T.~Karim}
\affiliation{Department of Astronomy \& Astrophysics, University of Toronto, Toronto, ON M5S 3H4, Canada}
\email{tanveer.karim@utoronto.ca}

\author[orcid=0000-0002-8828-5463,gname=David,sname=Kirkby]{D.~Kirkby}
\affiliation{Department of Physics and Astronomy, University of California, Irvine, 92697, USA}
\email{dkirkby@uci.edu}

\author[orcid=0000-0001-6356-7424,gname=Anthony,sname=Kremin]{A.~Kremin}
\affiliation{Lawrence Berkeley National Laboratory, 1 Cyclotron Road, Berkeley, CA 94720, USA}
\email{akremin@lbl.gov}

\author[orcid=0000-0002-1134-9035,gname=Ofer,sname=Lahav]{O.~Lahav}
\affiliation{Department of Physics \& Astronomy, University College London, Gower Street, London, WC1E 6BT, UK}
\email{o.lahav@ucl.ac.uk}

\author[gname=Andrew,sname=Lambert]{A.~Lambert}
\affiliation{Lawrence Berkeley National Laboratory, 1 Cyclotron Road, Berkeley, CA 94720, USA}
\email{arlambert@lbl.gov}

\author[orcid=0000-0003-1838-8528,gname=Martin,sname=Landriau]{M.~Landriau}
\affiliation{Lawrence Berkeley National Laboratory, 1 Cyclotron Road, Berkeley, CA 94720, USA}
\email{mlandriau@lbl.gov}

\author[orcid=0000-0001-7178-8868,gname=Laurent,sname={Le Guillou}]{L.~Le~Guillou}
\affiliation{Sorbonne Universit\'{e}, CNRS/IN2P3, Laboratoire de Physique Nucl\'{e}aire et de Hautes Energies (LPNHE), FR-75005 Paris, France}
\email{llg@lpnhe.in2p3.fr}

\author[orcid=0000-0002-1125-7384,gname=Aaron,sname=Meisner]{A.~Meisner}
\affiliation{NSF NOIRLab, 950 N. Cherry Ave., Tucson, AZ 85719, USA}
\email{aaron.meisner@noirlab.edu}

\author[gname=Ramon,sname=Miquel]{R.~Miquel}
\affiliation{Instituci\'{o} Catalana de Recerca i Estudis Avan\c{c}ats, Passeig de Llu\'{\i}s Companys, 23, 08010 Barcelona, Spain}
\affiliation{Institut de F\'{i}sica d'Altes Energies (IFAE), The Barcelona Institute of Science and Technology, Edifici Cn, Campus UAB, 08193, Bellaterra (Barcelona), Spain}
\email{rmiquel@ifae.es}

\author[orcid=0000-0002-2733-4559,gname=John,sname=Moustakas]{J.~Moustakas}
\affiliation{Department of Physics and Astronomy, Siena University, 515 Loudon Road, Loudonville, NY 12211, USA}
\email{jmoustakas@siena.edu}

\author[orcid=0000-0001-9070-3102,gname=Seshadri,sname=Nadathur]{S.~Nadathur}
\affiliation{Institute of Cosmology and Gravitation, University of Portsmouth, Dennis Sciama Building, Portsmouth, PO1 3FX, UK}
\email{seshadri.nadathur@port.ac.uk}

\author[orcid=0000-0002-4637-2868,gname=Enrique,sname=Paillas]{E.~Paillas}
\affiliation{Instituto de Estudios Astrof\'isicos, Facultad de Ingenier\'ia y Ciencias, Universidad Diego Portales, Av. Ej\'ercito Libertador 441, Santiago, Chile}
\affiliation{Steward Observatory, University of Arizona, 933 N. Cherry Avenue, Tucson, AZ 85721, USA}
\email{enrique.paillas@udp.cl}

\author[orcid=0000-0002-0644-5727,gname=Will,sname=Percival]{W.~J.~Percival}
\affiliation{Department of Physics and Astronomy, University of Waterloo, 200 University Ave W, Waterloo, ON N2L 3G1, Canada}
\affiliation{Perimeter Institute for Theoretical Physics, 31 Caroline St. North, Waterloo, ON N2L 2Y5, Canada}
\affiliation{Waterloo Centre for Astrophysics, University of Waterloo, 200 University Ave W, Waterloo, ON N2L 3G1, Canada}
\email{will.percival@uwaterloo.ca}

\author[orcid=0000-0001-6979-0125,gname=Ignasi,sname={P\'erez-R\`afols}]{I.~P\'erez-R\`afols}
\affiliation{Departament de F\'isica, EEBE, Universitat Polit\`ecnica de Catalunya, c/Eduard Maristany 10, 08930 Barcelona, Spain}
\email{ignasi.perez.rafols@upc.edu}

\author[orcid=0000-0001-7145-8674,gname=Francisco,sname=Prada]{F.~Prada}
\affiliation{Instituto de Astrof\'{i}sica de Andaluc\'{i}a (CSIC), Glorieta de la Astronom\'{i}a, s/n, E-18008 Granada, Spain}
\email{fprada@iaa.es}

\author[orcid=0000-0002-3500-6635,gname=Corentin,sname=Ravoux]{C.~Ravoux}
\affiliation{Universit\'{e} Clermont-Auvergne, CNRS, LPCA, 63000 Clermont-Ferrand, France}
\email{corentin.ravoux@clermont.in2p3.fr}

\author[gname=Graziano,sname=Rossi]{G.~Rossi}
\affiliation{Department of Physics and Astronomy, Sejong University, 209 Neungdong-ro, Gwangjin-gu, Seoul 05006, Republic of Korea}
\email{graziano@sejong.ac.kr}

\author[orcid=0000-0002-1609-5687,gname=Lado,sname=Samushia]{L.~Samushia}
\affiliation{Abastumani Astrophysical Observatory, Tbilisi, GE-0179, Georgia}
\affiliation{Department of Physics, Kansas State University, 116 Cardwell Hall, Manhattan, KS 66506, USA}
\email{lado@phys.ksu.edu}

\author[orcid=0000-0002-9646-8198,gname=Eusebio,sname=Sanchez]{E.~Sanchez}
\affiliation{CIEMAT, Avenida Complutense 40, E-28040 Madrid, Spain}
\email{eusebio.sanchez@ciemat.es}

\author[orcid=0000-0002-0408-5633,gname=Christoph,sname=Saulder]{C.~Saulder}
\affiliation{Max Planck Institute for Extraterrestrial Physics, Gie\ss enbachstra\ss e 1, 85748 Garching, Germany}
\email{csaulder@mpe.mpg.de}

\author[gname=David,sname=Schlegel]{D.~Schlegel}
\affiliation{Lawrence Berkeley National Laboratory, 1 Cyclotron Road, Berkeley, CA 94720, USA}
\email{djschlegel@lbl.gov}

\author[gname=Michael,sname=Schubnell]{M.~Schubnell}
\affiliation{Department of Physics, University of Michigan, 450 Church Street, Ann Arbor, MI 48109, USA}
\affiliation{University of Michigan, 500 S. State Street, Ann Arbor, MI 48109, USA}
\email{schubnel@umich.edu}

\author[orcid=0000-0003-3449-8583,gname=Ray,sname=Sharples]{R.~Sharples}
\affiliation{Centre for Advanced Instrumentation, Department of Physics, Durham University, South Road, Durham DH1 3LE, UK}
\affiliation{Institute for Computational Cosmology, Department of Physics, Durham University, South Road, Durham DH1 3LE, UK}
\email{r.m.sharples@durham.ac.uk}

\author[orcid=0000-0002-3461-0320,gname={Joseph Harry},sname=Silber]{J.~Silber}
\affiliation{Lawrence Berkeley National Laboratory, 1 Cyclotron Road, Berkeley, CA 94720, USA}
\email{jhsilber@lbl.gov}

\author[orcid=0000-0002-2949-2155,gname={Małgorzata},sname=Siudek]{M.~Siudek}
\affiliation{Institute of Space Sciences, ICE-CSIC, Campus UAB, Carrer de Can Magrans s/n, 08913 Bellaterra, Barcelona, Spain}
\affiliation{Instituto de Astrof\'{\i}sica de Canarias, C/ V\'{\i}a L\'{a}ctea, s/n, E-38205 La Laguna, Tenerife, Spain}
\email{msiudek@iac.es}

\author[orcid=0000-0003-1704-0781,gname=Gregory,sname={Tarl\'{e}}]{G.~Tarl\'{e}}
\affiliation{University of Michigan, 500 S. State Street, Ann Arbor, MI 48109, USA}
\email{gtarle@umich.edu}

\author[gname={Benjamin Alan},sname=Weaver]{B.~A.~Weaver}
\affiliation{NSF NOIRLab, 950 N. Cherry Ave., Tucson, AZ 85719, USA}
\email{benjamin.weaver@noirlab.edu}

\author[orcid=0000-0001-5381-4372,gname=Rongpu,sname=Zhou]{R.~Zhou}
\affiliation{Lawrence Berkeley National Laboratory, 1 Cyclotron Road, Berkeley, CA 94720, USA}
\email{rongpuzhou@lbl.gov}

\begin{abstract}

Using 10.8 million ``blank'' sky spectra from the DESI Year 3 dataset, we measure the diffuse galactic light (DGL) spectrum in the optical at spectral resolution $R \sim 4000$ by correlating with far-infrared emission from IRAS. Subdividing the sky into 54 deg$^2$ pixels (HEALPix, NSIDE = 8), we map the variation of the DGL correlation spectrum and nebular emission lines across the DESI footprint in the high-Galactic-latitude sky. The increased data volume over previous SDSS-based analyses enables several new detections in the DGL, including scattering both onto and out of the line of sight from neutral interstellar sodium and potassium. We further detect direct emission from ro-vibrational transitions of molecular hydrogen in the near-infrared with an absolute radiance of $0.65\substack{+0.13 \\ -0.12}\times10^{-9}~{\rm erg\, cm^{-2}\,s^{-1}\,sr^{-1}}$, roughly consistent with theoretical expectations, but with an apparent ortho-to-para line ratio that is lower by a factor of $0.60\substack{+0.28 \\ -0.27}$. We confirm previous detections of extended red emission (ERE) in the DGL and map its spatial variation. Our spatially resolved DGL maps provide important observational constraints for the radiative transfer efforts that are now possible with recent 3D models of the Milky Way.

\end{abstract}


\section{Introduction} \label{sec:intro}

Measuring the spectrum of light in the Milky Way that is scattered off of dust grains in the interstellar medium (ISM) provides important insights about both the sources of radiation and the scatterers. Because interstellar dust drives most of the scattering, the shape of the scattered light spectrum places constraints on the wavelength dependence of that scattering, and thus on dust grain size distribution and composition models \citep{Mathis_1973_ApJ,Savage_1979_ARA&A,Schiminovich_2001_ApJL,Draine_2003_ApJ,Sujatha_2005_ApJ}. This spectrum also constrains the properties and local environments of the dominant emission sources being scattered, such as the interstellar radiation field \citep{Murthy_1995_ApJ,Witt_1997_ApJ} and the temperature and density of emitting regions (e.g.,~\ion{H}{2} regions), as well as the properties of the intervening diffuse interstellar medium through which the light is propagating. 

\DeclareDocumentCommand{\yea}{s O{2.6cm} m}{{\color{tabgreen}\large\ding{51}}\ %
  \IfBooleanTF{#1}{{\scriptsize #3}}{\parbox[c]{#2}{\centering\scriptsize #3}}}
\DeclareDocumentCommand{\nay}{s O{2.6cm} m}{{\color{tabred}\large\ding{55}}\ %
  \IfBooleanTF{#1}{{\scriptsize #3}}{\parbox[c]{#2}{\centering\scriptsize #3}}}

\begin{deluxetable}{lcc}
\tabletypesize{\footnotesize}
\tablecaption{Summary of the origins of the signals detected in the
sky-fiber spectra via correlation with FIR dust emission.
\label{tab:signal_summary}}
\tablewidth{0pt}
\tablehead{
  \colhead{\raisebox{0.55em}[0pt][0pt]{Signal}} &
  \colhead{\raisebox{0.55em}[0pt][0pt]{Direct}} &
  \colhead{\shortstack{\rule{0pt}{2.6ex}Single Scattering\\ (Source $\times$ Scatterer)}}
}
\startdata
\noalign{\vskip 4pt}
Stars &
  \nay{suppressed by lack of FIR correlation} &
  \yea*{stars $\times$ dust} \\[14pt]
\parbox[c]{1.7cm}{\raggedright Nebular emission lines} &
  \nay{suppressed by lack of FIR correlation} &
  \yea*{H\,\textsc{ii} regions $\times$ dust} \\[14pt]
Na/K &
  \nay{collisional excitation and radiative recombination rates too low} &
  \yea*{stars $\times$ Na/K} \\[20pt]
H$_2$ &
  \yea{H$_2$ formation correlates with dust grain density} &
  --- \\[16pt]
ERE\rule[-3ex]{0pt}{0pt} & 
  \yea{expect carriers to correlate with dust} &
  --- \\
\enddata
\tablecomments{We neglect higher-order scattering processes, which
incur reductions in strength, after finding a process that could
explain our signal.}
\end{deluxetable}

Integrated broadly over many lines of sight, this spectrum can be used as a proxy for understanding the scattered component in the spectrum of external galaxies, where we do not have a handle on the detailed 3D distribution of stars and dust. However, if we are further able to access a spatially resolved map of the scattered light, this would be a measurement that serves as a quantitative test of the radiative transfer attempts to couple the 3D positions of O/B stars and 3D dust maps, that are just now becoming possible \citep{porter2018interstellar,McCallum_2025_MNRAS}.

Isolating this light scattered off of dust grains is hard because all measurements of the sky inherently include multiple components. Thus, modeling astrophysical observations of the sky is inherently a component separation problem, generally framed in terms of modeling point sources (either resolved or unresolved) and diffuse components (in absorption, scattering, or emission) that arise from different origins along the line of sight. The relative importance of each component depends strongly on the observed wavelength.

Components from the Earth's atmosphere are generally taken to be diffuse, but time-variable, and include strong ``telluric'' absorption and emission components, the latter often being called ``sky lines.'' In addition to planetary bodies, important contributions from the Solar system are both scattered and thermal emission from interplanetary dust ($\sim$200 K) called ``zodiacal light.'' Important Galactic components include direct emission from stars, gas, and dust in the interstellar medium. In addition to this direct emission, another diffuse component is the light scattered onto/off-of the line of sight by the interstellar medium. Extragalactic components include direct emission from either resolved or unresolved galaxies (e.g.,~the cosmic infrared background) and diffuse components like the cosmic microwave background. In this work, we are interested in the Galactic, scattered light component. We are specifically interested in improving the spectral and spatial resolution of these measurements in the optical.

\begin{deluxetable*}{lcccccccc}[t]
\tablewidth{0pt}
\tablecaption{DGL Sky Fiber Survey Comparison \label{tab:sky_surveys}}
\tablehead{
\colhead{\shortstack{Survey \\ Data Source}} & \colhead{\shortstack{Number of \\ Sky Spectra}} & \colhead{\shortstack{Field of View \\ (deg$^2$)}} & \colhead{\shortstack{Fibers/ \\ FoV}} & \colhead{\shortstack{Sky Fibers/ \\ FoV}} & \colhead{\shortstack{Fiber \\ Diameter}} & \colhead{\shortstack{Wavelength \\ Range (\AA)}} & \colhead{\shortstack{Spectral \\ Resolution ($R$)}} & \colhead{Citation}
}
\startdata
SDSS DR7  &  $9.20 \times 10^4$  &  7  &  640  &  32  &  $2.96''$  &  3800--9200  &  1850--2200  &  \citet{Brandt_2012_ApJ} \\
BOSS DR12  &  $2.39 \times 10^5$  &  7  &  1000  &  80  &  $2''$  &  3600--10400  &  1560--2650  &  \citet{Chellew_2022_ApJ} \\
DESI Y3  &  $1.08 \times 10^7$  &  $0.8$\tablenotemark{a}  &  $500$\tablenotemark{a}  &  $61$\tablenotemark{a}  &  $1.52''$  &  3600--9824  &  2000--5500  &  Saydjari et al. 2026 (this work) \\
\enddata
\tablenotetext{a}{While the full DESI spectrograph has an 8\,deg$^2$ FoV and 5000 fibers, of which 610 are sky fibers, we infer the DGL from correlations within each of 10 petals, so we must divide by 10 for a fair comparison.}
\end{deluxetable*}

We will isolate our component of interest, the spectrum of scattered light, by extracting only signals that correlate with dust as traced by its far-infrared thermal emission, which we describe in more detail below. For the purposes of this work, we will call the sum of all such signals the diffuse Galactic light (DGL). We summarize the main signals we detect in the DGL and the origin we ascribe them to in Table \ref{tab:signal_summary}. Direct emission from some of the brightest radiation sources, such as stars and hot gas, is suppressed by construction because of their lack of spatial correlation with the FIR dust emission. However, direct emission from species that have strong spatial correlations with dust can be captured in the DGL. One such species is molecular hydrogen, which has a strong spatial correlation with dust because its formation is catalyzed on dust grains and it is destroyed quickly in the ISM \citep{Jura_1975_ApJ,Draine_1996_ApJ}. Similarly, we might expect to detect extended red emission (ERE), which is attributed to fluorescence and phosphorescence of excited, dust-tracing molecular carriers (e.g.,~PAHs) \citep{Gordon_1998_ApJ}.

While we have defined all signals that correlate with the FIR dust as the DGL, we might view some unexpected signals as ``contamination.'' For example, previous work has shown that the FIR dust maps do not completely remove the cosmic infrared background, introducing correlations between dust maps and extragalactic large scale structure \citep{Chiang_2019_ApJ,Chiang_2023_ApJ}. Thus, our correlation with FIR dust maps may admit small contributions from external galaxies. In addition, the location of stars and \ion{H}{2} regions relative to dust are not entirely uncorrelated, which may allow a small contribution of direct emission from these sources. Beyond these, for each of the DGL components we describe, analogous contributions involving additional scattering events off of dust grains likely also contribute, though at a significantly lower amplitude. For now, we neglect all of these contributions as higher-order than those considered here, but future work to place quantitative constraints on their contributions would be valuable.

Beyond the constraints it provides on the sources and scatterers, the DGL is also important from a precision measurement standpoint: it universally contributes to spectra, and it is therefore necessary to measure and characterize this component, and its spatial variation between different lines of sight. Accurate measurements of the DGL, and especially its spatial variation, are also critical for measuring and understanding the extragalactic background light (EBL), for which the DGL is a bright foreground that must be modeled and subtracted, typically on the basis of only small, localized fields \citep{Lauer_2022_ApJL,Windhorst_2022_AJ}.

The DGL has been measured from the UV to the near-infrared using ground-based, rocket, and space-based platforms. Early detections came from ground-based photometry at $\sim$4500~\AA\ \citep{Elvey_1937_ApJ,Henyey_1941_ApJ,Elsasser_1960_ZA} and a sounding rocket \citep{Wolstencroft_1966_Natur}; satellite campaigns then extended these into the vacuum UV via the OAO at 1500--4200~\AA\ \citep{Lillie_1976_ApJ}, the TD-1 satellite \citep{Morgan_1978_MNRAS}, Voyager \citep{Murthy_1991_BAAS,Murthy_2012_ApJS}, and a series of additional experiments \citep{Henry_1981_ApJL,Zvereva_1982_A&A,Martin_1990_ApJ,Hurwitz_1991_ApJ,Sasseen_1995_ApJ,Schiminovich_2001_ApJL}, among which \citet{Martin_1990_ApJ} made the first detection of UV-pumped H$_2$ fluorescence in the DGL. 

More recently, the SPEAR/FIMS spectrograph ($R\sim550$, 1370--1720~\AA) \citep{Seon_2011_ApJS}, and GALEX broadband imaging in the FUV (1344--1786~\AA) and NUV (1771--2831~\AA) \citep{Hamden_2013_ApJ,Murthy_2014_ApJS} mapped the far-UV background over most of the sky. At optical wavelengths, Pioneer~10 and~11 provided zodiacal-light-free measurements, detecting ERE in the diffuse ISM \citep{Gordon_1998_ApJ} and constraining the broadband DGL \citep{Matsuoka_2012_IAUS}, while HST/FOS spectra across 54 sky fields covered 0.2--0.7~$\mu$m in 8 photometric bands \citep{Kawara_2017_PASJ}. In the near-infrared, the DGL has been characterized via COBE/DIRBE broadband photometry at 1.25--240~$\mu$m \citep{Arendt_1998_ApJ,Sano_2015_ApJ,Sano_2016_ApJ}, AKARI spectroscopy ($R\sim20$, 1.8--5.3~$\mu$m) with detection of the 3.3~$\mu$m PAH feature \citep{Tsumura_2013_PASJ}, CIBER spectroscopy ($R\sim15$--30, 0.95--1.65~$\mu$m) \citep{Arai_2015_ApJ}, and MIRIS broadband photometry at 1.1 and 1.6~$\mu$m \citep{Onishi_2018_PASJ}.

Measuring the DGL is difficult because the scattered light is much weaker than direct star light, but both are present over the whole sky. However, the scattered and direct emission from excited, diffuse gas only differ by a factor $\sim$2$\times$ (see Figure \ref{fig:whamHalpha}). The correlation-based approach described above was introduced at optical wavelengths by \citet{Brandt_2012_ApJ} and \citet{Chellew_2022_ApJ}, hereafter \citetalias{Brandt_2012_ApJ} and \citetalias{Chellew_2022_ApJ}, who isolated DGL spectra at $R\sim10^3$ using correlations of sky fiber flux in large spectroscopic surveys with far-infrared dust emission. These past works used sky spectra from SDSS DR7 and BOSS DR12, respectively, while we use DESI Y3, which is a bigger sample and the newest generation of these large, optical, spectroscopy surveys. The properties of the datasets from \citetalias{Brandt_2012_ApJ} and \citetalias{Chellew_2022_ApJ} are compared to this work in Table \ref{tab:sky_surveys}. \citetalias{Brandt_2012_ApJ} focused their analysis on one all-sky DGL spectrum, while \citetalias{Chellew_2022_ApJ} had sufficient signal to subdivide by two, into a Galactic north/south spectrum, and reported a marginal detection of ERE.

While an increase in the number of sky spectra is a primary factor in increasing the signal-to-noise ratio of the DGL measurement, we also list other key factors of the spectroscopic surveys for DGL measurements in Table \ref{tab:sky_surveys}. Importantly, as fiber-fed multi-object spectrographs move toward targeting smaller, more distant galaxies, fiber diameters have consistently decreased. However, this reduces the photometric aperture for extended surface brightness measurements like for the DGL. Similarly, multi-object spectrographs have now become sufficiently large that their fields of view are often subdivided between many different spectrographs. This can introduce systematics between spectrographs, and limits the usable field of view over which we can perform the correlation for extracting the DGL. Thus, while we do see an increase in signal-to-noise ratio from the massive increase in the number of sky spectra available from DESI compared to past surveys, the evolution of multi-object spectrograph designs has counteracted some of the gains associated with the increase in sample size.

In this work, we use 10.8 million sky fibers from the DESI \citep{DESICollaboration_2022_AJ} Year 3 dataset (DR2), which measure light in the optical ($\lambda\lambda3600-9824\AA$), together with the 100~$\mu$m IRIS far-infrared emission maps \citep{Miville-Deschenes_2005_ApJS} based on data from the Infrared Astronomical Satellite (IRAS) \citep{Neugebauer_1984_IRAS} to derive a higher signal-to-noise ratio and spectral resolution ($R\sim4000$) optical DGL spectrum. With the increased data volume/sky fiber densities (per square-degree) available from using DESI in this work, we can go beyond a north-south split to map the spatial variation of the DGL and detect new spectral features. Our paper is structured as follows. In Section \ref{sec:data}, we introduce the spectroscopic and far-infrared datasets we used, our statistical methods for extracting the DGL, and methods for fitting spectroscopic lines. We present the all-sky DGL correlation spectrum in Section \ref{sec:allskylines} and describe the DGL nebular emission lines both on average and in their spatial variation in Section \ref{sec:neblines}. We also discuss signatures from neutral interstellar atoms normally seen in absorption (Section \ref{sec:atomlines}) and molecular hydrogen (Section \ref{sec:h2lines}). In Section \ref{sec:continuum}, we describe the DGL continuum and evidence of ERE. We describe the data availability in Section \ref{sec:dataavil} before concluding in Section \ref{sec:conc}. In the Appendices, we show detailed calibration and validation plots.

\section{Data and Methods} \label{sec:data}

\subsection{DESI Sky Spectra} \label{sec:desisky}

\begin{figure}[htb!]
    \centering
    \includegraphics[width=\linewidth]{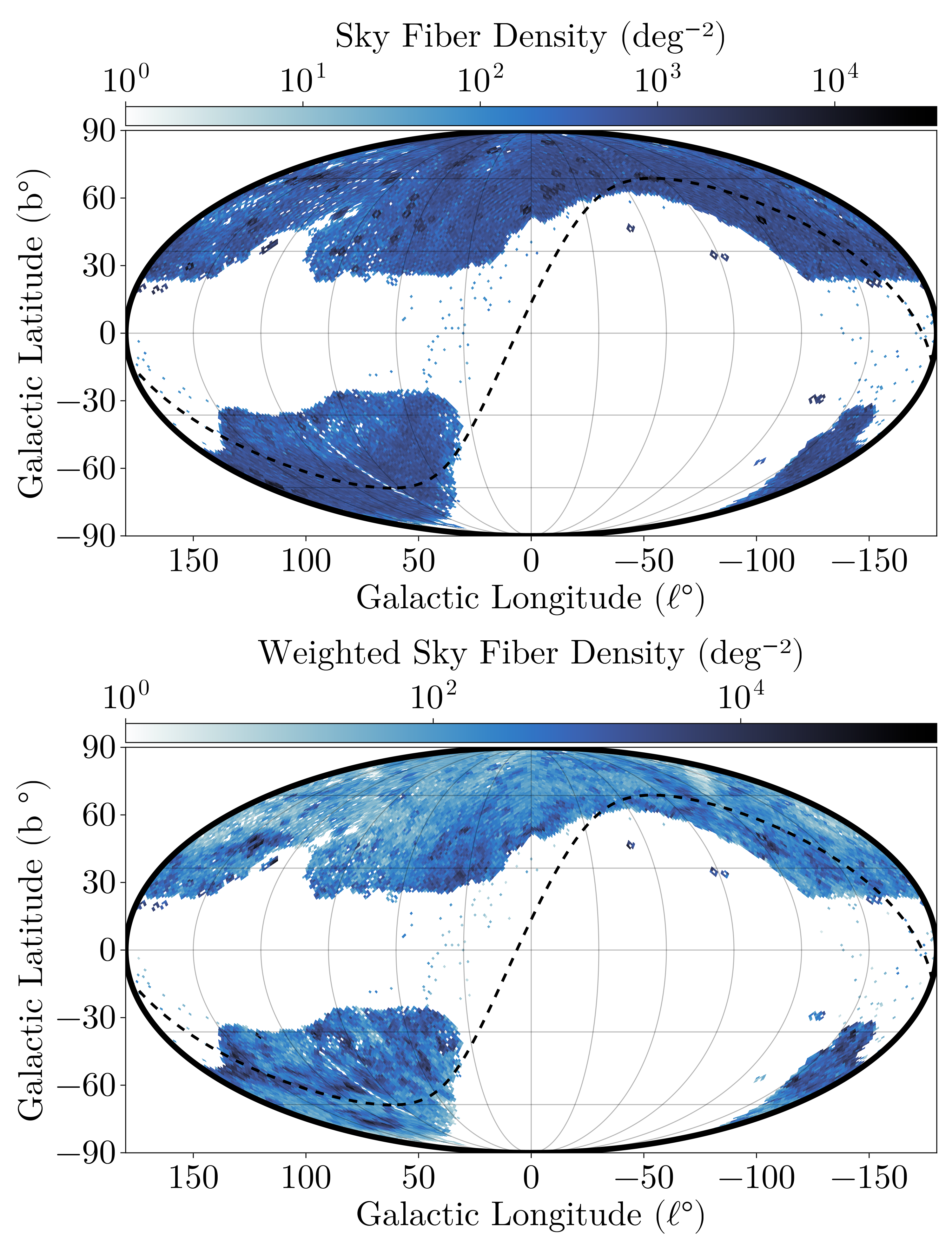}
    \caption{\textbf{Top:} Density of DESI sky fibers per sq. degree in Galactic coordinates used to measure the diffuse Galactic light. \textbf{Bottom:} Version of the top panel weighted by the variance of the far infrared emission across a petal-exposure ($p$) compared to the average per-petal variance in our sample, $\sigma^2_{100 \mu m, \, p}/\langle \sigma^2_{100 \mu m, \, p} \rangle$. This weight influences how significantly a given set of sky fibers impacts our global diffuse Galactic light spectrum. Both are Mollweide projections with a dashed line showing the ecliptic plane.}
    \label{fig:coverage_map}
\end{figure}

\paragraph{DESI Sample} The Dark Energy Spectroscopic Instrument (DESI) is a fiber-fed, multi-object optical spectrograph \citep{DESI2016b_Instr} mounted on the 4-meter Mayall telescope \citep{Corrector_Miller_2023} on Kitt Peak (I'oligam Du'ag) in Arizona, USA. It has robotic positioners \citep{FiberSystem_Poppett_2024} to place 5,000 science fibers, which are 107~$\mu$m or approximately 1.52 arcsec in diameter, in its 3.2$\degree$ diameter field of view and is conducting an eight-year survey of about 17,000 deg$^2$ of the sky \citep{SurveyOps_Schlafly_2023}. The focal plane is subdivided into 10 petals that are 36$\degree$-wide, pie-shaped wedges, each containing 500 fibers that feed one of 10 replicates of the DESI spectrograph design. These spectrographs have 3 arms (b, r, z), with slightly overlapping wavelength coverage ($3600 - 5800$, $5760 - 7620$, $7520 - 9824~$\r{A}) with resolutions increasing with increasing wavelength ($R = \lambda/\Delta\lambda \approx 2000-3200$, $3200-4100$, $4100-5000$). For more details on the instrumentation, see \citet{DESICollaboration_2022_AJ}. The raw spectra are extracted from 2D CCD images to 1D spectra on a uniform, linearly spaced, vacuum wavelength grid ($3600 - 9824$~\r{A}, $0.8$~\r{A} steps) using an implementation of ``spectroperfectionism'' \citep{Bolton+Schlegel_2010_PASP}. Throughout this work, ``wavelength'' refers to rest-frame vacuum wavelength (denoted $\lambda_{\rm vac}$ on figure axes), and quoted line rest frame wavelengths are vacuum values. For more information on the extraction and calibrations, see \citet{Guy_2023_AJ}. For the most recent public data release (DR1) and representative cosmological results, see \citet{DESICollaboration_2026_AJ}, \citet{DESI2024_VII_KP7B}, and \citet{AbdulKarim_2025_PhRvD}.

We use spectra from the ``Loa'' internal DESI data release, which is the Year 3 data release (DR2) candidate. We require the spectra be \texttt{OBJTYPE} $=$ \texttt{"SKY"}, \texttt{FIBERSTATUS} to have at most bits 1 (stuck positioner), 3 (restricted motion positioner), and 10 (fiber 30-100 $\mu$m off target) thrown. We will refer to data from a given petal during a given exposure as a ``petal-exposure''. For 19 exposures during the Survey Validation phase, the infrared (z) spectrographs for one or more petals failed to read out, even though the blue (b) and red (r) spectrograph data exist; we exclude these petal-exposures entirely. We need flux calibrations estimated from the guiding, focusing, and alignment (GFA) cameras (see Section \ref{sec:desisky}) and thus also exclude 77 exposures (732 petal-exposures) where these flux calibrations are not available.

The three main programs under which spectra are obtained represent the DESI survey \citep{SurveyOps_Schlafly_2023,Cooper_2023_ApJ} switching from fainter to brighter targets as observing conditions worsen, where \texttt{DARK} are observed in the ``best conditions'' (low sky brightness, high transparency, low extinction, good seeing) and \texttt{BACKUP} is the worst \citep{Dey_2026_AJ}. Because our spectra of interest, sky fibers, are the faintest targets, even a small fraction of fiber trace cross-talk, where light in the wings of the point-spread function (PSF) for one target leaks into the 1D spectrum extracted for another fiber nearby on the detector, can lead to significant contamination of the sky spectra. This contamination occurs especially when neighboring targets are bright. For this reason, we exclude $\sim20\%$ of all sky fibers by excluding exposures for programs with very bright targets (\texttt{backup}, \texttt{backup1}, \texttt{m31}, \texttt{m33}, \texttt{tertiary13}, \texttt{tertiary14}, and \texttt{tertiary32}).

At the per spectrum level, we also exclude (1) spectra that come through the DESI pipeline as all NaNs or (2) spectra at locations where the far infrared emission at 100 $\mu$m is greater than or equal to 10 MJy/sr (see Section \ref{sec:DGLModel} for justification). After these per spectrum cuts, we reject petal-exposures with 1 or fewer sky spectra remaining, which drops $\sim1.5\%$ of petal-exposures.

After these cuts, we obtain the dataset used in this study that contains 17,314 exposures, 169,890 petal-exposures, and 10,815,818 individual sky spectra. On average, there are 61 sky fibers per petal-exposure. These sky fiber observations are predominantly split between the \texttt{DARK} (5.2 M) and \texttt{BRIGHT} (4.2 M) programs, with the remaining 1.4 M coming from assorted smaller and survey validation programs. The distribution of these sky fibers on the sky is shown in Figure \ref{fig:coverage_map}. The bottom panel shows the same distribution, but including the relative weighting of the sky fibers as they enter into the all-sky DGL correlation spectrum, as described in Section \ref{sec:linfit}.

\paragraph{DESI Calibration} Some of the steps used in calibrating DESI spectra for the public data releases \citep{Guy_2023_AJ} are appropriate for our use case while others must be modified. Unfortunately, the sky fibers in the standard, sky-subtracted, calibrated, science outputs of the DESI pipeline for the ``Loa'' release (\texttt{cframe} files and subsequent summaries and coadds) are treated as if they were point-sources. For example, they are flux-calibrated using a set of standard stars observed by other fibers. A comparison of our recalibration approach with alternatives starting from other DESI pipeline data products is presented in Appendix \ref{sec:calCompare}.

Because our methodology for extracting the DGL relies on a cross-correlation with FIR emission (described in more detail in Section \ref{sec:DGLModel}), simple per exposure additive calibration errors will not impact our results. However, our DGL measurement \textbf{is} sensitive to multiplicative calibration errors or any calibration errors that spatially vary in a way that is correlated with the FIR emission across the field-of-view. This is what motivates the care we take with respect to the DESI sky fiber calibration.

Here we summarize our recalibration of the DESI 1D sky spectra, to treat them as the locally uniform surface-brightness sources that they are (private comm., J.~Guy). First, we divide the flux in the \texttt{frame} files by the ``fiber flats'' in the corresponding \texttt{fiberflatexp} files (matrix per petal-exposure).\footnote{See data models for these files here: \url{https://desidatamodel.readthedocs.io/en/25.3/DESI_SPECTRO_REDUX/SPECPROD/exposures/NIGHT/EXPID/index.html}.} This ideally homogenizes the response across fibers and spectrographs to isotropic illumination, including accounting for variations in the angular scale of fibers as a function of location in the focal plane and gain variations across and between different spectrographs. An example \texttt{fiberflatexp} is shown in Appendix \ref{sec:calEx}.

To account for the average telluric atmospheric absorption features and overall transmission function, we divide by the average \texttt{fluxcalib} for all of the sky fibers on all spectrographs in a given exposure (vector per exposure). While the \texttt{fluxcalib} estimated from DESI can include interstellar absorption features, by using the average estimated over all spectrographs, we prevent spatial variations in the \texttt{fluxcalib} that may be due to astrophysical variation in the sky from removing those variations from our data products. As we will discuss in Section \ref{sec:DGLModel}, it is these variations/correlations that we will leverage and need to preserve. An example \texttt{fluxcalib} is shown in Appendix \ref{sec:calEx}.

In addition, averaging over the different spectrographs builds up the signal-to-noise ratio for the \texttt{fluxcalib} estimate, which can be estimated from as few as 10 standard stars per petal. Unfortunately, this \texttt{fluxcalib} is still a calibration to stars and thus includes throughput variations due to wavelength-dependent PSF variations, which are relevant for stars but not for uniform surface-brightness sources. At the current time, this is an unavoidable limitation given the currently available data products from the DESI pipeline.

Next, we account for the seeing and overall throughput for a given exposure by multiplying by the \texttt{FIBER\_FRACFLUX\_GFA} (scalar per exposure) to obtain spectra with the standard DESI flux units of 10$^{-17}$ erg\,s$^{-1}$\,cm$^{-2}$\,\r{A}$^{-1}$. The \texttt{FIBER\_FRACFLUX\_GFA} is obtained by integrating the PSF from the guide cameras over the nominal 107 $\mu$m diameter fiber. Since we are considering constant surface brightness ``sources,'' we normalize this flux by the average fiber solid angle, that subtended by a 1.52 arcsec diameter fiber (to which \texttt{fiberflatexp} is normalized). When working with spectra throughout, we use $\lambda I_{\lambda}$ so we multiply the spectra by the wavelength grid. Finally, we convert to units of 10$^{-4}$ erg\,s$^{-1}$\,cm$^{-2}$\,sr$^{-1}$, having appropriately transformed the per-pixel inverse-variances by the usual Gaussian propagation of errors throughout the process. These \texttt{frame} files are \emph{not} ``sky-subtracted,'' meaning that we need to explicitly subtract the mean sky spectrum in Equation \ref{eq:linDGLmodel}.  This is in contrast to the corresponding data products from SDSS (\citetalias{Brandt_2012_ApJ}, \citetalias{Chellew_2022_ApJ}).  

\subsection{FIR Emission Map} \label{sec:FIRmap}

The other key dataset for extracting the DGL is the far infrared (FIR) emission that we correlate with sky fiber observations. We use the 100 $\mu$m emission from IRAS \citep{Neugebauer_1984_IRAS} as processed by \citet{Miville-Deschenes_2005_ApJS}, which is called ``IRIS.'' IRIS revisits the previous popular processing by \citet{Schlegel_1998_ApJ} (``SFD''). IRIS maintains a higher angular resolution (4 arcmin) and has improved destriping and diffuse emission calibration at scales smaller than 1$\degree$. Specifically, we use the version of the ``IRIS'' data products that are designed to match SFD at scales larger than 1$\degree$ and have infilled point sources (see Section \ref{sec:dataavil}). We do not apply any masks from SFD or IRIS, which for example indicate where small gaps in IRAS led to falling back to the use of DIRBE. The 100 $\mu$m emission from SFD is on average 0.2\% higher than that from IRIS in the data product we use over our survey footprint, and we adjust IRIS by a factor of 1.002 to be consistent with the SFD zeropoint.

\subsection{DGL Model} \label{sec:DGLModel}

In order to extract the spectrum of light scattered off of dust from the sky fiber observations, we search for contributions to the sky fibers that correlate with FIR dust emission. In the optically thin limit, we expect FIR dust emission to be the product of the dust column density and the incident starlight intensity. That is, we use the FIR emission, and not just the dust density, because we are interested in both where there is more dust to do the scattering and more incident starlight to be scattered. In detail, the proportionality to incident starlight may depend on temperature, because the temperature dependence of thermal dust emission at 100 $\mu$m and the dust temperature as a function of the strength of the incident starlight intensity do not exactly cancel (c.f.~Equation (1) of \citetalias{Brandt_2012_ApJ}).



However, in practice, our current temperature maps are an order of magnitude lower resolution than our $I_{100\mu\text{m}}$ maps (0.7 $\deg$ vs.~4 arcmin, see Section \ref{sec:FIRmap}) and so $T_{\text{dust}}$ does not measurably vary over a petal (0.8 $\deg^2$). More precise temperature corrections might be required in future work for surveys with larger fields of view and would be enabled by high angular resolution, precision temperature measurements from a future FIR mission like PRIMA \citep{Glenn_2025_JATIS}. For now, we proceed with the approximation that the DGL is linearly correlated with $I_{100\mu\text{m}}$. In addition to the scattered light component, this correlation spectrum could include direct emission from species with densities that correlate with the dust being traced by the FIR emission. These components will be further discussed in Sections \ref{sec:atomlines} and \ref{sec:h2lines}.

\subsection{Linear Fits} \label{sec:linfit}

\paragraph{Pure Linear Model} 
The simplest model to extract a DGL spectrum is to correlate optical intensity with FIR intensity assuming a direct underlying proportionality; this is the model used by \citetalias{Brandt_2012_ApJ}. Extracting the DGL under such a linear model requires three key assumptions: (1) the dust probed is optically thin, (2) $I_{100\mu\text{m}}$ accurately traces the product of the dust column and the incident starlight, and (3) no confounding components are correlated with dust FIR emission. Because we do not trust inter-spectrograph calibrations at the level necessary for this work, and we wish to mitigate systematics between different exposures, we only ask what signal in each petal-exposure is correlated with changes in the FIR dust emission across that petal-exposure. Mathematically, this can be expressed as
\begin{multline} 
\label{eq:linDGLmodel}
    \lambda I^{\rm sky, DESI}_{\lambda,j,p} 
        - \langle\lambda I^{\rm sky, DESI}_{\lambda,j,p}\rangle_{p} = \\
        \alpha_{\lambda} \nu [I^{100 \mu{\rm m, IRIS}}_{\nu,j,p}
        - \langle I^{100 \mu{\rm m, IRIS}}_{\nu,j,p} \rangle_{p}]
\end{multline}
which we abbreviate as
\begin{align}
    y_{\lambda,j,p} &= \alpha_{\lambda} x_{j,p} .
    \label{eq:linDGLmodel_abbrev}
\end{align}
In Equation \eqref{eq:linDGLmodel}, $\lambda I^{\rm sky, DESI}_{\lambda,j,p}$ are the sky fiber measurements from DESI in units of 10$^{-4}$ erg\,s$^{-1}$\,cm$^{-2}$\,sr$^{-1}$, calibrated as described in Section \ref{sec:desisky}, where $j$ indexes each of the individual sky fibers for a given petal-exposure $p$ and $\langle\lambda I^{\rm sky, DESI}_{\lambda,j,p}\rangle_{p}$ is the mean value over a given petal-exposure. $I^{100 \mu{\rm m, IRIS}}_{\nu,j,p}$ is the FIR emission at 100 $\mu$m in MJy\,sr$^{-1}$ at each fiber location from the IRIS processing of the IRAS data with $\nu = 3$ THz (100 $\mu$m). $\langle I^{100 \mu{\rm m, IRIS}}_{\nu,j,p} \rangle_{p}$ is the average value of the FIR emission over all fibers being used for the correlation in a given petal-exposure $p$. The slope in this linear relationship is the DGL spectrum we are after, $\alpha_{\lambda}$, and is a unitless ratio of optical and FIR surface brightnesses $\lambda I_{\lambda}/\nu I_{\nu}$. As in \citetalias{Brandt_2012_ApJ} and \citetalias{Chellew_2022_ApJ} we call $\alpha_{\lambda}$ a ``correlation spectrum.'' In Equation \eqref{eq:linDGLmodel_abbrev} we group terms in the left-hand and right-hand side of Equation \eqref{eq:linDGLmodel}, excluding the slope, into $y_{\lambda,j,p}$ and $x_{j,p}$, respectively, for ease of notation. This reformulation also emphasizes that under the approximation that the dust probed is optically thin, $x_{j,p}$ is independent of wavelength (in contrast to the non-linear model described below).

While there are many ways to fit a line, it is extremely useful in our case to compute the correlations from each petal-exposure and combine them later, both because of data volume (we want to process millions of spectra from hundreds of thousands of petal-exposures), and because we want to build maps of spatial variations in the DGL with variable spatial binning schemes. We can estimate the maximum likelihood $\alpha_{\lambda}$ via 
\begin{ceqn}
\begin{align} \label{eq:MLalpha}
    \alpha_{\lambda} &= \left(\sum_{j,p} \frac{y_{\lambda,j,p} x_{j,p}}{\sigma^2_{y,\lambda,j,p}} \right) \left(\sum_{j,p} \frac{x^2_{j,p}}{\sigma^2_{y,\lambda,j,p}} \right)^{-1},
\end{align}
\end{ceqn}
and we use jackknifing as described in Section \ref{sec:errors} to estimate the uncertainty on $\alpha_{\lambda}$.

Here $\sigma^2_{y,\lambda,j,p}$ are the variances on $y_{\lambda,j,p}$ coming from the DESI pipeline. Note that Equation \ref{eq:MLalpha} is simply the ratio of two independent sums. Computationally, we choose to sum over all $j$ in a given petal-exposure $p$, storing the numerator and denominator of Equation \ref{eq:MLalpha} per petal-exposure. This (1) affords us massive parallelization over $\lambda$ and $p$ and (2) allows us to compute arbitrary subsets (specifically spatial subdivisions) for the DGL by limiting which petal-exposures are included in the sum over p later during our analysis. We release the correlations per petal-exposure in Section \ref{sec:dataavil} so custom subsets can be made by the community.

\paragraph{Non-Linear Correction} In order to build signal from higher extinction sightlines, relaxing assumption (1) above, we need to add in the usual optical-depth correction $\beta_{\lambda}$ to account for self-absorption of the DGL by the dust scattering it. This is described in \citetalias{Chellew_2022_ApJ} as
\begin{ceqn}
\begin{align} \label{eq:betalambda}
    \beta_{\lambda} &= \frac{1-\exp\left[-\tau_{\lambda}\right]}{\tau_{\lambda}} .
\end{align}
\end{ceqn}
Our expected DGL signal is no longer proportional to $I^{100 \mu{\rm m, IRIS}}_{\nu,j,p}$, but is instead proportional to $\beta_{\lambda,j,p} I^{100 \mu{\rm m, IRIS}}_{\nu,j,p}$. In the optically thin limit ($\tau_{\lambda} \rightarrow 0$), $\beta_{\lambda} = 1$ recovering Equation \ref{eq:linDGLmodel}. In the optically thick limit ($\tau_{\lambda} \rightarrow \infty$), $\beta_{\lambda} = \tau_{\lambda}^{-1}$ and the DGL signal per unit FIR emission is reduced because some of the DGL is absorbed by the dust after scattering and does not make it to the observer. For context, only $\sim3\%$ of sky spectra passing our cuts have $\beta<0.8$ in the bluest wavelength bin, so this correction is only significant for a small fraction of the sample.

At the cost of making $x$ wavelength dependent, we can still use Equation \ref{eq:linDGLmodel} for higher optical depth sightlines by applying this correction, which can be expressed as changing to using the $x_{\lambda,j,p}$ defined by
\begin{align} 
\label{eq:xwavedepend}
    x_{\lambda,j,p} = \nu [I^{100{\rm \mu m, IRIS}}_{\nu,j,p}\beta_{\lambda,j,p} - \langle I^{100 \mu{\rm m, IRIS}}_{\nu,j,p} \beta_{\lambda,j,p}  \rangle_{p}] .
\end{align}
Throughout this article we report $\alpha_{\lambda}$ measured using Equation \ref{eq:MLalpha} with $x_{\lambda,j,p}$ from Equation \ref{eq:xwavedepend} instead of $x_{j,p}$. This only slightly modifies our interpretation of $\alpha_{\lambda}$ when predicting the absolute DGL signal, because it now depends on $\beta_{\lambda} I^{100 \mu{\rm m, IRIS}}_{\nu}$ or variations in that quantity.

Applying this correction requires us to correct the FIR emission using a wavelength-dependent optical depth to handle the nonlinear regime. To do so, we rely on the conversions established by \citet{Schlegel_1998_ApJ}. First, we multiply by the ``X-factor'' produced by SFD from DIRBE 100 $\mu$m and 240 $\mu$m emission to correct the observed emission for temperature variations. This converts the emission at 100 $\mu$m to the emission we would expect if all of the dust were at a temperature of 18.2 K, assuming the dust obeys modified black-body emission with $\beta=2$. However, DIRBE only has an angular resolution of 42 arcmin, which is much lower resolution than IRAS (6.1 arcmin in SFD, 4 arcmin in IRIS), and thus we cannot account for small angular-scale variations in temperature.

Next, we convert from MJy/sr to $E(B-V)$ by multiplying by 0.0184 \citep{Schlegel_1998_ApJ}. To convert to extinction in $A(V)$, we multiply $E(B-V)$ in SFD units by 2.742 according to the calibrations of \citet{Schlafly_2011_ApJ}, which assumes the extinction curve from \citet{Fitzpatrick_1999_PASP} (F99) with $R(V) = 3.1$. Finally, we use this same extinction curve to obtain the wavelength-dependent extinction and convert to optical depth by change of logs (Equation \ref{eq:optical_depth}),
\begin{ceqn}
\begin{align} \label{eq:optical_depth}
    \tau(\lambda) = \frac{\ln{10}}{2.5} \times \left[ \frac{A(\lambda)}{A(V)} \right]_{F99} \times A(V) .
\end{align}
\end{ceqn}

\subsection{Uncertainties} \label{sec:errors}

Jackknife resampling is a statistical technique to estimate and correct the bias of a given statistic $f(x_1,...,x_N)$ on a fixed sample of size $N$ by computing that statistic over all subsets of the sample that leave out one observation or set of related observations. By comparing the mean of that statistic over all subsets $\widetilde{\theta}_{\rm{jack}}$ (Equation \ref{eq:jack_average}) with the statistic on the full dataset $\hat{\theta}$, one obtains a bias-reduced estimate for that statistic $\hat{\theta}_{\rm{jack}}$ \citep[Equation \ref{eq:jack_bias_cor}, ][]{berger2006adjusted}. Similarly, one can estimate the covariance of the jackknifed estimate $\hat{C}(\hat{\theta}_{\rm{jack}})$ for a statistic given the covariance of that statistic on the leave-one-out subsets \citep[Equation \ref{eq:jack_covar},][]{berger2005jackknife}.

\begin{ceqn}
\begin{align} \label{eq:jack_average}
    \widetilde{\theta}_{\rm{jack}} &= \frac{1}{N} \sum_{i=1}^{N} f(x_1,...,x_{i-1},x_{i+1},...,x_N) \\
    \label{eq:jack_bias_cor}
    \hat{\theta}_{\rm{jack}} &= N\hat{\theta}-(N-1)\widetilde{\theta}_{\rm{jack}} \\
    \label{eq:jack_covar}
    \hat{C}(\hat{\theta}_{\rm{jack}}) &= \frac{\left(N-1\right)^2}{N} \frac{\left(\widetilde{\theta}_{\rm{jack}} - \hat{\theta}_{\rm{jack}} \right)^{T}\left(\widetilde{\theta}_{\rm{jack}} - \hat{\theta}_{\rm{jack}} \right)}{N-1}
\end{align}
\end{ceqn}

We estimate uncertainties in all versions of the DGL correlation spectrum reported by jackknifing over exposures, not petal-exposures. Many systematics, like seeing, sky brightness, and cloudiness, are strongly correlated over all petals of an exposure. By jackknifing over exposures rather than petals, leaving out all petals associated with a given exposure, we can better account for these systematics. Unless otherwise noted, we use the diagonal approximation to this covariance matrix for the uncertainties in spectral fitting throughout the paper. However, we report and make available the full covariance matrix for use in future work.

This jackknifed uncertainty still only reflects our uncertainty on the mean spectrum, either the all-sky spectrum or the spectrum in a given HEALPix pixel. Broadly, the uncertainty (standard deviation) increases by an average factor of 1.6 for the all-sky spectrum as a result of jackknifing, though the change in uncertainties does have significant variation with wavelength. Our final reported uncertainty depends on the signal to noise of the DGL components in our sky spectra, the underlying variability of the samples around the mean (for example due to spatial variations in the DGL we might be averaging over), and the sample size. In general, we find spatial variability to be large relative to our uncertainties on the all-sky spectrum, which we discuss in more detail in Section \ref{sec:spatialvarylines}.

\subsection{Systematics} \label{sec:systematics}

\paragraph{Slope Bias from FIR Scatter} In the linear fitting procedure described in Section \ref{sec:linfit}, we neglect uncertainties on the independent variable. Neglecting scatter in the independent variable tends to bias the inferred slope low. Several fitting procedures to avoid this bias exist when the x-uncertainties are known \citep{Hogg_2010_arXiv,Saydjari_2026_ApJ}. When the x-uncertainties are not known well, one way to estimate the amplitude of this bias is to repeat the linear fit after cuts limiting the input data to a higher ``x'' signal-to-noise ratio subset of the data, which asymptotically approaches an unbiased estimate of the slope. 

\begin{figure}[htb]
    \centering
    \includegraphics[width=\linewidth]{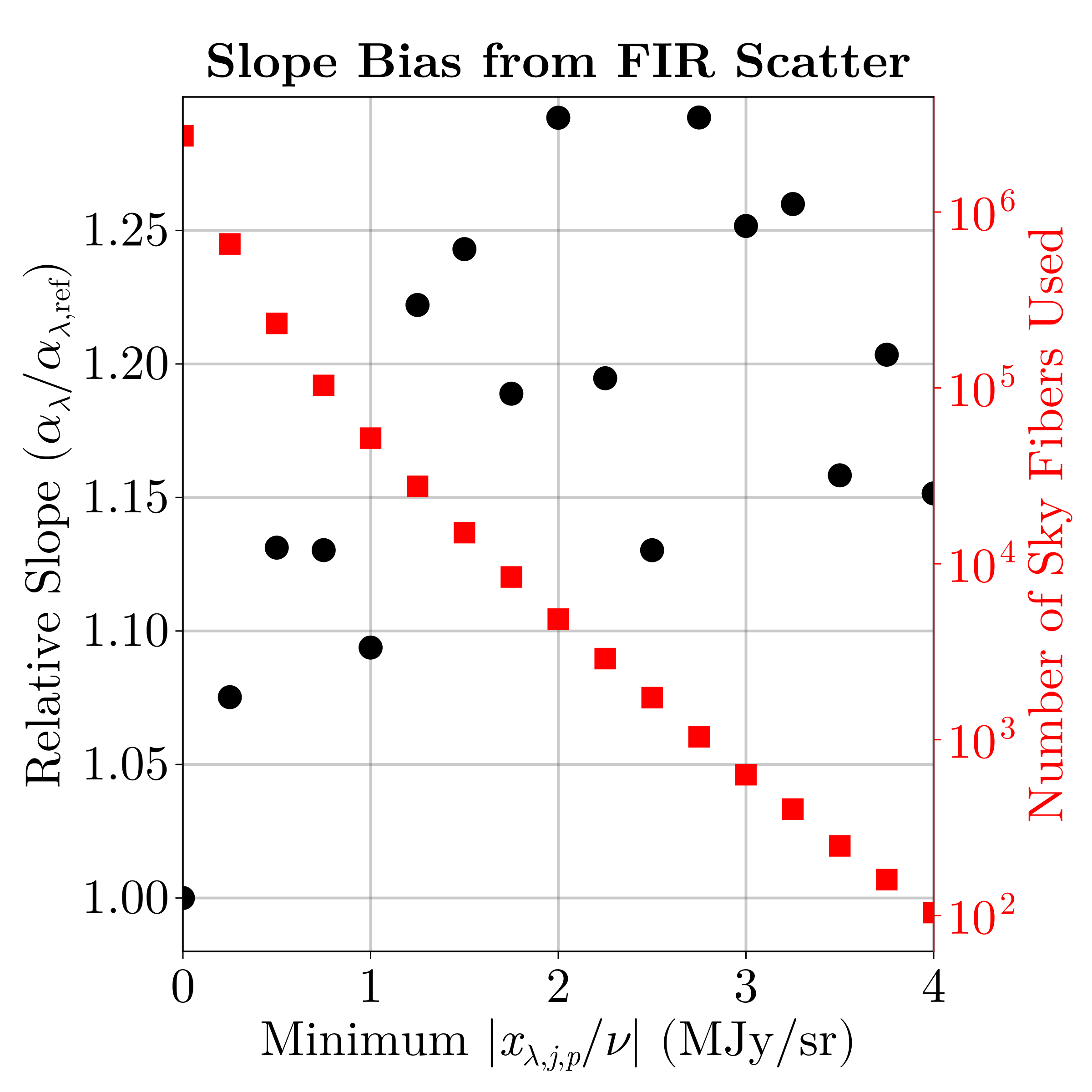}
    \caption{Ratio of mean $\alpha_{\lambda}$ for $6600 < \lambda < 6700$ \r{A} inferred only using data for which $|x_{\lambda,j,p}/\nu| > x_{\text{min}}$ MJy sr$^{-1}$ to reference value inferred with no $x_{\text{min}}$ cut (black circles, left y-axis). We use the values with $x_{\lambda,j,p}/\nu > 2$ MJy sr$^{-1}$ to estimate the central value and scatter of the ``correction factor'' $C$ to account for a bias in the linear fit slope due to unmodeled scatter in the FIR emission (``x'' in the linear fit). The total number of sky fibers used in the DGL spectrum measurement as a function of the increasingly strict FIR emission cut is also shown (red squares, right y-axis).}
    \label{fig:Csystematic}
\end{figure}

We repeat the procedure described in \citetalias{Brandt_2012_ApJ}, computing the mean $\alpha_{\lambda}$ for $6600 < \lambda < 6700$ \r{A}, but only using fibers for which $|x_{\lambda,j,p}/\nu| > x_{\text{min}}$ MJy sr$^{-1}$ for $0 < x_{\text{min}} < 5$ MJy sr$^{-1}$, in 0.3 MJy sr$^{-1}$ steps. For computational ease, we restricted this test to an arbitrary 40k petal-exposure subset of the full $\sim170$k sample. We found similar results when changing the subset and considering wavelength ranges in the other DESI spectrograph arms ($4500 < \lambda < 4600$ \r{A}, $8100 < \lambda < 8200$ \r{A}). The ratio of $\alpha_{\lambda}$ to the value measured at $x_{\text{min}}=0$ is shown in Figure \ref{fig:Csystematic} as a function of $x_{\text{min}}$. By considering fits limited to $x_{\lambda,j,p}/\nu > 2$ MJy sr$^{-1}$, we find a median ``correction factor'' C = $1.2\pm0.07$, with the uncertainty estimated from the spread over the same x-range. 

This value is significantly lower than those found by \citetalias{Brandt_2012_ApJ} and \citetalias{Chellew_2022_ApJ} of $C=1.9-2.1$, who also found that these values of $C$ estimated as above were consistent with corrections relative to theoretical radiative transfer calculations. Our value of $C$ is far more consistent with biases in $\alpha_{\lambda}$ one expects based on the estimated noise level in the IRIS map (formally 0.06 MJy\,sr$^{-1}$, though practically limited by zodiacal light subtraction) and filamentary structure in the FIR emission at the scale of our fibers (1$.\!\!''$52) unresolved by IRIS (4$'$), which are expected to be $\sim10\%$ effects. This difference in correction factors between BOSS and DESI could be the result of better fluxing (e.g.,~accounting for fiber positioning errors) or better calibrations for uniform surface brightness measurements (e.g.,~Section \ref{sec:desisky}). Higher angular resolution and more sensitive FIR emission maps, like those possible with a PRIMA all-sky survey \citep{Saydjari_2025_prim}, should further mitigate this source of bias and drive these correction factors closer to 1.

Throughout this paper, we apply our correction factor of $C$ in all plots and values reported. This correction only matters for absolute measurements, when converting $\alpha_{\lambda}$ to an estimated optical surface brightness for a given FIR surface brightness; however we include it throughout for self consistency. Because of the $\sim10\%$ uncertainty on $C$, our true uncertainty on absolute measurements should have 10\% uncertainties added in quadrature (which we have not added to reported error bars). Because these errors come from the FIR map, they are perfectly correlated between optical DGL wavelengths. These facts emphasize the value of relative measurements (e.g.,~line equivalent widths, or EWs).

\paragraph{Corrections for Line Ratios} The precise interpretation of spectra that are composites of populations of stars and emission lines (e.g.,~galaxy spectra) can require calibration after individual lines are measured. We discuss two such corrections here: (1) stellar absorption and (2) extinction.

At low spectral resolution, equivalent width measurements of Balmer series lines can be viewed as the sum of a positive emission line component from \ion{H}{2} regions and a negative absorption component from the hydrogen lines in stellar spectra. \citetalias{Brandt_2012_ApJ} and \citetalias{Chellew_2022_ApJ} apply a correction to their measured equivalent widths for H$\alpha$ and H$\beta$ to remove the stellar absorption contribution. They predict the stellar contribution from a linear relationship with $\delta_{4000}$ (see Section \ref{sec:d4000}) fit to model spectra. Unfortunately, this simple additive picture for the emission and absorption EW is incomplete at higher spectral resolution, when the narrower interstellar emission is resolved within the broader stellar absorption trough. 

\begin{figure*}[htb!]
    \centering
    \includegraphics[width=\linewidth]{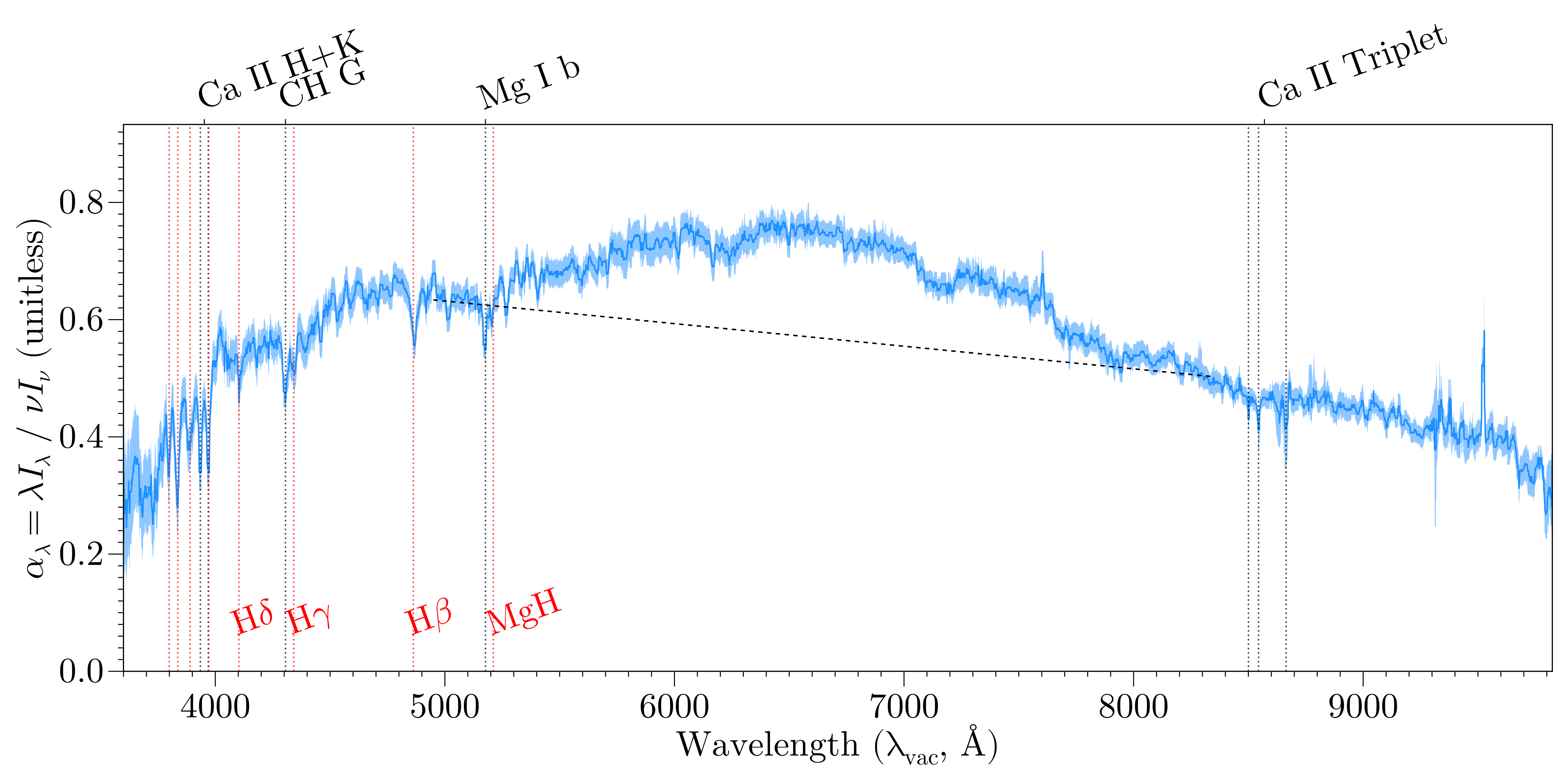}
    \caption{DGL continuum obtained by masking emission lines and smoothing with the 21 pixel (16.8 \r{A}) wide moving median kernel. Solid line represents central value with shading representing $3\sigma$ uncertainties. Vertical guidelines indicate strong stellar absorption lines that are clearly observed. Red guidelines are used to denote the Balmer series, where only H$\beta$ through H$\delta$ are labeled to avoid label crowding. Dashed guideline highlights red emission in excess of expected stellar continuum, a detection of ``extended red emission.''}
    \label{fig:continuum}
\end{figure*}

The H$\beta$ (and higher Balmer series) stellar absorption features persisting after masking the narrow emission lines in Figure \ref{fig:continuum} demonstrate that we are in this regime. In this limit, stellar absorption can cause an apparent increase in the Balmer line equivalent width by reducing the continuum in the denominator of the equivalent width calculation. Directly applying the $\delta_{4000}$ from \citetalias{Brandt_2012_ApJ} to our data set leads to Balmer ``decrements'' (see above) that imply negative extinction for the radiation field illuminating the dust, which is clearly nonphysical. Therefore, we do not apply a correction for stellar absorption throughout this work. Rather than degrade our resolution to make the aforementioned correction work, we suggest that future work should leverage our improved measurements of the DGL correlation spectrum to perform more precise Galaxy spectral modeling of the DGL, jointly fitting a stellar population synthesis and emission line model (for example, with \texttt{prospector}).\footnote{\url{https://prospect.readthedocs.io/en/stable/}}

Second, the preferential extinction by dust at blue compared to red wavelengths can alter observed emission line ratios. The current state-of-the-art has demonstrated that connecting observed to emitted emission line ratios correctly requires full 3D radiative transfer accounting for the location of the primary ionizing sources and the 3D distribution of dust \citep{Jin_2022_ApJL, Jin_2023_ApJ}. In our case, this is even harder because we are only measuring the scattered, not the direct, component. This means that any correction should include the wavelength dependence of dust scattering, not just the dust extinction. However, in extragalactic work, it is common to approximate the influence of dust as if it were a uniform screen between the observer and the galaxy. Then one can assume an extinction curve, learn an $A_V$ assuming a line ratio is known, and correct the remaining line ratios. 

The most common method is to use the ``Balmer decrement,'' assuming the ratio of H$\alpha$ and H$\beta$ is set by theoretical case B recombination for a given electron temperature and density. Here we assume $T_e = 9000$ K, $n_e = 100$ cm$^{-3}$ which gives a theoretical H$\alpha$/H$\beta= 2.89$. Then we can compute $A_V$ using Equation \ref{eq:balmerDec}, and de-redden our observed line surface brightnesses using the same extinction curve, for which we use CCM89 with $R_V=3.1$ \citep{Cardelli_1989_Extinction_Curve_CCM89}. Here $E(\lambda)$, with $\lambda$ in \r{A}, represents the extinction in magnitudes given by the assumed extinction law.
\begin{ceqn}
\begin{align} 
    \label{eq:balmerDec}
    A_V = \frac{-2.5 \left( \log_{10} \left(I_{\text{H}\alpha}/I_{\text{H}\beta} \right) - \log_{10} \left( 2.89 \right)\right)}{E(6564.61) - E(4862.72)}
\end{align}
\end{ceqn}
By default, we prefer to present the raw measured values in the text and only apply this extinction correction when necessary for interpretation. In those cases, we explicitly note its use.

\begin{deluxetable}{r@{\hskip 3pt}l r@{\hskip 3pt}l r@{\hskip 3pt}l}[t]
\setlength{\tabcolsep}{-2pt}
\small
\tablewidth{0pt}
\tablecaption{Line Fitting Priors \label{tab:priors}}
\tablehead{
\multicolumn{2}{c}{Single Line} & \multicolumn{2}{c}{Doublet} & \multicolumn{2}{c}{Multiple Lines}
}
\startdata
$D$ & $= [\lambda_0\pm 400]$ & $D$ & $= \cup_i [\lambda_{0_i}\pm 400]$ & $D_i$ & $= [\lambda_{0_i}\pm 400]$ \\
$\tfrac{\alpha}{\sqrt{2\pi}}$ & $= \max_D |I_\lambda|$ & $\tfrac{\alpha}{\sqrt{2\pi}}$ & $= \max_D |I_\lambda|$ & $\tfrac{\alpha_i}{\sqrt{2\pi}}$ & $= \max_{D_i} |I_\lambda|$ \\
$F$ & $\sim \mathcal{U}(0,5\alpha)$ & $F_i$ & $\sim \mathcal{U}(\pm5\alpha)$ & $F_i$ & $\sim \mathcal{U}(\pm5\alpha_i)$ \\
$\lambda$ & $\sim \mathcal{U}(D)$ & $\lambda$ & $\sim \mathcal{U}(D)$ & $\lambda_i$ & $\sim \mathcal{U}(\lambda_{0_i} \pm 100)$ \\
 & & $\tfrac{\Delta\lambda}{\Delta\lambda_0}$ & $\sim \mathcal{U}(0.85,1.15)$ & & \\
$\sigma$ & $\sim \mathcal{U}(0.2,2.0)$ & $\sigma$ & $\sim \mathcal{U}(0.8,1.2)$ & $\sigma$ & $\sim \mathcal{U}(0.2,1.5)$ \\
$c$ & $\sim \mathcal{U}(-5,10)$ & $c$ & $\sim \mathcal{U}(-5,10)$ & $c_i$ & $\sim \mathcal{U}(0,1)$ \\
$s$ & $\sim \mathcal{U}(-1,1)$ & $s$ & $\sim \mathcal{U}(-1,1)$ & & \\
$\varsigma^2$ & $\sim \mathcal{N}(0,2)[0.5,\infty]$ & $\varsigma^2$ & $\sim \mathcal{N}(0,2)[0.5,\infty]$ & $\varsigma^2$ & $\sim \mathcal{N}(0,2)[0.5,\infty]$ \\
\enddata
\tablecomments{All quoted wavelength ranges of the form $\lambda \pm$ are in terms of velocities in km\,s$^{-1}$. Units for parameters are defined in text.}
\end{deluxetable}

\subsection{Spectral Line Fitting} \label{sec:speclinfit}

We perform Gaussian fits to the DGL spectral lines by evaluating model-data comparisons using MCMC with a No U-Turn Sampler (NUTS) with a total acceptance ratio of 0.65 and 25,000 samples as implemented in \texttt{Turing.jl}.\footnote{\url{https://github.com/TuringLang/Turing.jl}} We report the median of the marginal posterior distribution for each parameter, with lower and upper uncertainties representing the difference between the median at 16$^\text{th}$ and 84$^\text{th}$ percentile of the marginal posterior, respectively. We convert the posterior samples for line central wavelengths ($\lambda$) to velocities ($v$) relative to their rest-frame vacuum wavelength ($\lambda_0$) via Equation \ref{eq:lambda2v}, where $c$ is the speed of light in km/s. 

\begin{ceqn}
\begin{align} 
    \label{eq:lambda2v}
    z &= \frac{\lambda - \lambda_0}{\lambda_0}
    \quad\quad\quad   v = \frac{(z+1)^2-1}{(z+1)^2+1}c
\end{align}
\end{ceqn}

Because our DGL correlation spectrum $\alpha_{\lambda} = \lambda I_{\lambda}/\nu I_{\nu}$ is unitless, we multiply our spectrum by $\nu/\lambda$ prior to line fitting in order to obtain line surface brightness in units of erg\,s$^{-1}$\,cm$^{-2}$\,sr$^{-1}$ for a FIR emission of 1 MJy sr$^{-1}$. In order to obtain equivalent widths, we divide by the posterior mean of the constant background term at the line center. When making spatial maps, we also report the median spectral signal-to-noise ratio on a narrower range around the line center $\lambda_0 \pm 200$ km/s for quality control purposes.

For a single spectral line, our model is that of a Gaussian profile with free parameters $F$, the area under the curve in erg\,s$^{-1}$\,cm$^{-2}$\,sr$^{-1}$, $\lambda$ the center wavelength of the Gaussian in \r{A}, and $\sigma$ the Gaussian width in \r{A}. We also include a background model with a constant term $c$ in erg\,s$^{-1}$\,cm$^{-2}$\,sr$^{-1}$\,\r{A}$^{-1}$ and slope $s$ in erg\,s$^{-1}$\,cm$^{-2}$\,sr$^{-1}$\,\r{A}$^{-2}$ pivoted around the rest frame central wavelength of the line. Our fits are restricted to a local domain $D$ that is within 400 km s$^{-1}$ of the rest frame line center. The prior on $F$ is positive, uniform (denoted by $\mathcal{U}$) up to $5\times$ the extremal value in the domain, $\lambda$ is uniform over the domain, and $\sigma$ is set to a reasonable range of $0.2-2$ \r{A}. The background terms are given wide uniform priors, $c$ from -5 to 10 erg\,s$^{-1}$\,cm$^{-2}$\,sr$^{-1}$\,\r{A}$^{-1}$ and $s$ from -1 to 1 erg\,s$^{-1}$\,cm$^{-2}$\,sr$^{-1}$\,\r{A}$^{-2}$. The priors on our model parameters are summarized in Table \ref{tab:priors}. 

In addition to these model parameters, we include an uninformative prior on a multiplicative rescaling $\varsigma$ of our measurement variances, which is a normal distribution $\mathcal{N}(0,2)$ (mean 0, variance 2) truncated at 0.5 to prevent any extreme reductions in measurement uncertainties. Variance priors provide important flexibility used during model fitting and also serve as a check on the uncertainty calibration of measurements; for more, see \citet{gelman2006prior}. While we provide a full jackknifed covariance matrix for the DGL, all of our spectral line fits use only the diagonal of this covariance matrix and thus assume each wavelength bin of the DGL is uncorrelated.

When fitting doublets, our model is modified to include two Gaussian profiles with the same $\sigma$, but independent line strengths. The centroids of the lines are parameterized by shifts of the mean wavelength of the doublet $\lambda$ and changes in the doublet separation $\Delta\lambda$, where we place a fractional prior on the latter compared to the nominal spacing of the doublet. Fitting the doublets jointly with MCMC is important if one wants to measure line ratios because it is necessary to capture covariances between the line amplitudes. In addition, many of the doublets we examine have significant overlap between the two lines (are closely separated). 

\begin{deluxetable}{ccc}[b]
\tablewidth{0pt}
\tablecaption{DGL Integrated Surface Brightness at $I_{100\mu\text{m}} =1$ MJy/sr\label{tab:contAbs}}
\tablehead{
\colhead{\shortstack{Photometric \\ Filter}} & \colhead{\shortstack{Surface Brightness \\ $\text{erg s}^{-1} \text{cm}^{-2}\text{sr}^{-1}$}} & \colhead{\shortstack{Flux ($1''$ seeing) \\ AB mag}}
}
\startdata
LSST g-band & $5.4 \times 10^{-6}$ & 26.7 \\
LSST r-band & $4.8 \times 10^{-6}$ & 26.2 \\
LSST i-band & $3.0 \times 10^{-6}$ & 26.2 \\
LSST z-band & $1.6 \times 10^{-6}$ & 26.3 \\
\enddata
\end{deluxetable}

Similarly, when interested in line ratios of multiple lines (as in Sections \ref{sec:spatialvarylines} and \ref{sec:h2lines}), we use joint fits. In these cases, we modify our model to be the union ($\cup_i$) of independent piecewise Gaussian line profiles, independent constant backgrounds, and independent priors. The only shared parameter between the lines is the Gaussian linewidth, which has a uniform prior from 0.2 to 1.5 \r{A}. For the per-pixel line-ratio measurements of Section \ref{sec:spatialvarylines}, we jointly fit the five lines [\ion{N}{2}] $\lambda\lambda$6550,6585, H$\alpha$ $\lambda$6565, and [\ion{S}{2}] $\lambda\lambda$6718,6733 in this way; the close spacing of the H$\alpha$ + [\ion{N}{2}] complex makes a joint fit preferable for deblending, and the joint posterior captures the covariances needed for rigorous line-ratio uncertainties.

\section{All-Sky DGL Correlation Spectrum} \label{sec:allskylines}

\begin{figure*}[htb!]
    \centering
    \includegraphics[width=0.8\linewidth]{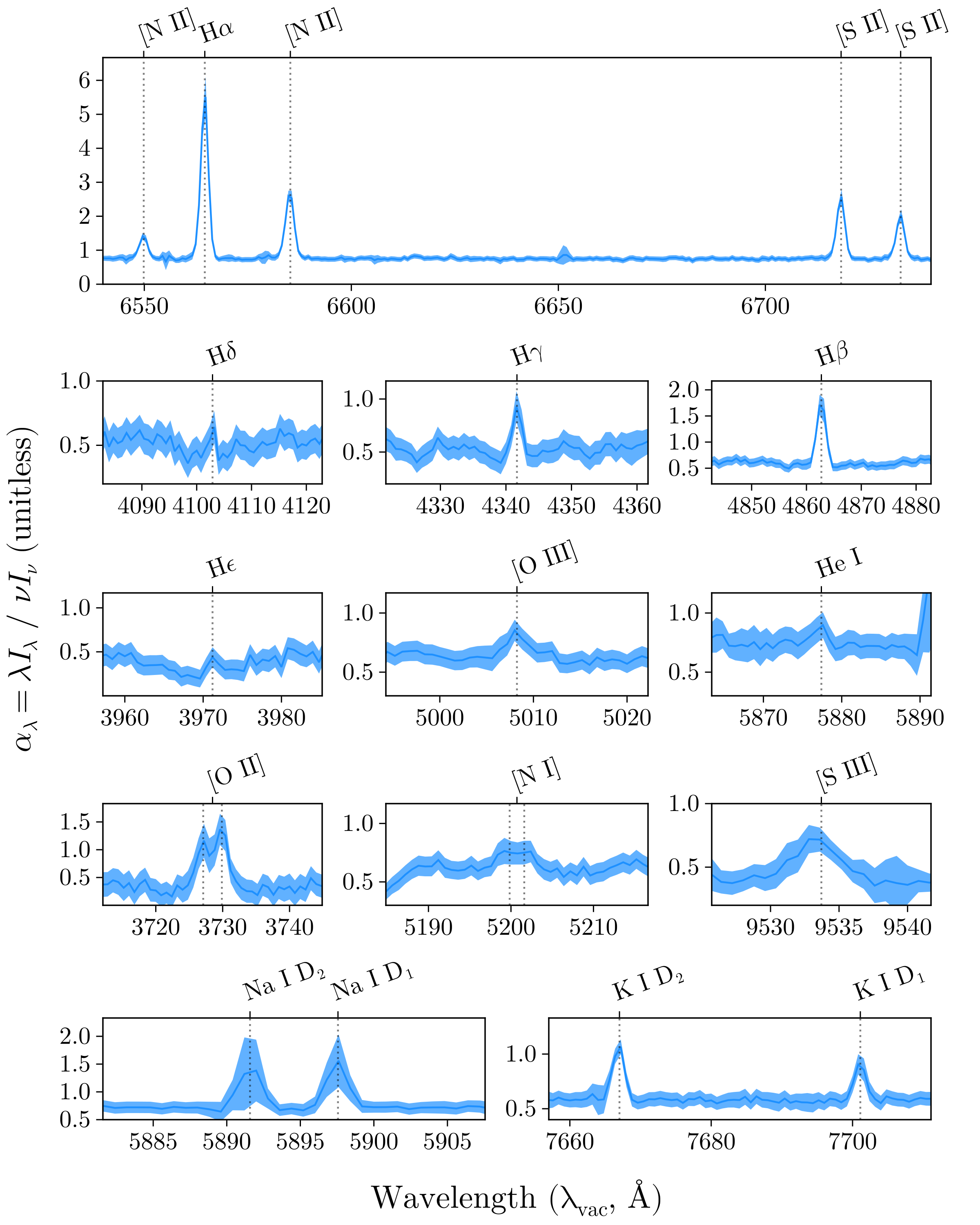}
    \caption{All-sky DGL correlation spectrum (solid line) with 3-$\sigma$ uncertainty contours (shaded). Sub-panels highlight lines of interest, with species labels and guidelines at the rest frame wavelengths. Spectrum is the unitless ratio of optical scattered light in DESI $\lambda I_\lambda$ and 100$\mu$m FIR emission from IRAS $\nu I_{\nu}$. Marginal detections of [\ion{O}{1}] $\lambda6302$, [\ion{S}{3}] $\lambda9071$, 5-2 S(1) $\text{H}_2$ and non-detections of [\ion{O}{3}] $\lambda4960$ and \ion{He}{1} $\lambda3890$ are shown in Appendix \ref{sec:margDetect}.}
    \label{fig:emission_lines}
\end{figure*}

We start by combining all of the data in the sample (Section \ref{sec:desisky}) into a single all-sky DGL correlation spectrum, where we show the continuum in Figure \ref{fig:continuum} and sub-panels focusing on specific lines of interest in Figure \ref{fig:emission_lines}. The shaded regions in Figures \ref{fig:continuum} and \ref{fig:emission_lines} represent 3-$\sigma$ contours from the diagonal approximation to the DGL uncertainty covariance matrix. The DGL correlation spectrum is unitless with a value that represents the surface brightness of the DGL as a fraction of the surface brightness of the 100 $\mu$m FIR dust emission.

Broadly, we see a continuum that resembles the spectrum resulting from stellar population synthesis models and the classic strong emission lines prevalent in \ion{H}{2} regions. We also discuss more subtle astrophysical features present such as extended red emission and interstellar neutral atomic lines in subsequent sections. At some wavelengths, such as 6555.0, 6578.5, and 9521.5~\r{A}, we see features in our DGL spectrum that can be attributed to the incomplete rejection of sky emission lines by the procedure described in Section \ref{sec:DGLModel}. However, this normally occurs with an accompanying increase in the reported uncertainties in the DGL spectrum and these can be easily identified by overplotting a mean sky spectrum. Rarely, we see inflated variance in the DGL spectrum for which we do not have a clear assignment, which occurs most notably at 6651.6~\r{A}.

\begin{deluxetable*}{lccccc}[t]
\tablewidth{0pt}
\tablecaption{All-Sky DGL Line Measurements \label{tab:neblines}}
\tablehead{
\multirow{2}{*}{Line (\r{A})} & \multirow{2}{*}{\shortstack{Surface Brightness\tablenotemark{a} \\ $10^{-9} \text{erg s}^{-1} \text{cm}^{-2}\text{sr}^{-1}$}} & \multirow{2}{*}{$\dfrac{I_{\lambda_{\rm long}}}{I_{\lambda_{\rm short}}}$} & \multicolumn{3}{c}{Equivalent Width (\r{A})} \\
\cline{4-6}
\colhead{} & \colhead{} & \colhead{} & \colhead{This Work} & \colhead{\citetalias{Brandt_2012_ApJ}} & \colhead{\citetalias{Chellew_2022_ApJ}}
}
\startdata
H$\delta$ $\lambda$4103  & $1.77\substack{+1.08 \\ -0.81}$ & & $0.55\substack{+0.34 \\ -0.25}$ & & \\
H$\gamma$ $\lambda$4342  & $5.37\substack{+0.59 \\ -0.59}$ & & $1.65\substack{+0.18 \\ -0.18}$ & & \\
H$\beta$ $\lambda$4863  & $15.47\substack{+0.81 \\ -0.82}$ & & $4.57\substack{+0.24 \\ -0.24}$ & $4.8\pm0.7$ & $3.9\pm0.3$ \\
\text{[\ion{O}{3}]} $\lambda$5008  & $4.73\substack{+0.92 \\ -0.90}$ & & $1.31\substack{+0.25 \\ -0.25}$ & $0.8\pm0.6$ & $1.0\pm0.3$  \\
\text{\ion{He}{1}} $\lambda$5877  & $1.75\substack{+0.40 \\ -0.37}$ & & $0.47\substack{+0.11 \\ -0.10}$ & $0.3\pm0.8$ & $0.2\pm0.3$ \\
\text{[\ion{N}{2}]} $\lambda$6550  & $7.60\substack{+0.37 \\ -0.36}$ & & $2.20\substack{+0.11 \\ -0.10}$  & $2.4\pm0.5$ & $1.4\pm0.3$ \\
H$\alpha$ $\lambda$6565  & $48.5\substack{+1.2 \\ -1.2}$ & & $14.18\substack{+0.34 \\ -0.35}$ & $12.5\pm0.5$ & $10.7\pm0.5$  \\
\text{[\ion{N}{2}]}  $\lambda$6585  & $22.12\substack{+0.48 \\ -0.47}$ & & $6.29\substack{+0.14 \\ -0.13}$ & $6.6\pm0.5$ & $5.6\pm0.4$  \\
\text{[\ion{S}{2}]}  $\lambda$6718  & $19.23\substack{+0.37 \\ -0.38}$ & & $5.70\substack{+0.11 \\ -0.11}$ & $5.7\pm0.4$ & $4.6\pm0.4$  \\
\text{[\ion{S}{2}]}  $\lambda$6733  & $13.97\substack{+0.42 \\ -0.42}$ & & $4.23\substack{+0.13 \\ -0.13}$ & $4.3\pm0.4$ & $3.3\pm0.3$  \\
\text{[\ion{S}{3}]} $\lambda$9534  & $3.78\substack{+0.49 \\ -0.47}$ & & $2.96\substack{+0.38 \\ -0.36}$ &  & \\
\hline
\text{[\ion{O}{2}]}  $\lambda\lambda$3727,3730  & $37.6\substack{+2.8 \\ -2.6}$ & $1.32\substack{+0.17 \\ -0.15}$ & $17.3\substack{+1.3 \\ -1.2}$ &  & $16\pm5$ \\
\text{[\ion{N}{1}]}  $\lambda\lambda$5199,5201  & $4.93\substack{+0.48 \\ -0.50}$ & $1.01\substack{+0.25 \\ -0.20}$ & $1.44\substack{+0.14 \\ -0.15}$ &  & \\
\hline
\text{\ion{Na}{1}} D $\lambda\lambda$5892,5898  & $13.4\substack{+1.4 \\ -1.5}$ & $1.11\substack{+0.28 \\ -0.20}$ & $3.75\substack{+0.39 \\ -0.41}$ &  & \\
\text{\ion{K}{1}} $\lambda\lambda$7667,7701  & $6.78\substack{+0.22 \\ -0.21}$ & $0.671\substack{+0.046 \\ -0.043}$ & $3.002\substack{+0.097 \\ -0.095}$ &  & \\
\hline
H$_2$ 5-2 S(2) $\lambda$9116  & $0.65\substack{+0.13 \\ -0.12}$ & & $0.452\substack{+0.093 \\ -0.082}$ &  & \\
H$_2$ 6-3 S(2) $\lambda$9758  & $0.69\substack{+0.16 \\ -0.15}$ & & $0.63\substack{+0.15 \\ -0.14}$ &  & \\
H$_2$ 9-5 S(1) $\lambda$9148  & $0.28\substack{+0.12 \\ -0.12}$ & & $0.202\substack{+0.087 \\ -0.089}$ &  & \\
H$_2$ 4-1 Q(2) $\lambda$9060  & $0.41\substack{+0.13 \\ -0.12}$ & & $0.275\substack{+0.087 \\ -0.083}$ &  & \\
\enddata
\tablenotetext{a}{Because we report for our DGL correlation spectrum a unitless ratio, $\lambda I_{\lambda}/\nu I_{\nu}$, we are reporting a notion of the surface brightness of the lines here as the surface brightness for a given location on the sky with an FIR emission of 1 MJy/sr at 100 $\mu$m. The table sections are, from top: nebular emission lines, nebular doublets, resonantly scattered atomic doublets (Section \ref{sec:atomlines}), and directly emitted H$_2$ lines (Section \ref{sec:h2lines}).}
\end{deluxetable*}

\subsection{Surface Brightness} \label{sec:absmag}

To begin, we consider how big an effect the DGL is, that is its absolute surface brightness. While our absolute calibration has uncertainties around $\pm10\%$  (see Section \ref{sec:systematics}), understanding the amplitude of the DGL is essential to evaluating when this component of the multi-component sky is significant relative to a given signal of interest. To convert $\alpha_{\lambda}$ to an interpretable surface brightness, we report the DGL surface brightness given 1 MJy\,sr$^{-1}$ of 100 $\mu$m emission. This requires dividing $\alpha_{\lambda}$ by $\lambda$ and multiplying by $3 \times 10^{-5}$ erg\,s$^{-1}$\,cm$^{-2}$\,sr$^{-1}$ before integration for a line or band flux. 

To demonstrate the amplitude, we integrate the DGL correlation spectrum against the LSST filter curves with matched wavelength coverage to DESI, which gives $4.8 \times 10^{-6}$ erg\,s$^{-1}$\,cm$^{-2}$\,sr$^{-1}$ for LSST r-band (see Table \ref{tab:contAbs} for more). We also convert to AB magnitudes by integrating the photon-weighted mean flux density with the LSST filter curves to obtain the last column of Table \ref{tab:contAbs}. Here we have assumed a fiducial value for the seeing, a $1''$ full width at half maximum (FWHM) PSF, for illustration purposes. For comparison, the LSST 10-year depths are expected to be 27.4, 27.5, 26.8, and 26.1 mag in g, r, i, and z-band respectively \citep{Bianco_2022_ApJS}. The expected r-band DGL is 26.2 mag (AB), 1.3 mag brighter than the LSST 10-year depth. Even though the DGL is much fainter than the typical night sky background \citep[$\sim21$ mag, ][]{Krisciunas_2007_PASP}, the DGL will have a detectable contribution in modern deep imaging surveys. Any PSF-scale fluctuations in the amount of dust $\gtrsim24$ mmag E(B-V) will likely cause spurious LSST detections, and any fluctuations $>5.5$ mmag E(B-V) will exceed the statistical LSST photometric uncertainties. Toward the Galactic plane, the DGL will be an even more significant contribution.

\citet{Wolstencroft_1966_Natur} measured the DGL in a field at ($\ell$,b)=(46$\degree$,$-$12.3$\degree$), finding a brightness equivalent to $\approx38$ 10$^{\text{th}}$ magnitude stars per square degree. Averaged over their 2.7 deg$^2$ beam, this sightline has $I_{100\mu\text{m}} = 7.4$ MJy\,sr$^{-1}$. Scaling our DGL correlation spectrum to this intensity, our B-band surface brightness is equivalent to $\approx15$ such stars per square degree, about a factor of 2 lower than their measurement. This is reasonable agreement, given the challenges of absolute calibration and contamination from airglow and unresolved starlight in early rocket photometry. The units from \citet{Wolstencroft_1966_Natur} are unusual in modern astronomy. However, these units motivate a related question of how the DGL surface brightness compares to the density of stars below the detection threshold in upcoming surveys.

If we integrate the spectrum of BD$-11\degree$3759, an M3.5V star, we find that the DGL surface brightness (for 1 MJy\,sr$^{-1}$ of 100 $\mu$m emission) in LSST r-band corresponds to $2\times$10$^7$ 27$^{\rm th}$ mag M-dwarfs per square degree. Simply extrapolating star counts per mag with a power law index of $0.35$ from existing surveys predicts 10$^8$-10$^9$ such M-dwarfs, which puts the DGL contribution only 0.5-1.5 orders of magnitude below backgrounds from faint undetected stars in crowded LSST fields. For regions with little spatial variation in the dust, the DGL contribution will likely be removed easily with a smooth ``sky'' background model. However, in regions with large spatial variations in the dust, the DGL could present a highly-structured scattered light background that is difficult to remove. 

\section{Nebular Lines: Dust Scattering} \label{sec:neblines}

We present the equivalent width and surface brightness of all of the strong nebular emission lines found in the DGL in \citetalias{Brandt_2012_ApJ} and \citetalias{Chellew_2022_ApJ} in the top section of Table \ref{tab:neblines}. In addition to those lines, our wavelength coverage extends further into the near-infrared compared to \citetalias{Brandt_2012_ApJ} and \citetalias{Chellew_2022_ApJ}, allowing us to detect [\ion{S}{3}] $\lambda$9534, which gives us access to multiple ionization states of the same species. We also have higher signal-to-noise ratio detections of lower amplitude lines in the Balmer series, detecting H$\gamma$, H$\delta$, and H$\epsilon$, reporting the EWs for the first two. As in \citetalias{Brandt_2012_ApJ} and \citetalias{Chellew_2022_ApJ}, we find no significant detection of [\ion{O}{3}] $\lambda$4960.

\begin{figure*}[htb]
    \centering
    \includegraphics[width=0.82\linewidth]{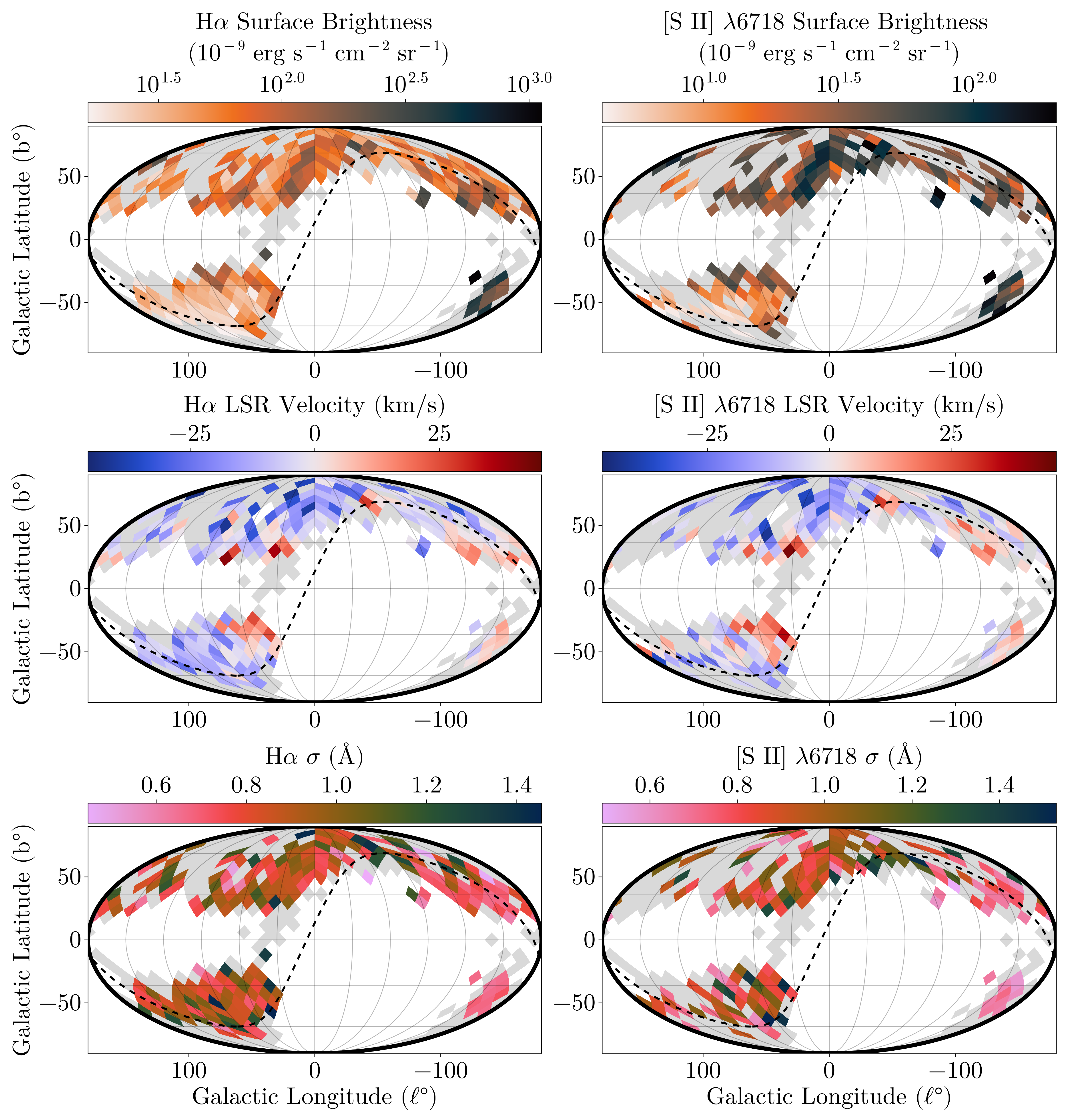}
    \caption{Maps of Gaussian fit parameters for emission lines in the DGL across the DESI footprint on a HEALPix grid with resolution NSIDE $=$ 8. Columns show different lines, left is H$\alpha$ and right is [\ion{S}{2}] $\lambda$6718. Rows show different moments of the Gaussian line profile, top is surface brightness for 1 MJy/sr FIR emission at 100 $\mu$m, middle is LSR velocity, and bottom is the line width. Pixels without any data are shown in white and pixels without significant detections (see text) are shown in light grey. Projections are Mollweide with a dashed line showing the ecliptic plane.}
    \label{fig:line_fits_columns}
\end{figure*}

The line fluxes for the nebular emission lines are presented in Table \ref{tab:neblines}. We can compare these to object line fluxes by multiplying by a typical fiber angular size ($4 \times 10^{-11}$ sr for DESI), which gives an H$\alpha$ line flux of $1.9 \times 10^{-18}$ erg\,s$^{-1}$\,cm$^{-2}$. While this might seem small, it is only a factor of 40 smaller than the 8-$\sigma$ detection limit on line fluxes in the main DESI survey. This argues that errors in ``sky'' subtraction of the DGL could present significant line confusion for upcoming ground-based spectroscopic surveys. Errors of the amplitude above could arise if reddening along the line-of-sight for sky fibers varies systematically from the target-of-interest by $E(B-V)$ $= 18.4$ mmag, and scales linearly with the amplitude of that reddening difference.

We compare equivalent width measurements for nebular lines in the all-sky DGL correlation spectrum to \citetalias{Brandt_2012_ApJ} and \citetalias{Chellew_2022_ApJ} in Table \ref{tab:neblines}. All measurements are consistent within error bars ($3\sigma$), and the uncertainties range from being comparable to a factor of $5\times$ improvement. An important source of uncertainty for these equivalent width measurements is spatial variability, which we demonstrate and explore in Section \ref{sec:spatialvarylines}. This is one explanation for how lines of similar strengths in Figure \ref{fig:emission_lines} can have significantly different relative uncertainties (e.g.,~Na and K).

For detected doublets, we report the surface brightness, equivalent width, and line ratio in the second section of Table \ref{tab:neblines}. The spectral resolution of DESI compared to the SDSS and BOSS spectrographs allows us to improve on the past detection of the [\ion{O}{2}] $\lambda\lambda3727,3730$ doublet and resolve it in the DGL for the first time. The line ratio $I_{\lambda 3730}/I_{\lambda 3727} = 1.32\substack{+0.17 \\ -0.15}$ is slightly suppressed relative to the value obtained from pure statistical weights of the upper levels, 1.5. This line ratio is often used as a diagnostic of the electron density, as it begins to be suppressed at electron densities greater than $\sim10^3$ cm$^{-3}$ \citep{Wang_2004_AA}. Within the current uncertainties, our measured $I_{\lambda 3730}/I_{\lambda 3727}$ is entirely consistent with the typical sources of light scattering off of the dust being in the low electron density limit $< 10^3$ cm$^{-3}$.

We also add a detection of the [\ion{N}{1}] $\lambda\lambda5199,5201$ doublet, which is usually seen in shock-excited regions. However, it is one of our weakest lines close to our detection limit. Our measured line ratio of $I_{\lambda 5201}/I_{\lambda 5199} = 1.01\substack{+0.25 \\ -0.20}$ is higher than literature measurements which range 0.55-0.65 \citep{Sharpee_2003_AAS}, though the difference is not statistically significant and the line ratio is difficult to fit because the line is barely resolved.

\subsection{Spatial Variation} \label{sec:spatialvarylines}

We next repeat the measurement of nebular line properties restricted to spatial subsets of the data.  This comes with a reduced signal-to-noise ratio, but allows us to map which parts of the sky contribute most to the observed global DGL correlation spectrum and to measure spatial variations in the emission line properties. We divide the data using a HEALPix (equal area, iso-latitude, hierarchical) grid with a resolution of NSIDE $=$ 8 (approximately 7.3$\degree$ wide pixels). We compute the mean (RA, Dec) of all of the sky-fibers in a given petal-exposure and assign the correlation from that petal-exposure to the HEALPix pixel containing those mean coordinates. Just as in Section \ref{sec:errors}, we then jackknife over exposures to obtain the mean, variance, and covariance for the DGL correlation spectrum for a given HEALPix pixel. Pixels with only one exposure within their boundary are excluded. These per-pixel DGL spectra are provided as a data product to the community (see Section \ref{sec:dataavil}).

\begin{figure*}[htb]
    \centering
    \includegraphics[width=\linewidth]{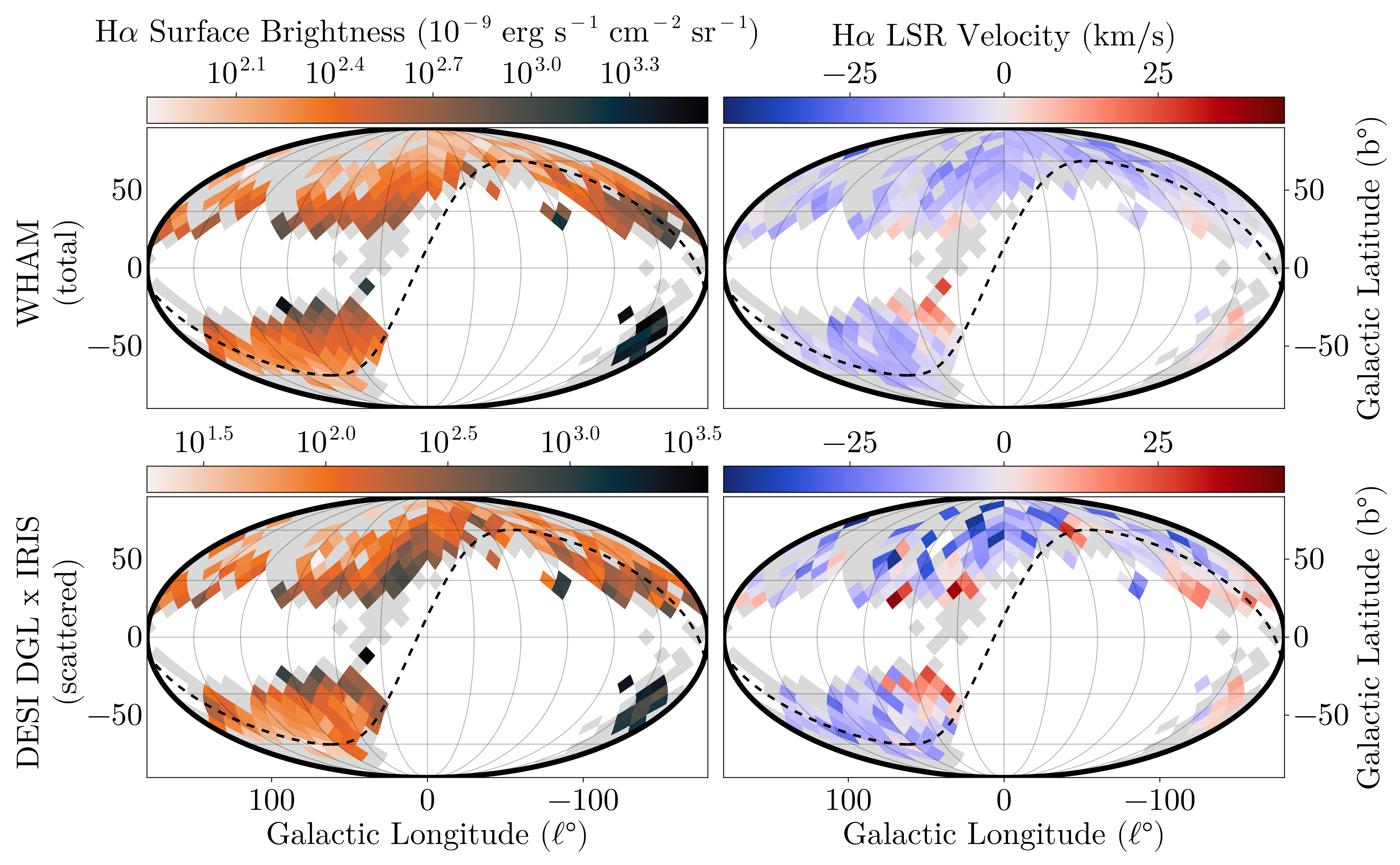}
    \caption{Comparison of total H$\alpha$ emission measured by WHAM (top row) at our resolution and the scattered H$\alpha$ from multiplying our spatially-resolved DESI DGL measurement times the average $I_{100\mu\text{m}}$ from IRIS  (bottom row). The first column shows the surface brightnesses, which differ by $\sim2\times$ on average. The second column is the LSR velocity. Pixels without any data are shown in white and pixels without significant detections (see text) are shown in light grey. Projections are Mollweide with a dashed line showing the ecliptic plane.}
    \label{fig:whamHalpha}
\end{figure*}

\begin{figure}[htb]
    \centering
    \includegraphics[width=\linewidth]{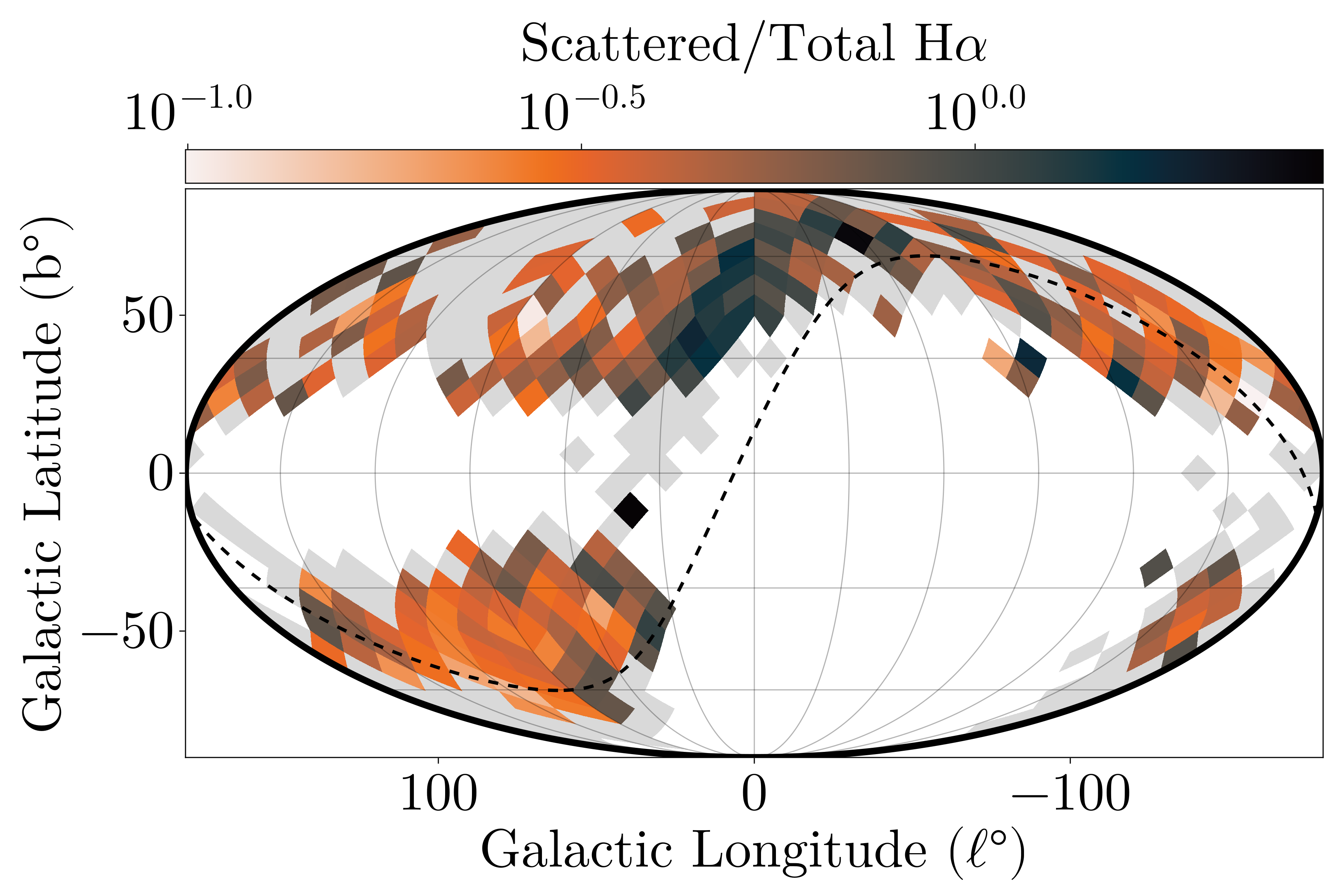}
    \caption{Map of the ratio of scattered to total H$\alpha$ emission. An enhancement is observed at low longitudes. Pixels without any data are shown in white and pixels without significant detections (see text) are shown in light grey. Projection is Mollweide with a dashed line showing the ecliptic plane.}
    \label{fig:ratioWHAM}
\end{figure}

\begin{figure}[htb]
    \centering
    \includegraphics[width=\linewidth]{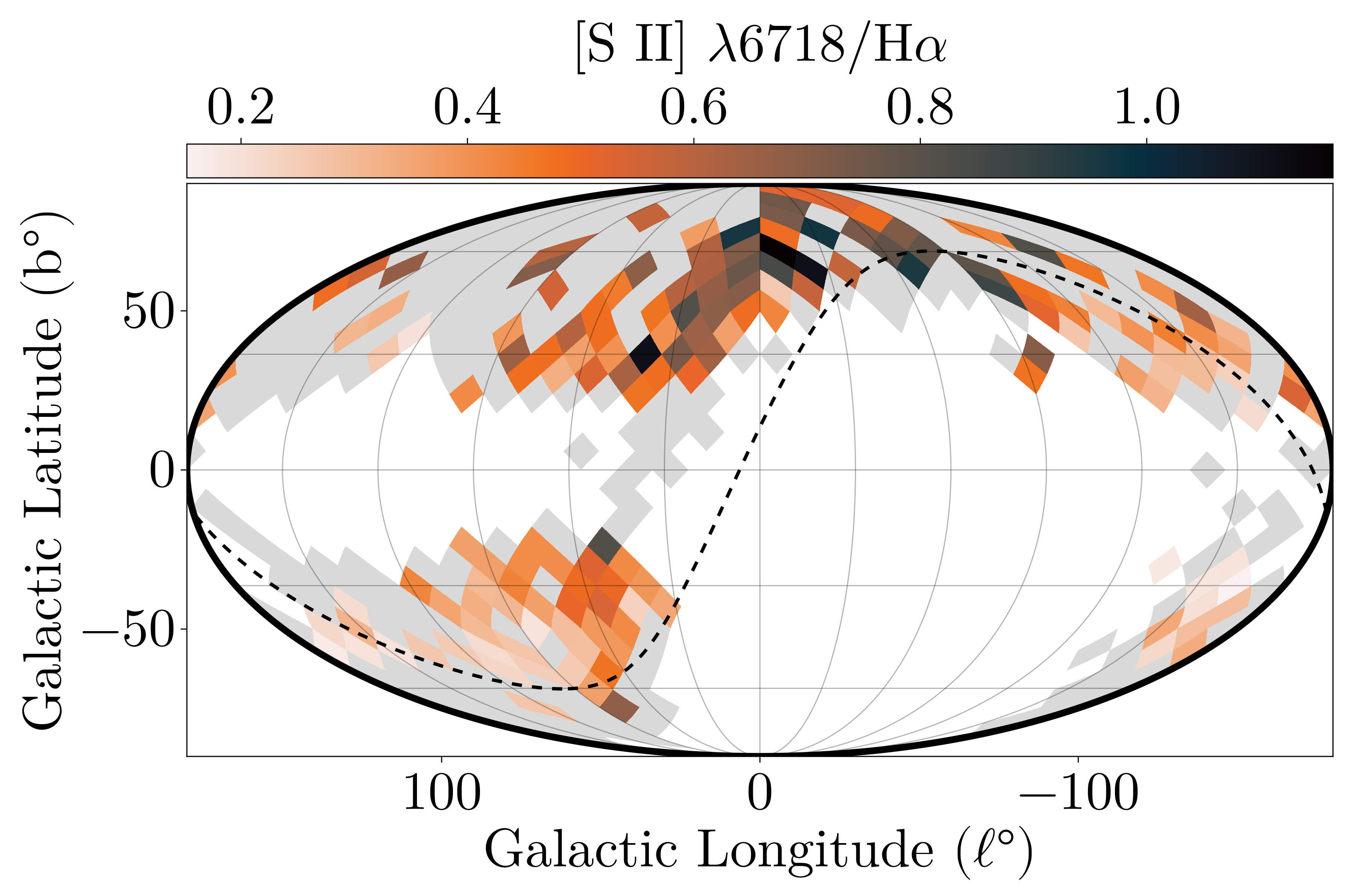}
    \caption{Ratio of the surface brightness for [\ion{S}{2}] $\lambda$6718 compared to H$\alpha$ in the DGL on a HEALPix map at a resolution of NSIDE $=$ 8. Pixels without any data are shown in white and pixels without significant detections (see text) are shown in light grey. Projection is Mollweide with a dashed line showing the ecliptic plane.}
    \label{fig:halpha_sii_ratio}
\end{figure}

Our choice of a resolution of NSIDE $=$ 8 was driven by trying to balance maintaining a signal-to-noise ratio per pixel that is informative and preserving the spatially varying structure of the DGL. Gaussian weighting schemes that fall-off from the HEALPix pixel center have been used previously \citep{Zucker_2025_ApJ} with spatial maps to increase their angular resolution and ensure smoothly varying values. We do not take that approach here because we are working so close to the low signal-to-noise ratio limit and the dynamic range is large (a few pixels have an out-sized contribution to the global signal). This choice leads to sharp discontinuities in the DGL maps, but we attribute these to real (unresolved) localized sources of emission (see Figure \ref{fig:line_fits_columns}) and thus prefer not to smooth over them.

For the DGL correlation spectrum in each HEALPix bin, we apply the same emission line fitting as in Section \ref{sec:allskylines} and summarize the lines by their three Gaussian moments: surface brightness, center, and linewidth. We convert their centers into a velocity offset compared to the vacuum wavelength of the line. We report local standard of rest (LSR) velocities using the LSR frame that assumes the ``classic'' solar motion of 20 km\,s$^{-1}$ toward (RA, Dec) = (18h, 30$^\circ$) at epoch 1900 to facilitate comparisons with WHAM \citep{Haffner_2003_ApJS} below.

The spatial maps of these fitted coefficients for two representative lines, H$\alpha$ and [\ion{S}{2}] $\lambda$6718, are shown in Figure \ref{fig:line_fits_columns}. Pixels without any data are shown in white and pixels without significant detections are shown in grey. For Figure \ref{fig:line_fits_columns}, our detection cuts require that (1) the signal-to-noise ratio of the surface brightness be greater than 2.8, as measured by the mean surface brightness compared to the 16$^{\rm{th}}$ and 84$^{\rm{th}}$ percentiles, (2) the median spectral (per wavelength element) signal-to-noise ratio within 200 km/s of the rest-frame line center is greater than 0.7, (3) the uncertainty on the linewidth is not too large (84$^{\rm{th}}$ - 16$^{\rm{th}}$ percentile $<$ 0.45 \r{A}), and (4) that the linewidth is not too close to the prior upper bound (84$^{\rm{th}}$ percentile $<$ 1.85 \r{A}). 

If we had not imposed a detection cut, we would see an expected trend in the surface brightness toward zero as a function of signal-to-noise ratio for many of the pixels. However, we impose detection cuts in order to trust the higher-order moment maps (velocity and linewidth). As shown in Figure \ref{fig:line_fits_columns}, there are significant spatial variations in the surface brightness of H$\alpha$ in the DGL. While not unexpected given the strong spatial variation of direct H$\alpha$ emission \citep{Finkbeiner_2003_ApJS}, our work is even more sensitive to spatial variations because it is measuring the localized impact of isolated sources (e.g.,~\ion{H}{2} regions) after their emission propagates through the ISM and scatters off of dust. For H$\alpha$, the surface brightness for 1 MJy/sr FIR emission at 100 $\mu$m in pixels with detections ranges from $16.4\pm1.6$ to $1122\pm61$ $\times 10^{-9}$ erg\,s$^{-1}$\,cm$^{-2}$\,sr$^{-1}$, compared to the average (integrated) DGL correlation spectrum value of 48.5$\pm1.2$ $\times 10^{-9}$ erg\,s$^{-1}$\,cm$^{-2}$\,sr$^{-1}$.

Regions which are bright in both the H$\alpha$ and [\ion{S}{2}] $\lambda$6718 lines include (1) a patch at both negative Galactic latitude ($b \sim -50^{\circ}$) and longitude ($\ell \sim -120^{\circ}$) and (2) a ridge at positive Galactic latitude ($b \sim 25-60^{\circ}$) and longitude near zero ($\ell \sim 0 - 20^{\circ}$). The first is at a latitude similar to the LMC, but at more negative Galactic longitude. The second looks to be coincident with the edge of the Local Bubble as seen in 3D dust maps \citep{ONeill_2024_ApJ}. Our observed enhancement is coincident with the extent of a cloud detected in $\rm{H}_2$ fluorescence from FIMS/SPEAR \citep{Jo_2017_ApJS} on the surface of the Local Bubble, recently dubbed the ``Eos'' cloud \citep{Burkhart_2025_NatAs}. 

The second feature is also spatially coincident with the North Polar Spur (NPS), a feature seen in X-ray and radio emission, but with low H$\alpha$ surface brightness in direct emission \citep{Sofue_2023_MNRAS}. While the origin of the NPS has been debated, with theories including Galactic center feedback and associations with the Fermi bubbles \citep{Su_2010_ApJ}, recent work has suggested the NPS is at least partially due to interactions in the more local Scorpius–Centaurus association, the nearest OB star association to the sun \citep{Das_2020_MNRAS}. In this picture, the enhancement of the DGL along the NPS would be naturally expected and attributed to scattered light from the Scorpius–Centaurus association.

Turning to the second row of Figure \ref{fig:line_fits_columns}, we see predominantly negative (blue-shifted) velocities at high Galactic latitudes. Pixels with positive velocities tend to be near the plane and are consistent with the usual quadrupole from Galactic rotation. The velocities measured for both of the emission lines shown agree well, including for pixels that are slightly anomalous relative to the global pattern. Quantitative modeling of the velocity structure of H$\alpha$ in the DGL will require accounting for the relative velocity between the sources of emission and the dust that is responsible for the scattering. These dust velocities are now becoming available with next-generation kinematic dust maps \citep{Tchernyshyov_2018_AJ,Saydjari_Frank_KT_unpub} that leverage large, all-sky diffuse interstellar band catalogs \citep{Saydjari_2023_ApJ, Saydjari_APOGEE_DIB_unpub}. These considerations go one step beyond recent work at reproducing the H$\alpha$ with radiative transfer through 3D dust geometries by \citet{McCallum_2025_MNRAS}, which does keep track of both the direct and scattered light components.

The best literature comparison for the H$\alpha$ velocity (and surface brightness) maps is the Wisconsin H$\alpha$ Mapper (WHAM) survey \citep{Haffner_2003_ApJS}, which we show in Figure \ref{fig:whamHalpha}. WHAM measured H${\alpha}$ emission with 1$\degree$ spatial resolution, 12 km\,s$^{-1}$ spectral resolution, and $3.6 \times 10^{-8}$ erg\,s$^{-1}$\,cm$^{-2}$\,sr$^{-1}$ 3$\sigma$-sensitivity. Their measurement captures the total H${\alpha}$ emission (direct and scattered). To compare with our DGL measurements, we perform an intensity-weighted average of the moment-1 WHAM maps from their native $1\degree$ resolution to HEALPix NSIDE$=8$ ($\sim7\degree$) and apply our detection mask to only display data in the same spatial pixels. We also multiply our spatially-resolved DGL spectrum by the average IRIS $I_{100\mu\text{m}}$ per pixel so we can directly compare expected observed surface brightnesses.

On average, we find the fraction of H$\alpha$ emission that is scattered is $\leq0.47 \pm 0.27$, consistent with the $\sim$0.5 measured for high-latitude cirrus by \citet{Witt_2010_ApJ}, but larger than the value inferred by \citetalias{Brandt_2012_ApJ} ($0.19 \pm 0.04$). Here our error bars represent the spread of measurements from different lines of sight. We can also ask if there is coherent structure on the sky by making the spatial map, which is shown in Figure \ref{fig:ratioWHAM}. We clearly see an enhancement in the fraction of scattered light near low Galactic longitudes, coincident with the NPS feature in the DGL maps. Using this to infer variations in dust properties would require contextualizing this measurement relative to the location of \ion{H}{2} regions in the disk and should be the subject of future work.

The total H$\alpha$ emission is also much smoother than the scattered component. It closely resembles a smooth decrease with increasing Galactic latitude at this resolution, and is without the NPS-like feature, which is stronger in normalization of Figure \ref{fig:line_fits_columns} but is still clearly visible after taking a product with the average FIR intensity as in Figure \ref{fig:whamHalpha}. This suggests that the scattered light component is more sensitive to the 3D structure of the dust and could be a stronger constraint on the aforementioned 3D radiative transfer modeling of the solar neighborhood. Similarly, we attribute the increased dynamic range of the scattered light component ($10^2$) compared to the total H$\alpha$ emission ($10^1$) to the fact that the scattered component is a product of the density distribution of both dust and nearby H$\alpha$ emitters. 

The velocity maps for WHAM and the DGL closely resemble each other, which we take as a validation of our measurement. We interpret this similarity to also suggest that the scattered light is dominated by nearby sources with similar kinematics. The DGL velocity map is understandably noisier, but does appear to have slightly more positive velocities in the upper third quadrant, which may warrant further investigation in the future.

\begin{figure}[b]
    \centering
    \includegraphics[width=\linewidth]{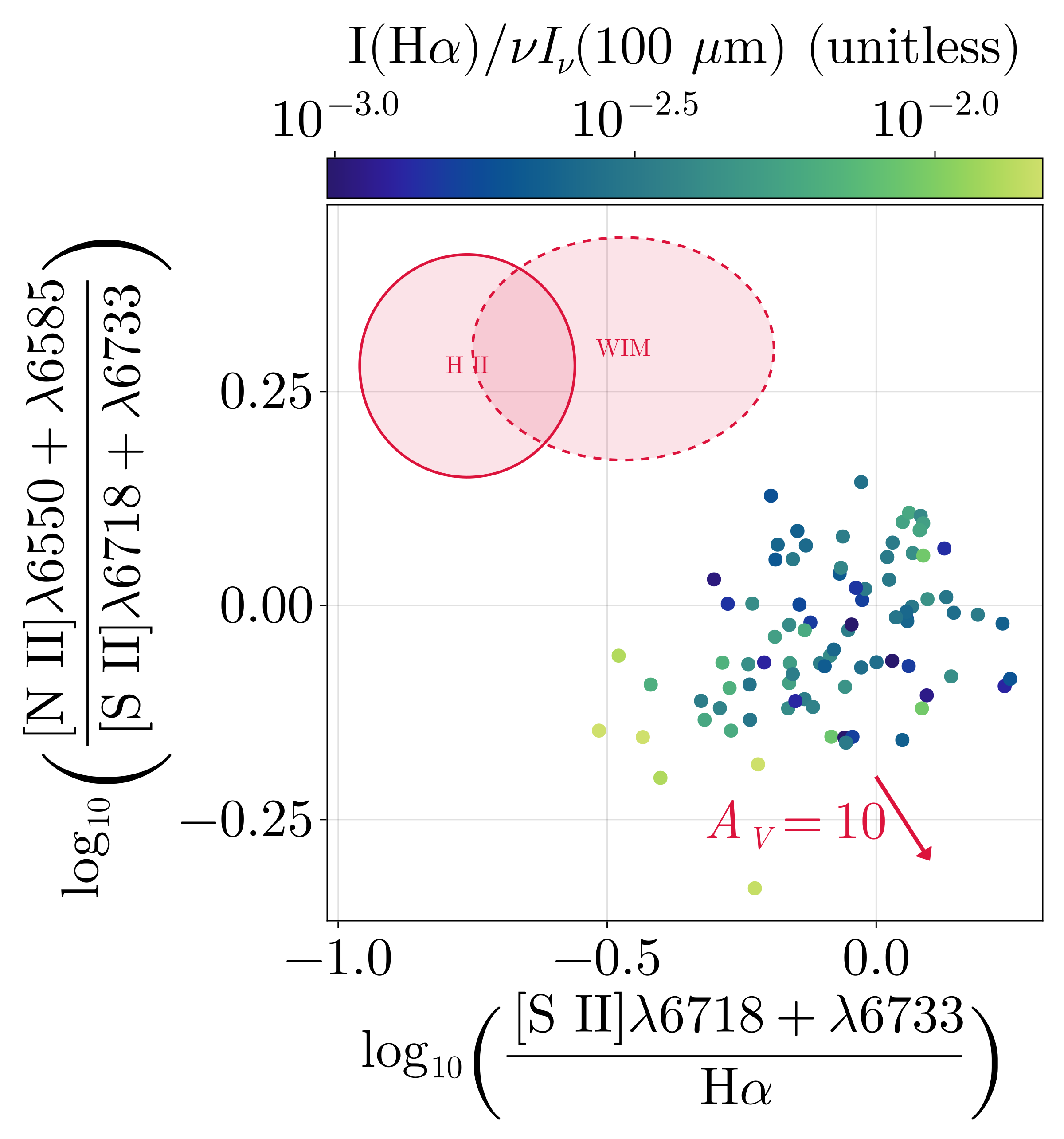}
    \caption{Line ratios of [\ion{N}{2}]/[\ion{S}{2}] versus [\ion{S}{2}]/H$\alpha$ from a joint five-line fit ([\ion{N}{2}] $\lambda\lambda$6550,6585, H$\alpha$, [\ion{S}{2}] $\lambda\lambda$6718,6733; Section \ref{sec:speclinfit}) to the DGL correlation spectrum for different lines of sight at HEALPix resolution NSIDE = 8. The colorbar is the dimensionless ratio of the H$\alpha$ line surface brightness to the FIR surface brightness, $I(\text{H}\alpha)/\nu I_{\nu}(100~\mu\text{m})$.}
    \label{fig:lineratioscatter}
\end{figure}

The line width, reported in $\sigma$ as opposed to FWHM (${\rm FWHM} \approx 2.35\sigma$ for a Gaussian), is shown in the bottom row of Figure \ref{fig:line_fits_columns}. The line widths of both emission lines shown broadly agree and are correlated (Spearman's rank correlation coefficient $\rho =$ 0.5) across the spatial footprint. The clearest feature in the line width maps is that the third quadrant has narrower line widths of $\sim 0.7$ \r{A} (median) compared to the rest of the sky which has line widths of $\sim 0.9$ \r{A}. One explanation for this observation could be that there are two main sources of emission that drive the DGL (e.g., Sco OB2 and Cygnus OB2), with different velocity dispersions. If true, this suggests that attempts to theoretically forward model the spatial variations in the DGL could leverage a ``few source'' approximation.

When comparing the columns of Figure \ref{fig:line_fits_columns}, it is clear the relative surface brightness of H$\alpha$ and [\ion{S}{2}] $\lambda$6718 changes. This change is especially clear when comparing the two bright regions of interest we discussed. We can more easily explore this via the line ratio [\ion{S}{2}] $\lambda$6718/H$\alpha$ map in Figure \ref{fig:halpha_sii_ratio}, which varies by an order of magnitude across the footprint from 0.16 to 1.18. On average, the line ratio is highest near zero Galactic longitude, decreasing toward the Galactic anti-center. The line ratio does not depend strongly on the overall surface brightness. For example, at negative Galactic latitudes, both the high surface brightness region at negative Galactic longitudes and the low surface brightness region at positive Galactic longitudes have [\ion{S}{2}] $\lambda$6718/H$\alpha$ $\sim 0.3$.

The most rigorous way to use these emission line ratios to learn about the physical conditions of the emitting regions in the ISM contributing to the DGL would require modeling both the effect of attenuation (including geometric effects, not just a simple extinction correction) and scattering on processing the source spectrum into the DGL. However, we will neglect these effects for now and map the variation in these line ratios in the DGL into 2D line ratio plots. Ideally, these line ratios would probe different ionization states like classic ``BPT'' \citep{Baldwin_1981_PASP} diagrams ([\ion{O}{3}] $\lambda5007$/H$\beta$ vs. [\ion{N}{2}] $\lambda6583$/H$\alpha$) so they can be used to trace variations in the ``hardness'' of the source. However all of our doubly ionized lines are low signal to noise such that we can only report such ratios for the all-sky spectrum. We leave mapping these lines to future work, with larger sample sizes.

The strongest lines in the dust-correlated DGL are H$\alpha$, [\ion{N}{2}], and [\ion{S}{2}]. In Figure \ref{fig:lineratioscatter} we plot $\log_{10}(\text{[\ion{N}{2}]/[\ion{S}{2}]})$ versus $\log_{10}(\text{[\ion{S}{2}]/H$\alpha$})$ from a per-pixel joint fit of the five lines (Section \ref{sec:speclinfit}). The ratios in the dust-correlated DGL are well-separated from values (shaded pink loci) observed in the WIM \citep{Haffner_1999_ApJ}. Figure \ref{fig:lineratioscatter} includes a reddening vector showing the displacement produced by $A_V = 10$~mag. It is clear that the observed line ratios in the dust-correlated DGL cannot be due to reddening of emission from either \ion{H}{2} regions or the WIM: the emission would be totally obscured by the required extinction $A_V > 20$~mag.

Much of the observed [\ion{S}{2}] in the dust-correlated DGL must originate in the ``Warm Neutral Medium'': \ion{H}{1} regions (where the S is photoionized by starlight, but N is neutral) that have been heated to $T \gtrsim 5000$ K by photoelectric heating, mechanical heating (shocks or turbulent dissipation), or perhaps some other mechanism. This would naturally produce low ratios of [\ion{N}{2}]/[\ion{S}{2}] and high ratios of [\ion{S}{2}]/H$\alpha$. If this ``phase'' of the interstellar medium is preferentially located away from the midplane, and is correlated with dust emitting at 100 $\mu$m, it will appear in our DGL correlation spectrum. These regions must also contribute to the integrated emission from galaxies, but presumably make a subdominant contribution to the overall [\ion{S}{2}] emission.

Further studies to identify specific regions with elevated [\ion{N}{2}]/[\ion{S}{2}] and elevated [\ion{S}{2}]/H$\alpha$ will enable the conditions in such regions to be better characterized, to test our prediction that these must be unusually warm \ion{H}{1} regions, and clarify the heating mechanisms.

\section{Atomic Lines: Resonant Scattering} \label{sec:atomlines}

\begin{figure}[htb]
    \centering
    \includegraphics[width=\linewidth]{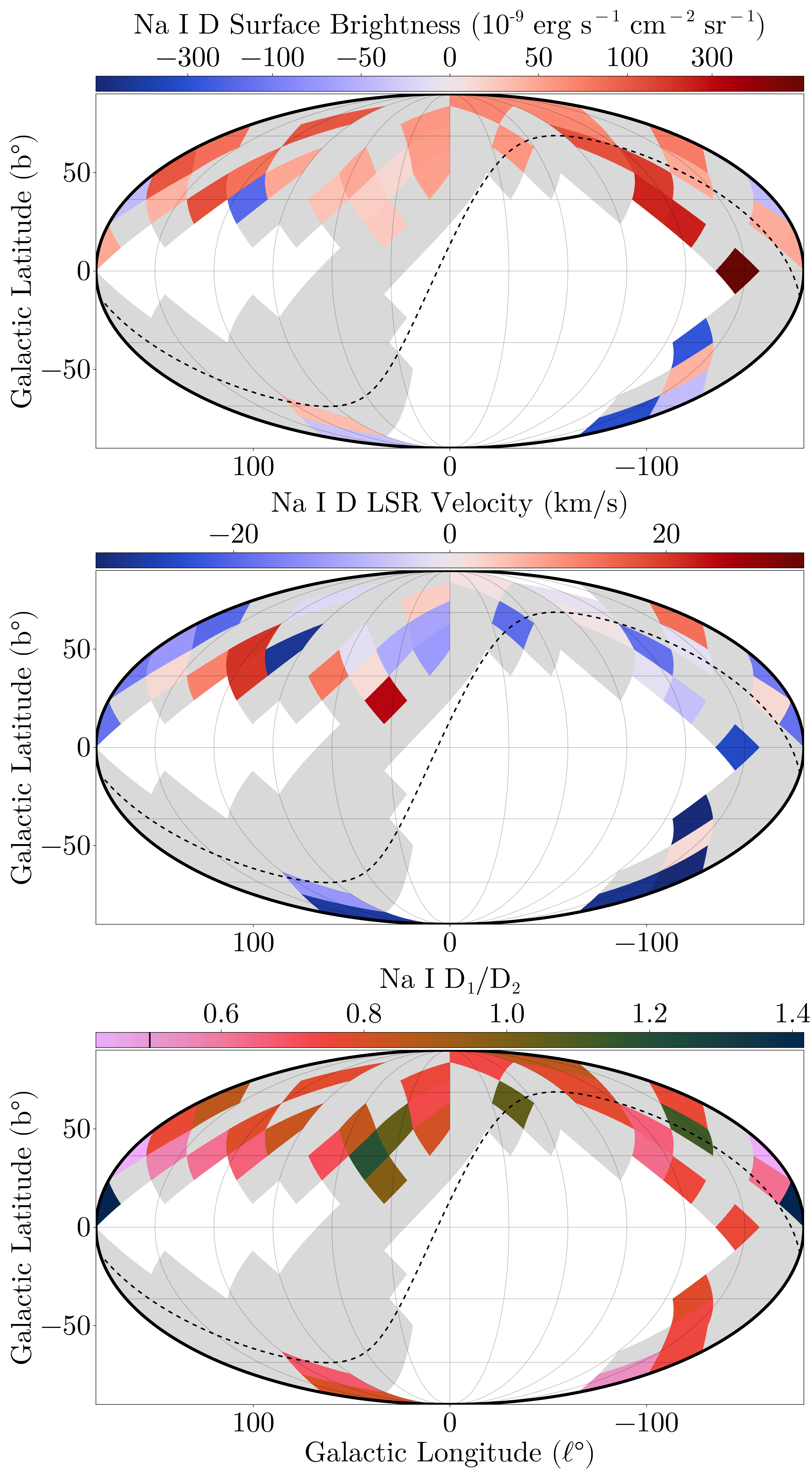}
    \caption{Maps of Gaussian fit parameters for \ion{Na}{1} D doublet lines in the DGL across the DESI footprint on a HEALPix grid with resolution NSIDE $=$ 4. The first row shows surface brightness for 1 MJy/sr FIR emission at 100 $\mu$m, the second shows LSR velocity, and the last shows the flux ratio between the \ion{Na}{1} D lines ($D_1/D_2)$. Pixels without any data are shown in white and pixels without significant detections (see text) are shown in light grey. Projections are Mollweide with a dashed line showing the ecliptic plane.}
    \label{fig:naD_map}
\end{figure}

In addition to the nebular emission lines discussed in Section \ref{sec:neblines}, the bottom row of Figure \ref{fig:emission_lines} also shows the classic ``D'' doublet resonance lines of \ion{Na}{1} (D$_2$ 5891.58 \r{A}, D$_1$ 5897.56 \r{A}) and \ion{K}{1} (D$_2$ 7667.02 \r{A}, D$_1$ 7701.10 \r{A}). In stellar spectra, the lines appear as interstellar absorption features, but in the dust-correlated DGL, they are seen in \emph{emission} (Figure \ref{fig:emission_lines}).

In the all-sky DGL correlation spectrum, the Na doublet is stronger ($13.4\substack{+1.4 \\ -1.5}$ $\times 10^{-9}$ erg\,s$^{-1}$\,cm$^{-2}$\,sr$^{-1}$) compared to the K doublet ($6.78\substack{+0.22 \\ -0.21}$ $\times 10^{-9}$ erg\,s$^{-1}$\,cm$^{-2}$\,sr$^{-1}$), but only by a factor of $\sim2$ (Table \ref{tab:neblines}). This strength ratio does not match the 15$\times$ lower abundance of K compared to Na. Interestingly, despite the strength of the Na doublet, the error bars on the DGL correlation spectrum are much larger for the Na doublet compared to those of the K doublet.

It is easy to show that the observed \ion{Na}{1} D and \ion{K}{1} D emission cannot be the result of either radiative recombination of \ion{Na}{2} or \ion{K}{2}, or collisional excitation of \ion{Na}{1} or \ion{K}{1}. The observed emission can instead be understood to be starlight that has been scattered (i.e., absorbed and re-emitted) by \ion{Na}{1} and \ion{K}{1} in the ISM. When we observe a star directly, we see \ion{Na}{1} and \ion{K}{1} in ``absorption'', but the photons haven't been destroyed -- they have just been redirected. The ``sky fibers'' have specifically been positioned to avoid direct starlight, but will capture these scattered photons. Because the scattering \ion{Na}{1} and \ion{K}{1} atoms are correlated with dust, we can expect to see the \ion{Na}{1} and \ion{K}{1} D features appearing as emission features in the dust-correlated DGL. 

Of course, dust also scatters starlight; the light illuminating the dust will include direct starlight with absorption features from interstellar \ion{Na}{1} and \ion{K}{1}, as well as photons scattered by interstellar \ion{Na}{1} and \ion{K}{1}. Thus we anticipate that the dust-correlated DGL could show \ion{Na}{1} and \ion{K}{1} features either in emission or absorption, depending on the distribution of stars, absorbing gas, and dust affecting the radiation field in a given direction. The net positive contribution in the all sky DGL (Figure \ref{fig:emission_lines}) makes sense because (1) the sky fibers are deliberately placed to avoid direct starlight, and (2) the DESI survey footprint is predominantly at high Galactic latitudes, while the continuum sources (stars) are concentrated near the Galactic plane.

The abundance of interstellar \ion{Na}{1} is high enough for the \ion{Na}{1} D lines to become optically thick, in which case the scattering no longer varies linearly with column density. The \ion{K}{1} D lines also become optically thick on some sightlines \citep[e.g., ][]{Hobbs_1974_ISM_Na_Ca_K_absorption}, but this is less common than for \ion{Na}{1} D due to the 15$\times$ lower abundance of K. The greater tendency for the Na D lines to become optically thick presumably accounts for the larger uncertainties in the DGL correlation spectrum for the Na D doublet compared to the K D doublet. However, it is possible that this enhanced Na variability could be in part due to the lack of telluric Na correction as compared to the presence of a correction for telluric K in the DESI \texttt{fluxcalib} vector (discussed further below).

Whether the dust-correlated spectrum will show the \ion{Na}{1} and \ion{K}{1} D lines in emission or absorption will be determined by the particular geometry of the stellar sources, absorbing gas, and dust grains doing the scattering.

To explore the possibility of spatial variation, we subdivide the sky as in Section \ref{sec:spatialvarylines} and fit the Na doublet in each HEALPix pixel independently (Figure \ref{fig:naD_map}). We use the same detection cuts as before, with the line flux being replaced with the total line flux of the doublet, and additionally require each component of the doublet to be individually detected (component flux $|$S/N$| > 2$). We use a lower resolution (NSIDE = $4$, 15$\degree$ wide pixels) to compensate for the lower signal-to-noise ratio of the Na D doublet compared to the nebular emission lines. 

Our first observation from the surface brightness map (top panel of Figure \ref{fig:naD_map}) is that we see the Na D doublet with both positive and negative contributions to the DGL. The pixels where the strongest Na D scattering out of the line of sight is observed correspond to regions with the highest surface brightness in H$\alpha$ (Figure \ref{fig:line_fits_columns}). Pixels where Na D scatters light onto the line of sight tend to dominate at positive Galactic latitudes. 

This observation of Na D with both positive and negative contributions to the DGL in our spatial maps is consistent with the predictions from the resonant scattering process outlined above. Thus, we interpret positive flux from the D doublets in the DGL as the scattering of stellar emission onto the line of sight by interstellar sodium and potassium. Because the source spectra for the DGL are not smooth, the observed strength of the D doublets will depend in detail on the strength of the Na and K absorption in the stellar atmospheres that dominate the DGL source spectrum for a given line of sight.

A real concern for this measurement is that Na in the Earth's atmosphere (``telluric'' Na) might be confounding astrophysical signals. First, and most importantly, it is unlikely that telluric Na would correlate with gradients in the FIR dust emission from distant astrophysical sources (Milky Way dust clouds). Second, an obvious way to get a ``false'' positive flux from Na would be if a calibration applied during the spectral reduction pipeline ``over-corrected'' for telluric Na. However, the \texttt{fluxcalib} vectors that the DESI pipeline fits for this purpose (see Section \ref{sec:desisky}) have explicitly masked out the Na D doublet and Ca H and K lines (3934.77, 3969.59 \r{A}).\footnote{\url{https://github.com/desihub/desispec/blob/9e833aed8c10d120981564edf718df687eeb3adc/py/desispec/fluxcalibration.py\#L1374}} The \texttt{fluxcalib} is linearly interpolated over $\pm6$ \r{A} from each of these four line centers, which is approximately $\pm300$ km/s for the Na D lines. Thus, no telluric Na correction is being applied to the DESI spectra we are using and so no ``over-correction'' is possible.

Finally, we can inspect the kinematics recovered in the middle panel of Figure \ref{fig:naD_map}. The observed Na D velocities broadly agree with those recovered from the nebular emission lines (Figure \ref{fig:line_fits_columns}), which supports their predominantly astrophysical, and not telluric, origin. If the Na signal were telluric, we would expect it to have kinematics consistent with the average barycentric velocity correction applied to the exposures used to construct each HEALPix pixel. A map of expected kinematics from telluric species is shown in Appendix \ref{sec:baryCor}. Since the velocity map in Figure \ref{fig:naD_map} does not resemble the velocity map in Appendix \ref{sec:baryCor}, we conclude the Na must not be of primarily telluric origin. The positive velocity cluster in the northern second quadrant of Figure \ref{fig:naD_map} is the most anomalous compared to the nebular emission line kinematics and merits further investigation. 

\begin{figure}[tb!]
    \centering
    \includegraphics[width=\linewidth]{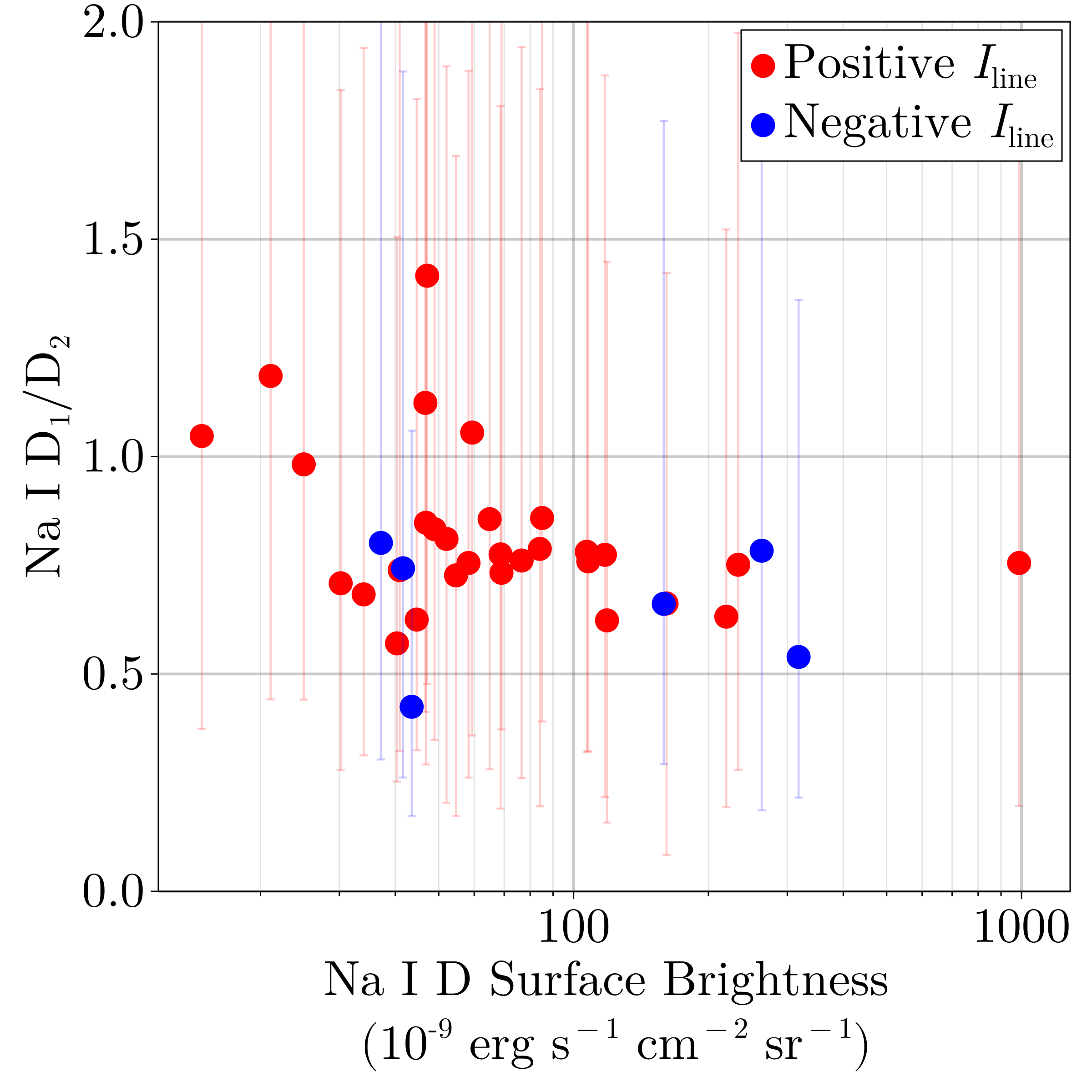}
    \caption{Scatter plot of the Na D$_1$/D$_2$ ratio as a function of the line surface brightness where scattering onto the line of sight dominates (red) or absorption dominates (blue). Error bars represent $16^{\text{th}}$ and $84^{\text{th}}$ percentiles ($\sim 1\sigma$ uncertainties).
    }
    \label{fig:nacog}
\end{figure}

\begin{figure*}[tb!]
    \centering
    \includegraphics[width=\linewidth]{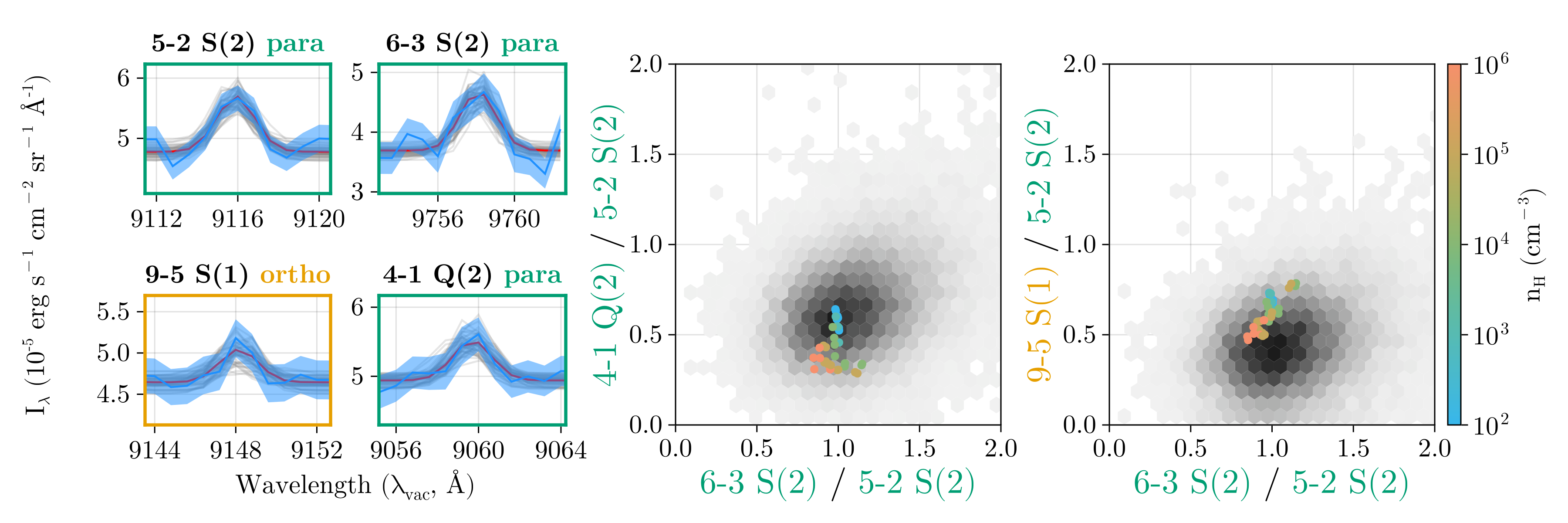}
    \caption{UV-pumped H$_2$ rovibrational lines. Left subpanels show DGL correlation spectrum centered on the wavelength of four of the H$_2$ lines we simultaneously fit. Shading represents 1$\sigma$ for diagonal approximation to the uncertainties. The mean Gaussian line profile for each subplot is shown in red, and 50 draws from the posterior are shown in light grey. Right subpanels show our posterior for three line ratios in a greyscale 2D histogram. Different models from \citetalias{Bertoldi_1996_ApJ} colored by hydrogen atom density are overplotted. The disagreement with \citetalias{Bertoldi_1996_ApJ} only on the line ratio including transitions from both ortho and para hydrogen means we measure a different apparent ortho-to-para ratio than \citetalias{Bertoldi_1996_ApJ}.}
    \label{fig:h2lines}
\end{figure*}

If the \ion{Na}{1} and \ion{K}{1} absorption lines were optically thin, the dust-correlated DGL features would also have the optically-thin doublet line ratio D$_1$/D$_2$ $=$ 1/2. However, the abundance of \ion{Na}{1} in the ISM is high enough that the interstellar Na D lines are often optically thick, resulting in absorption equivalent width ratios $0.5 < W_1/W_2 < 1$, which would naturally have counterparts in the D$_1$/D$_2$ ratio in the apparent emission features. In the all-sky DGL correlation spectrum, we find D$_1$/D$_2$ $=$ $1.11\substack{+0.28 \\ -0.20}$ for Na and $0.671\substack{+0.046 \\ -0.043}$ for K (Table \ref{tab:neblines}).

For Na, the D$_1$/D$_2$ appears close to the limiting value of 1, corresponding to saturation of both lines. However the interpretation is complicated by the fact that the starlight itself has absorption features arising in the stellar atmospheres; while these are typically broader than the interstellar absorption features on the direct sightline to a star, the velocity dispersion of interstellar gas makes it difficult to distinguish the contributions of stellar atmospheres and interstellar scattering to the \ion{Na}{1} D features in the dust-correlated DGL.

The bottom row of Figure \ref{fig:naD_map} shows a map of Na D$_1$/D$_2$ from our spatially-resolved doublet fits, with D$_1$/D$_2$ varying from $0.43\substack{+0.21 \\ -0.17}$ to $1.42\substack{+0.94 \\ -0.48}$, with typical (median) uncertainties of 0.36. For Na the all-sky D$_1$/D$_2$ is 3$\sigma$ away from the optically-thin limit of 0.5. Measurements in the bottom row of Figure \ref{fig:naD_map} are at most $1.9\sigma$ different from 0.5 because of the decreased signal-to-noise ratio from the spatial subdivision.

Figure \ref{fig:nacog} shows that D$_1$/D$_2$ for \ion{Na}{1} exhibits considerable scatter, consistent with varying between $\sim0.5$ and 1 when seen in either emission or absorption, with no clear trend with \ion{Na}{1} D surface brightness. Note that the error bars in Figure \ref{fig:nacog} tend to significantly exceed the scatter between the different sightlines. This may be in part due to the significantly non-Gaussian shape of the posterior in the space of the line width versus Na D$_1$/D$_2$ ratio, or may indicate significant variation on scales smaller than our admittedly coarse spatial resolution.

On any given sightline, the \ion{Na}{1} surface brightness (either in ``emission'' or ``absorption'') is presumably the result of 3-D resonance line radiative transfer, with a distribution of stellar continuum sources (some with intrinsic Na D absorption features). The radiative transfer problem requires (1) the 3-D distribution of the stellar sources (and the velocities of stars with Na I D absorption features), (2) the 3-D distribution and velocities of Na I atoms, and (3) the 3-D distribution (and velocities) of dust grains scattering the radiation.  At this time, all we can say is that the observations of Na I D lines and K I D lines appear to be consistent with this general picture, but quantitative comparison would require ambitious radiative transfer modeling.


\section{H$_2$ Lines: Direct Emission} \label{sec:h2lines}

So far, we have only considered signatures associated with atomic species, either emission lines scattered off of dust (Section \ref{sec:neblines}) or continuum scattered off of interstellar neutrals correlated with dust (\ion{Na}{1}, \ion{K}{1} in Section \ref{sec:atomlines}). We might also expect contributions from molecular species correlated with dust. Specifically, molecular hydrogen (H$_2$) is abundant, its formation is catalyzed on dust grains, and when UV-pumped, emits near-infrared ro-vibrational lines in the DESI wavelength range.

To look for these H$_2$ lines, we investigate the 20 strongest lines in the DESI wavelength range as predicted by a photo-dissociation region (PDR) model from \citet{Bertoldi_1996_ApJ} (\citetalias{Bertoldi_1996_ApJ}) with conditions most similar to the diffuse ISM ($n_{\text{H}} = 10 ^2$ cm$^{-3}$, T = 300 K, 1 Habing ($G_0$) FUV radiation intensity).\footnote{am3d, \url{https://www.astro.princeton.edu/~draine/pdr.html}}

After avoiding regions that overlap with Paschen lines or have clear telluric contamination, we find four lines in our DGL correlation spectrum at wavelengths corresponding to the 9116.0 \r{A} 5-2 S(2), 9757.8 \r{A} 6-3 S(2), 9148.1 \r{A} 9-5 S(1), and 9059.7 \r{A} 4-1 Q(2) transitions of H$_2$ (Figure \ref{fig:h2lines}). These are some of the strongest lines according to the PDR model, and they are predicted to have line strengths corresponding to 51\%, 50\%, 37\%, and 32\%, respectively, relative to the strongest predicted line in the DESI wavelength range, which is the 5-2 S(1) transition (which was not confidently detected due to blending with a Paschen line; see Appendix \ref{sec:margDetect}). Our detected lines have surface brightnesses of $0.65\substack{+0.13 \\ -0.12}$, $0.69\substack{+0.16 \\ -0.15}$, $0.28 \pm 0.12$, and $0.41\substack{+0.13 \\ -0.12}$ $\times10^{-9}{\rm erg\, cm^{-2}\, s^{-1}\, sr^{-1}}$ respectively (Table \ref{tab:neblines}).

We can compare the absolute surface brightness of these lines to theoretical estimates based on H$_2$ formation, destruction, and emission mechanisms. H$_2$ formation on grain surfaces is balanced by photodissociation. On the average, for each photodissociation there are $\sim$6 UV-pumping events that do not lead to photodissociation, but which populate the excited vibrational levels of the electronic ground state \citep{Draine_2011_piim}. Spontaneous decay of the vibrationally-excited levels produces a rich emission spectrum, with $\Delta v=1$ emission in the K band (e.g.,~1-0 S(1) 2.1218$\mu$m) and weaker $\Delta v=3$, and $\Delta v=4$ emission lines in the far-red, first observed by \citet{Burton_1992_MNRAS} in NGC 2023.

The fluorescent surface brightness in a line at wavelength $\lambda$ is
\begin{equation}
I_{\rm line}=\frac{Rn_{\rm H} N({\rm H})}{f_{\rm diss}}~
\frac{\epsilon_{\rm line}}{4\pi}~ \frac{hc}{\lambda}
\end{equation}
where $f_{\rm diss}\approx0.15$ is the probability that a UV pumping event will lead to photodissociation, $R\approx 3\times10^{-17}{\rm cm^3\,s^{-1}}$ is the empirical rate coefficient for H$_2$ formation on grains \citep{Jura_1975_ApJ}, and $\epsilon_{\rm line}$ is the probability that one UV pumping event will lead to emission in this line. \citet{Draine_1996_ApJ} (\citetalias{Draine_1996_ApJ}) modeled the photoexcitation of H$_2$ in PDRs. For the 5-2 S(2) 9116.0\AA\ line, \citetalias{Draine_1996_ApJ} found $\epsilon_{\rm line}=0.0021$.

Thus, for an \ion{H}{1} cloud we expect surface brightness
\begin{multline}\label{eq:I_line}
I_{\rm line}\approx 1.4\times10^{-10} {\rm erg\, cm^{-2}\, s^{-1}\, sr^{-1}} \\
\times \left(\frac{N_{\rm H}}{10^{20}\,{\rm cm}^{-2}}\right)
\left(\frac{n_{\rm H}}{20\,{\rm cm}^{-3}}\right)
\left(\frac{\epsilon_{\rm line}}{0.002}\right)
\left(\frac{0.9\,\mu{\rm m}}{\lambda}\right)
\end{multline}
For cloud density $n_{\rm H}\approx20\,{\rm cm}^{-3}$ and column density $N_{\rm H}\approx 1.2\times10^{20}\,{\rm cm}^{-2}$
(corresponding to $A_V\approx 50$\,mmag, $E(B-V)\approx 18.4$\,mmag, $I_{\nu(100\mu m)}\approx 1$\,MJy\,sr$^{-1}$) Equation (\ref{eq:I_line}) predicts $I_{\rm line}\approx 0.18\times10^{-9}\,{\rm erg\, cm^{-2}\, s^{-1}\, sr^{-1}}$. Our observed strength for the 5-2 S(2) 9116.0\AA\ line was $0.65\substack{+0.13 \\ -0.12}\times10^{-9}\,{\rm erg\, cm^{-2} \,s^{-1}\, sr^{-1}}$. Given the roughness of this theoretical estimate, agreement within a factor of a few supports our line assignments and measurement.

Setting aside the absolute strength of the lines, we can also compare our observed line ratios to those expected from \citetalias{Bertoldi_1996_ApJ} models. We form three line ratios, shown in the right two panels of Figure \ref{fig:h2lines}, all with the strongest line in the denominator. All four lines are fit simultaneously with MCMC, similar to previous sections. Here we model all lines as having the same linewidth, with a uniform prior from 0.2 to 1.5 \r{A}. The prior on each line center is uniform over $\pm100$ km/s and are independent. The prior on each amplitude is uniform over $\pm5\times$ the amplitude of the largest value in the search window.

First we form line ratios from the three lines from para-hydrogen, which has even rotational quanta, in this case $J=2$ for all three lines. In the middle panel of Figure \ref{fig:h2lines}, our posterior for the line ratios is shown in a greyscale 2D histogram. Different models from \citetalias{Bertoldi_1996_ApJ} colored by hydrogen atom density are overplotted. Our posterior agrees well with the lowest density ($n_{\text{H}} = 10 ^2$ cm$^{-3}$) models, as we would expect for H$_2$ in the diffuse ISM.

We have one transition, 9-5 S(1), from ortho-hydrogen. Ratios of lines comparing ortho and para hydrogen are sensitive to variations in the ortho-to-para ratio of H$_2$. Because nuclear spins are weakly coupled to electromagnetic radiation making ortho-to-para conversion very slow, the ortho-to-para ratio probes the H$_2$ formation mechanism(s), which may also depend on the local ISM conditions. A completely nuclear spin agnostic formation pathway would lead to a 3-to-1 ratio based on degeneracy factors, despite thermodynamic considerations favoring the more stable para-hydrogen. 

In Figure \ref{fig:h2lines}, we show that our observed ortho-to-para line ratios are lower than those predicted by \citetalias{Bertoldi_1996_ApJ}, which assumed a 3-to-1 ortho-to-para ratio at formation, and included ortho-para conversion by collisions with H$^+$, H atoms, and cold grain surfaces. By comparing our line ratio with the \citetalias{Bertoldi_1996_ApJ} model with conditions most analogous to the diffuse ISM, we find our observed ortho-to-para line ratio is lower by a factor of $0.60\substack{+0.28 \\ -0.27}$. Unfortunately, because the UV pumping lines tend to be saturated, determining the true ortho-to-para ratio from the observed fluorescent line ratios would require detailed modeling of the UV pumping process for the inhomogeneous ISM and its population of OB stars. In general, the UV pumping lines of ortho-hydrogen are more highly saturated than para-hydrogen for ortho-to-para ratios $> 1$, so the resulting observed ratio of ortho-hydrogen fluorescent lines to para-hydrogen lines will be lower than the actual ortho-to-para ratio.

In addition to the four transitions discussed above, we also find a peak near the wavelength of the ortho-hydrogen 5-2 S(1) transition (9231.7 \r{A}), which is the strongest predicted line in the DESI wavelength range (see Appendix \ref{sec:margDetect}, Figure \ref{fig:marginalDetect}). However, this transition overlaps with Pa(9) which is at 9231.5 \r{A}. Using theoretical case B line ratios for hydrogen recombination lines with an electron temperature of 9000 K and $n_e = 100$ cm$^{-3}$, we estimate Pa(9) should be $0.47\times10^{-9}{\rm erg\, cm^{-2} s^{-1} sr^{-1}}$ based on our H$\alpha$ surface brightness, after applying corrections for reddening.

Modeling the observed peak with only a single Gaussian, we find a surface brightness of $0.96\substack{+0.11 \\ -0.11}\times10^{-9}\,{\rm erg\, cm^{-2}\, s^{-1}\, sr^{-1}}$ centered at $9231.80 \pm 0.10$ \r{A}. Scaling our measured 9-5 S(1) line by the theoretical 5-2 S(1)/9-5 S(1) ratio, we expect a 5-2 S(1) surface brightness of $\sim 0.8 \times10^{-9}\,{\rm erg\, cm^{-2} \,s^{-1}\, sr^{-1}}$; because both transitions arise from ortho-hydrogen, this prediction is insensitive to the ortho-to-para ratio. Combined with our Pa(9) prediction, the total is $1.2\times10^{-9}\,{\rm erg\, cm^{-2}\, s^{-1}\, sr^{-1}}$, with the observed feature only $\sim$20\% below this simple prediction. Since the predicted 5-2 S(1) line is $\sim$1.6$\times$ stronger than the predicted Pa(9), the H$_2$ line is likely the larger contributor to the observed feature. We leave more quantitative comparisons requiring joint fits of 5-2 S(1) and Pa(9) to future work with higher signal-to-noise ratio and/or spectral resolution.

\section{DGL Continuum} \label{sec:continuum}

To obtain a clearer picture of the shape of the DGL continuum, we smooth $\alpha_{\lambda}$ by a moving median with kernel size of 21 pixels (16.8 \r{A}) with symmetric tapering.\footnote{\url{https://github.com/Firionus/FastRunningMedian.jl\#taperings-visualized}} Before median filtering, we also mask $\pm3$\r{A} around all emission lines in Figure \ref{fig:emission_lines}, discussed in Section \ref{sec:neblines} (whether or not they were significantly detected), and lines in Appendix \ref{sec:margDetect}-- Pa(9) $\lambda9232$, and two skylines $\lambda5579$ and $\lambda6302$. This yields Figure \ref{fig:continuum}, with the central value shown as a solid line and shading representing the $3\sigma$ uncertainty on the moving median. We compute the uncertainty on the moving median as $\sigma_{\text{median}}=\text{median}_i(\sigma_i)\sqrt{\pi/2n}$ where the median is over all $n = 21$ points $i$ in the moving median window.

In the continuum, we observe several broad stellar absorption lines associated with a range of stellar types, consistent with the interpretation that the DGL is predominantly scattered star light. The infrared Ca II triplet and Ca II H+K are both clearly present in Figure \ref{fig:continuum}. We also detect the Mg I b band and a coincident feature that is likely the MgH band $\lambda5211$ that often appears with the Mg I b band triplet. In terms of molecular features, we also see the CH G-band at $4306$ \r{A}. The dips red-ward of 7000\r{A} and between 7500-8000\r{A} may be broad TiO and TiO+VO molecular absorption bands, respectively, but we are not confident enough in the baseline calibration of DESI to claim their detection.

We see absorption lines from the Balmer series all the way up to the $2 \rightarrow 10$ transition (red vertical guidelines). Because H$\epsilon$ is coincident with Ca II H, we cannot distinguish it. The H$\alpha$ absorption feature is also not immediately identifiable, but we expect this is due to the strength of the H$\alpha$ emission and because the wings of the stellar H$\alpha$ absorption are being masked as part of the proximal [\ion{N}{2}] emission line masking. Future work could inspect our high resolution (unsmoothed) DGL correlation spectrum for narrower stellar absorption lines.

In the near-infrared there is a prominent peak near 9521\r{A} that we attribute to a skyline, seen in the mean sky spectrum (Appendix \ref{sec:calEx}), which appears to leak through into the DGL spectrum. 

\begin{figure}[htb]
    \centering
    \includegraphics[width=\linewidth]{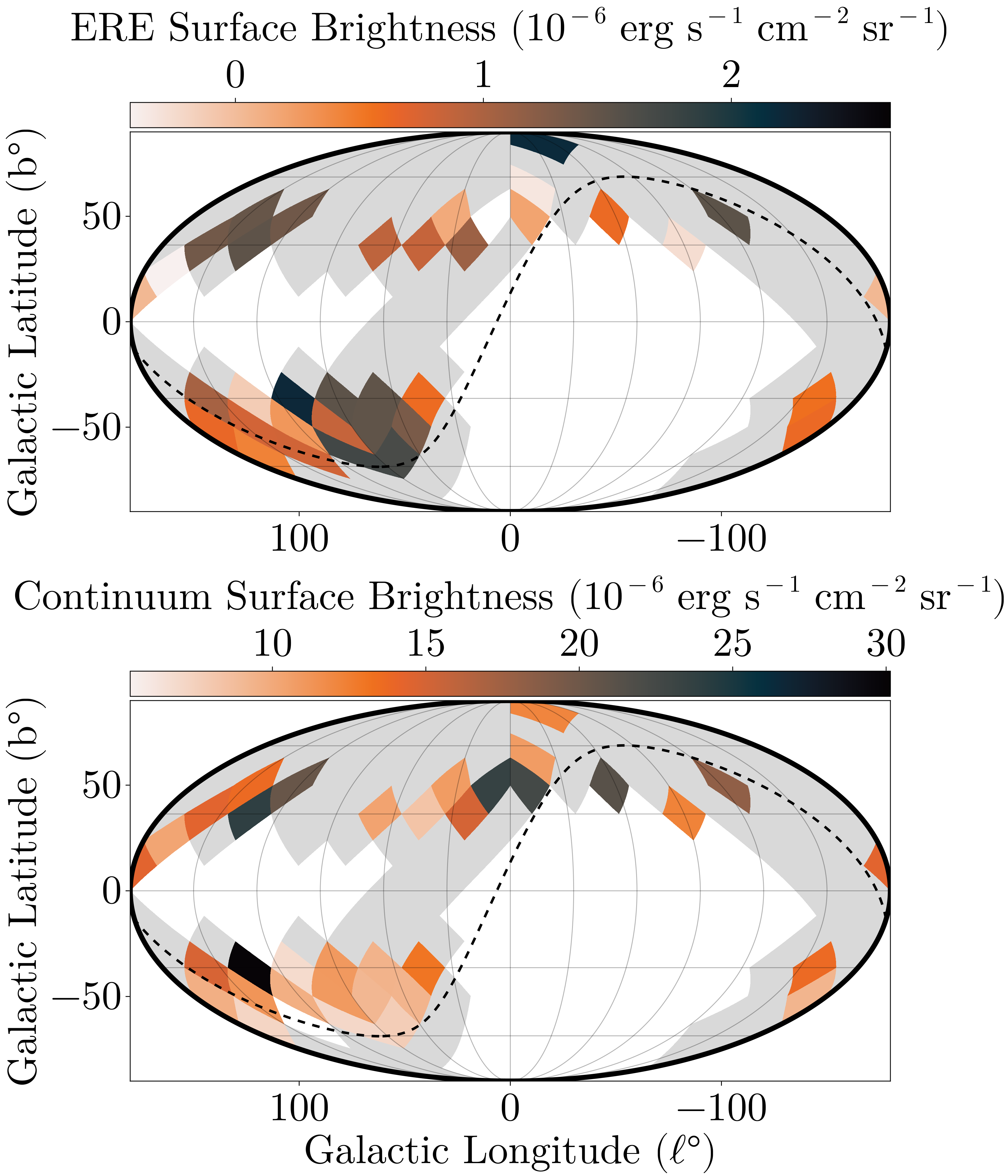}
    \caption{Maps of ERE measurement in the DGL across the DESI footprint on a HEALPix grid with resolution NSIDE $=$ 4. Both rows show surface brightness for 1 MJy/sr FIR emission at 100 $\mu$m, with the top row showing the ERE and the bottom row showing the continuum across the ERE wavelength range for comparison. Pixels without any data are shown in white and pixels without significant detections (see text) are shown in light grey. Projections are Mollweide with a dashed line showing the ecliptic plane.}
    \label{fig:ere}
\end{figure}

\subsection{Extended Red Emission} \label{sec:ERE}

Several past measurements have detected extended red emission (ERE), often attributed to fluorescence or phosphorescence of molecules in the ISM, in the DGL \citep{Szomoru_1998_ApJL,Gordon_1998_ApJ,Chellew_2022_ApJ}. The ERE appears as an excess in flux between $\sim$5400-8000\r{A} compared to the stellar continuum, with the prototypical detection being in the Red Rectangle nebula. \citetalias{Chellew_2022_ApJ} reported a difference in the DGL between the northern and southern Galactic hemispheres, which they interpreted as being due to contributions from young, hot stars and more ERE in the south. Motivated by this, we investigated the DESI DGL measurement for possible spatial variations in ERE. We do this without referring to any models by taking the median over 4700-5200\r{A} and 8100-8600\r{A}, wavelength ranges bracketing the peak of the ERE feature, and measuring the excess relative to a line fit to those two median values (e.g.,~dashed guideline in Figure \ref{fig:continuum}).

We show the map of spatial variation in the ERE measured this way in Figure \ref{fig:ere} as a HEALPix NSIDE$=4$ (14.6$\degree$ diameter pixels). We exclude pixels without any data (white) and those with a median per wavelength bin S/N in the integration window for the ERE (4950.4-8349.6\r{A}) less than 2.5 (grey). In the top panel of Figure \ref{fig:ere}, we show the ERE surface brightness corresponding to 1 MJy/sr of FIR emission at 100 $\mu$m. To demonstrate that the observed variation is not the result of overall variations in the DGL continuum surface brightness (integrated over the same wavelength range) across the sky, we show this quantity in the bottom panel of Figure \ref{fig:ere}, which if anything slightly anti-correlates with the ERE surface brightness variations. For the all-sky DGL correlation spectrum, we find for 1 MJy sr$^{-1}$ of 100 $\mu$m emission, $I_{\text{ERE}} = 925\pm12 \times10^{-9}{\rm erg\, cm^{-2} s^{-1} sr^{-1}}$ and $I_{\text{cont,scattered}} = 9591\pm20 \times10^{-9}{\rm erg\, cm^{-2} s^{-1} sr^{-1}}$, which corresponds to $I_{\text{ERE}}/I_{\text{cont,scattered}}\approx 10\%$. For comparison, \citetalias{Chellew_2022_ApJ} report an integrated ERE intensity of $\approx 0.28\times10^{-5}$ erg\,s$^{-1}$\,cm$^{-2}$\,sr$^{-1}$ for their southern footprint at a mean 100 $\mu$m intensity of 2.6 MJy\,sr$^{-1}$, i.e., $\approx 1100\times10^{-9}$ erg\,s$^{-1}$\,cm$^{-2}$\,sr$^{-1}$ per MJy\,sr$^{-1}$ of FIR emission, consistent with our all-sky value at the $\sim15\%$ level despite the differing footprints and absolute calibration definitions. However we find that the $I_{\text{ERE}}/I_{\text{cont,scattered}}$ ratio varies spatially and reaches values as high as 30\% in Figure \ref{fig:ere}. 

On average the ``Southern'' Galactic fields have stronger ERE than ``Northern'' fields, in agreement with \citetalias{Chellew_2022_ApJ}. Our map goes further to localize regions on the sky contributing strongly to ERE in the DGL. By comparing Figures \ref{fig:ere} and \ref{fig:halpha_sii_ratio}, we see that ERE appears to positively correlate in strength with [\ion{S}{2}] $\lambda6718$/H$\alpha$ until the ratio exceeds $0.8$, after which we do not detect co-local ERE. This correlation may provide evidence to help constrain the ionization state of ERE carriers or place limits on the interstellar radiation field strengths they can tolerate. For example, we might interpret this positive correlation as evidence that ERE carriers are more abundant in softer ionization environments typically associated with higher [\ion{S}{2}] $\lambda6718$/H$\alpha$ ratios. Further interpreting this correlation will require more careful correction for spatially varying extinction.

\begin{figure}[htb]
    \centering
    \includegraphics[width=\linewidth]{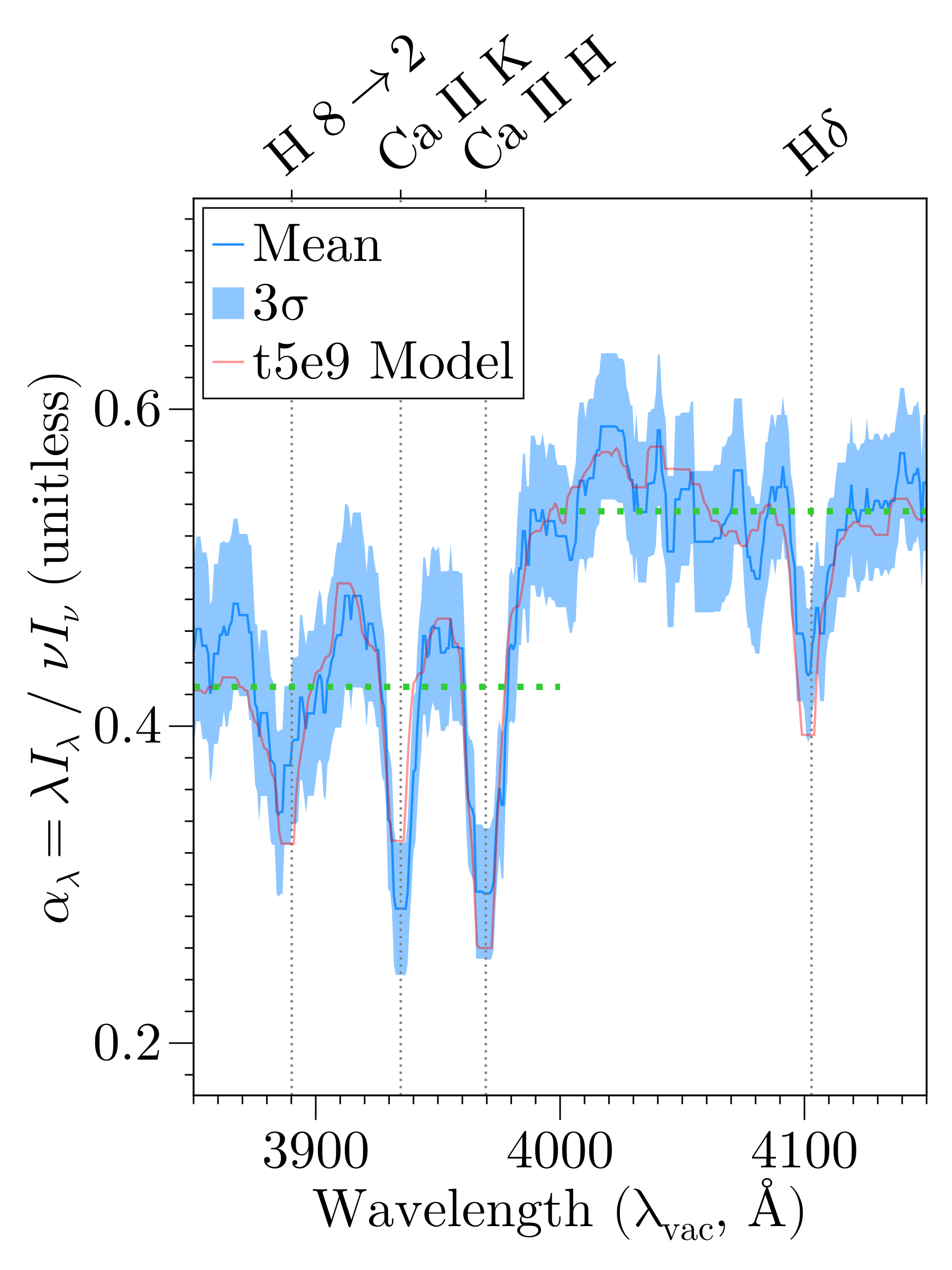}
    \caption{All-sky DGL correlation spectrum in region around the 4000\r{A} break used to measure $\delta_{4000}$. Spectrum is smoothed with the 11 pixel (8.8 \r{A}) wide moving median kernel. Solid line represents central value with shading representing $3\sigma$ uncertainties. Dotted green horizontal segments show the mean continuum levels over $3850-4000$\r{A} and $4000-4150$\r{A} used to define $\delta_{4000}$; labeled guidelines show strong hydrogen and Ca II stellar absorption lines. A stellar population spectral model spectrum (t5e9) smoothed with the same kernel is shown in red for comparison after a multiplicative scaling so its median matches that of $\alpha_{\lambda}$.   
    }
    \label{fig:d4000break}
\end{figure}

\subsection{4000\r{A} Break} \label{sec:d4000}

In Figure \ref{fig:d4000break} we show a view of the 4000\r{A} break in our DGL correlation spectrum. A stellar population synthesis (SPS) model from \citet{Bruzual_2003_MNRAS} (BC03) is shown for comparison after a multiplicative scaling so its median matches that of $\alpha_{\lambda}$ and applying the same moving median filter as we used for the DGL continuum. We use the t5e9 model which has an exponentially decreasing star formation history with a 5 Gyr time constant over a 12 Gyr history and is at solar metallicity simply as a representative example, following \citetalias{Brandt_2012_ApJ,Chellew_2022_ApJ}. A comparison between the DGL and t5e9 model shows impressive agreement. The largest discrepancy in the continuum is $\sim1-2\%$ at the bluest wavelengths. The other main difference is the depth of some of the strongest lines (e.g.,~Ca II H \& K), though the SPS model remains within the 3$\sigma$ contour of our DGL measurement for these lines.

Following \citetalias{Chellew_2022_ApJ}, we can define and measure $\delta_{4000}$ as the intensity ratio of the left $3850-4000$\r{A} to right $4000-4150$\r{A} side of the 4000\r{A} break (dotted green segments in Figure \ref{fig:d4000break}). For our all-sky DGL correlation spectrum, this is $\delta_{4000}=0.8233\pm0.0077$. This is within $3\sigma$ of the \citetalias{Chellew_2022_ApJ} result of $\delta_{4000}=0.73\pm0.04$ and moves the DGL measurement to be even more consistent with the DGL continuum coming purely from scattered starlight off of dust, which \citetalias{Chellew_2022_ApJ} showed corresponds to $\delta_{4000}=0.80-0.83$. This means our measurement does not suggest there is significant blue luminescence (BL) in the DGL, which is proposed to come from fluorescence of small, shielded, 3-4 ringed neutral polycyclic aromatic hydrocarbons \citep{Vijh_2005_ApJ}.

\section{Data/Code Availability} \label{sec:dataavil}

Jupyter notebooks and data products necessary to reproduce all plots in this work are released with this paper on Zenodo. In addition, we include all intermediate data at the petal-exposure correlation level, which dominates the archive volume, so that users can make DGL spectra for custom spatial bin choices. The total archive is 64 GB unzipped and is available at \dataset[doi:10.5281/zenodo.21709438]{https://doi.org/10.5281/zenodo.21709438}.

The SFD-matched, point-source removed version of IRIS was obtained from the Planck ESA archive\footnote{\url{http://pla.esac.esa.int/pla/aio/product-action?MAP.MAP_ID=IRIS_combined_SFD_really_nohole_nosource_4_2048.fits}} and was queried by bilinear interpolation implemented by the \texttt{Healpix.jl} package. The SFD data products (xmap) were obtained from the Harvard Dataverse\footnote{\url{https://dataverse.harvard.edu/dataset.xhtml?persistentId=doi:10.7910/DVN/EWCNL5}} (doi:10.1086/305772) and were queried by bilinear interpolation after converting coordinates to the native Lambert projection of SFD as implemented by the \texttt{DustExtinction.jl} package.

\section{Conclusion} \label{sec:conc}

We have measured the DGL with improved spectral resolution and over an order-of-magnitude more data than previous measurements by using 10.8 million ``blank'' sky fibers from the DESI Y3 observations. With our increased data volume, we have mapped spatial variations in the strength, central wavelength, and width of the nebular emission lines in the DGL at resolutions up to HEALPix NSIDE=8 (7.3$\degree$ diameter pixels), as well as the line ratio for doublets where applicable. We also detect neutral Na and K emission features in the DGL for the first time, and use their spatial variation to interpret the positive sign in the all-sky DGL correlation spectrum as the result of resonant scattering of these lines into our line of sight.

Our improved sensitivity enabled detections of direct emission from at least four H$_2$ ro-vibrational transitions, with absolute surface brightnesses that agree roughly with theoretical predictions. Comparing our measured H$_2$ ortho-to-para line ratios to theoretical models yielded ortho-to-para line ratios that were lower by a factor of $0.60\substack{+0.28 \\ -0.27}$. Measurements of the 4000\r{A} break are entirely consistent with predictions for the DGL continuum coming from purely scattered starlight off of dust, without a BL contribution. We report tentative detection of a correlation between the spatial variations in ERE and [\ion{S}{2}]$\lambda6718$/H$\alpha$, which could aid in constraining the conditions in which ERE carriers are excited and can survive.

With this work, we have pushed measurements of the DGL to the point where our spectra require and can test more advanced spectral modeling methods. Matching the spatial variations we observe will provide important validation for ongoing works attempting radiative transfer through 3D dust maps \citep{porter2018interstellar,McCallum_2025_MNRAS}. By exploring how sensitive the spatially varying DGL is to emission sources, dust scattering and extinction curves, and dust geometry, these investigations can clarify and contextualize approximations in extragalactic investigations.

\begin{acknowledgments}
A.K.S. acknowledges support for this work was provided by NASA through the NASA Hubble Fellowship grant HST-HF2-51564.001-A awarded by the Space Telescope Science Institute, which is operated by the Association of Universities for Research in Astronomy, Inc., for NASA, under contract NAS5-26555. 

We acknowledge helpful discussions with Adam Wheeler and the SNC-ers (Ilija Medan, Zach Way). A.K.S. acknowledges Sophia S\'{a}nchez-Maes for helpful discussions and much support. This work made use of the Cannon cluster supported by the FAS Division of Science Research Computing Group at Harvard University. The authors acknowledge Interstellar Institute's program ``With Two Eyes'' and the Paris-Saclay University's Institut Pascal for hosting discussions that nourished the development of the ideas behind this work.

This material is based upon work supported by the U.S. Department of Energy (DOE), Office of Science, Office of High-Energy Physics, under Contract No. DE–AC02–05CH11231, and by the National Energy Research Scientific Computing Center, a DOE Office of Science User Facility under the same contract. Additional support for DESI was provided by the U.S. National Science Foundation (NSF), Division of Astronomical Sciences under Contract No. AST-0950945 to the NSF’s National Optical-Infrared Astronomy Research Laboratory; the Science and Technology Facilities Council of the United Kingdom; the Gordon and Betty Moore Foundation; the Heising-Simons Foundation; the French Alternative Energies and Atomic Energy Commission (CEA); the National Council of Humanities, Science and Technology of Mexico (CONAHCYT); the Ministry of Science, Innovation and Universities of Spain (MICIU/AEI/10.13039/501100011033), and by the DESI Member Institutions: \url{https://www.desi.lbl.gov/collaborating-institutions}. Any opinions, findings, and conclusions or recommendations expressed in this material are those of the author(s) and do not necessarily reflect the views of the U. S. National Science Foundation, the U. S. Department of Energy, or any of the listed funding agencies.

The authors are honored to be permitted to conduct scientific research on I'oligam Du'ag (Kitt Peak), a mountain with particular significance to the Tohono O’odham Nation.

During the preparation of this manuscript, the authors used Claude (Fable 5; Anthropic) as an assistive tool to help develop and refine figure-plotting code, to search for additional relevant references, and for grammar and text editing. All AI-assisted output, including suggested references, was reviewed and verified by the authors, who take full responsibility for the content of this work.

\end{acknowledgments}

\begin{contribution}

A.K.S. performed the data calibration and analysis, produced the figures, and wrote the first draft of the manuscript. B.T.D. conceived of the project. B.T.D., T.D.B., E.F.S., A.D., and D.P.F. provided feedback during the analysis and detailed comments on the manuscript. All additional authors are co-authors through DESI builder status. All authors have read and agreed to the manuscript.

\end{contribution}

\facilities{Mayall (DESI)}

\software{Julia \citep{bezanson2017julia},
AstroTime.jl,
BenchmarkTools.jl \citep{chen2016benchmarktools},
BinnedStatistics.jl,
CairoMakie.jl \citep{danisch2021makie},
ColorSchemes.jl,
CSV.jl \citep{csvjl},
DataFrames.jl \citep{bouchetvalat2023dataframes},
DataStructures.jl,
Distributions.jl \citep{besancon2021distributions, distributions2019zenodo},
DustExtinction.jl,
EllipsisNotation.jl,
FastRunningMedian.jl,
FITSIO.jl \citep{Pence_2010_A_A},
FreqTables.jl,
Glob.jl,
HDF5.jl \citep{hdf5},
Healpix.jl \citep{tomasi2021healpix, gorski2005healpix},
ImageFiltering.jl \citep{imagefilteringjl},
ImageTransformations.jl,
Interpolations.jl \citep{interpolationsjl},
JLD2.jl,
KrylovKit.jl \citep{krylovkit},
LaTeXStrings.jl,
LsqFit.jl \citep{lsqfitjl},
Optim.jl \citep{mogensen2018optim},
PairPlots.jl \citep{thompson2023pairplots},
ParallelDataTransfer.jl,
PhotometricFilters.jl,
ProgressMeter.jl,
ShiftedArrays.jl,
SkyCoords.jl,
StatsBase.jl \citep{statsbasejl},
Suppressor.jl,
Turing.jl \citep{fjelde2025turing, ge2018turing},
WCS.jl \citep{greisen2002wcs, calabretta2002wcs}
}

\begin{figure*}[htb]
    \centering
    \includegraphics[width=\linewidth]{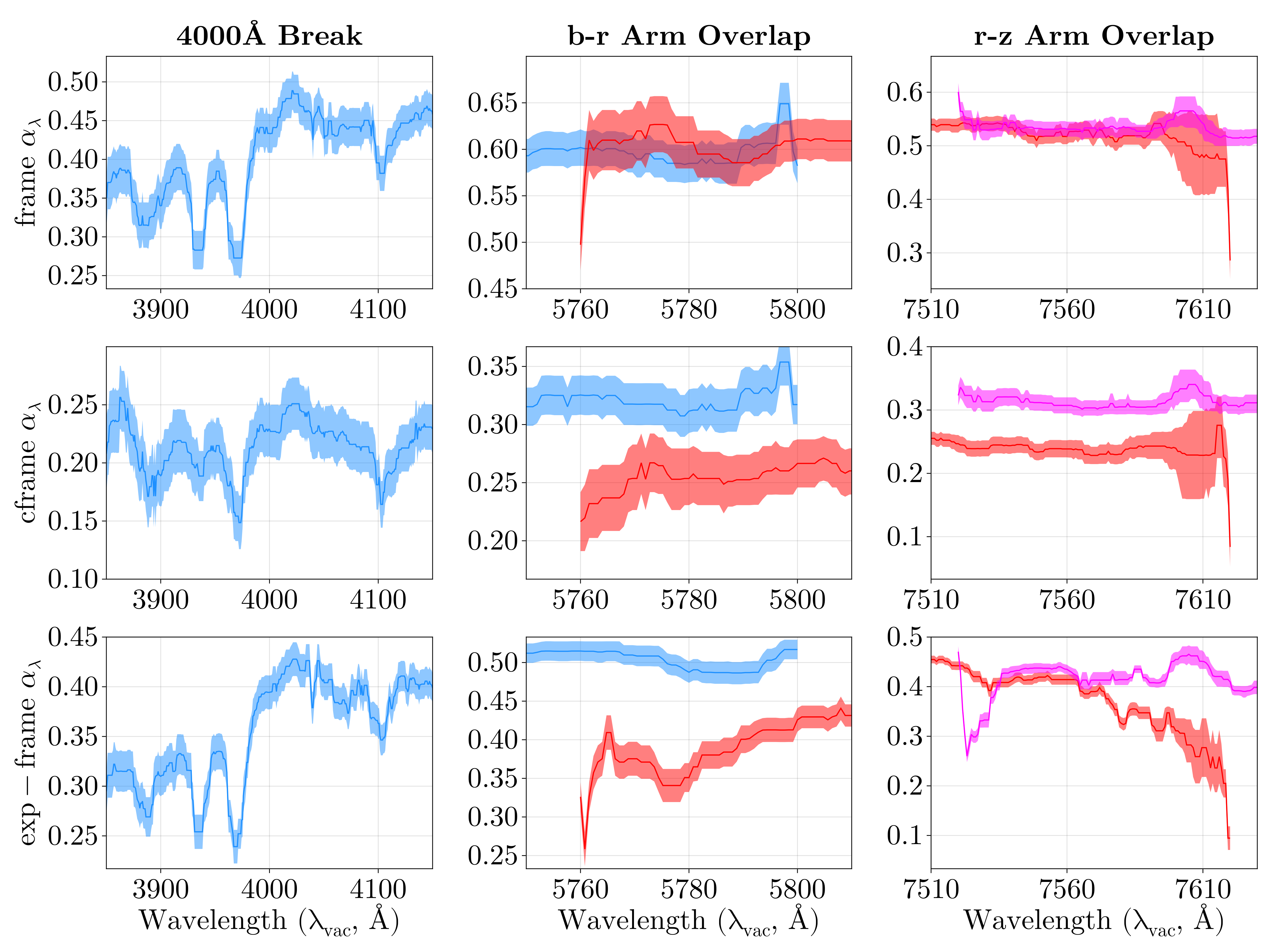}
    \caption{Comparison of the DGL correlation spectrum, with no calibrations or jackknifing applied, starting from \texttt{frame} files (top row), \texttt{cframe} files (middle row), or using \texttt{frame} files combining all spectra in an exposure from different petals (bottom row). Solid lines show central values and shaded regions represent $3\sigma$ uncertainties. These correlation spectra are prior to combining the different DESI spectrograph arms, which are shown in blue, red, and magenta for the b-,r-, and z-arm, respectively.}
    \label{fig:calCompare}
\end{figure*}

\appendix

\section{Comparing Calibrations} \label{sec:calCompare}

DESI sky fibers are calibrated to point source photometry \citep{Guy_2023_AJ}, which is inappropriate for the use case in this paper.  Section \ref{sec:desisky} details our approach for recalibrating the spectra to uniform surface brightness.  Here we discuss and compare alternatives.

Another approach to calibrating DESI sky fibers for uniform surface brightness science would be to start from the \texttt{cframe} files, the primary sky-subtracted, calibrated science spectra data products in a DESI release. This is where most users would think to start and is most similar to the calibration approach of \citetalias{Brandt_2012_ApJ} and \citetalias{Chellew_2022_ApJ}. First, one needs to ``undo'' the flux calibration step that accounts for the size and shape of the PSF relative to the fiber placement and diameter. We implement this by dividing by the \texttt{FLAT\_TO\_PSF\_FLUX} (a scalar per fiber) in the \texttt{FIBERCORR} extension of the \texttt{fluxcalib} file corresponding to a given exposure. This replaces the flatfielding and fluxcalibration steps when starting with the frame files described in Section \ref{sec:desisky}, though the subsequent calibration steps are identical. We calibrate for transparency variations by multiplying by \texttt{FIBER\_FRACFLUX\_GFA} (a scalar per exposure) to obtain flux units in $10^{-17}$ erg\,s$^{-1}$\,cm$^{-2}$\,\r{A}$^{-1}$, which we can then divide by the angular size of the fiber.

In Figure \ref{fig:calCompare}, we compare the DGL correlation spectrum obtained from the \texttt{frame} files (top row, used in the main text) with that obtained starting from the \texttt{cframe} files (middle row). In this figure we show the raw $\alpha_{\lambda}$ prior to combining spectrograph arms or applying the absolute calibration factor $C$. Here the uncertainties are the uncertainties on the moving median (11 pixel kernel for the left column, 21 pixel kernel for the other columns) based on the uncertainties on the maximum likelihood estimator only (not jackknifed). The central value is shown as a solid line and the 3$\sigma$ uncertainty range is shown as a shaded band. In the left column, we compare the 4000\r{A} break region, where the strength of the stellar lines and the break are suppressed in the \texttt{cframe} files DGL correlation spectrum compared to that from the \texttt{frame} files. The middle and right columns show the wavelength ranges where the DESI spectrograph arms overlap. While the overlap regions agree within uncertainties for the \texttt{frame} files analysis, the b- and z-arms are higher than the r-arm for the \texttt{cframe} files analysis by $\sim23\%$ and $\sim14\%$, respectively. These issues with starting from the \texttt{cframe} files along with the conceptually cleaner (fewer steps unrelated to our intended uniform surface brightness measurement) approach described in Section \ref{sec:desisky} motivate us to use the \texttt{frame} file analysis described in the main text.

\begin{figure}[htb]
    \centering
    \includegraphics[width=\linewidth]{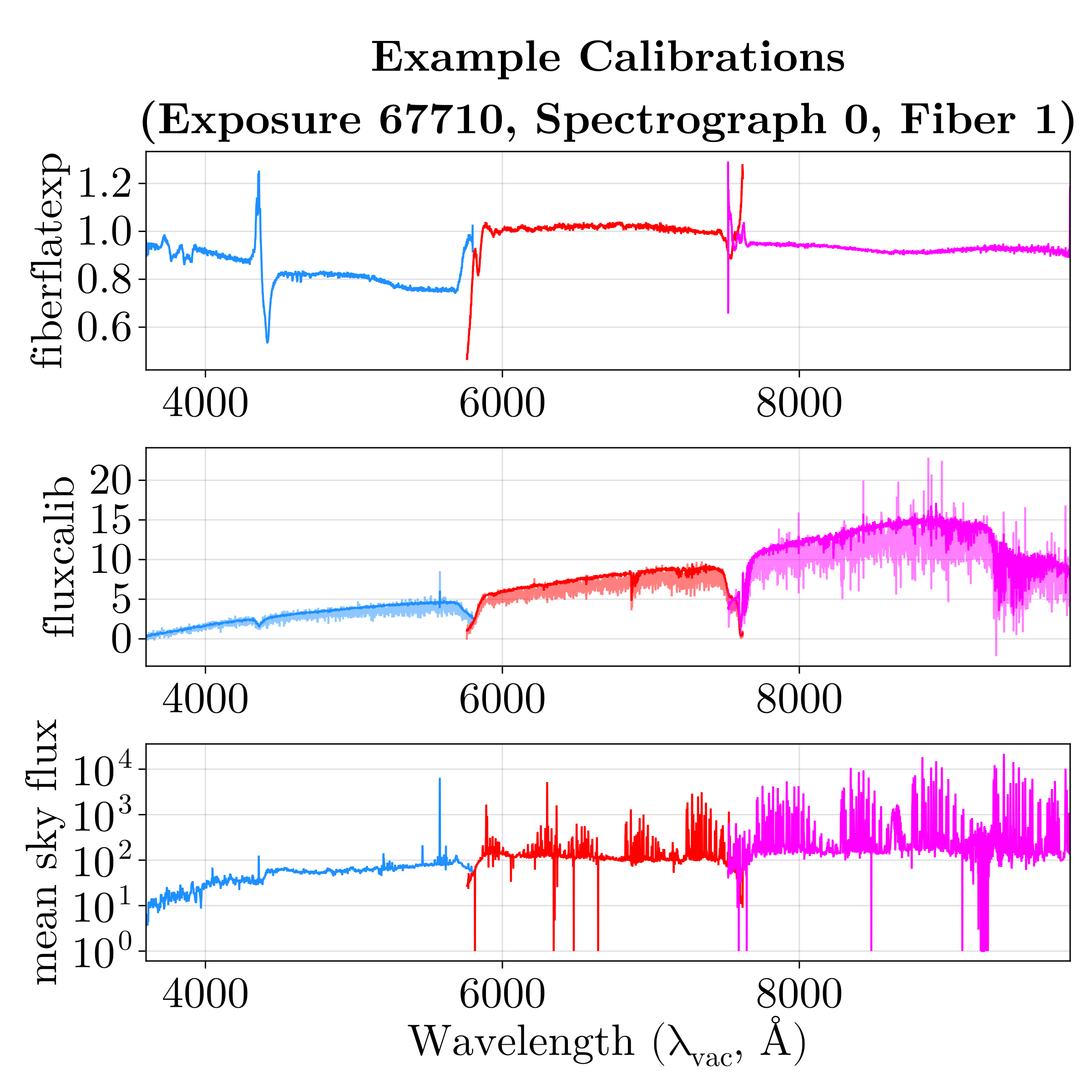}
    \caption{Example from an arbitrary fiber-spectrograph-exposure of the key calibration vectors from the DESI pipeline that we applied in calibrating the sky fibers for measuring the DGL (see Section \ref{sec:desisky}) as well as the mean sky spectrum for comparison. The different DESI spectrograph arms are shown in blue, red, and magenta for the b-,r-, and z-arm, respectively. For the \texttt{fluxcalib} vector, we show the mean over all sky fibers across all spectrographs for a given exposure (which is what we applied) in the darker line and the noisier per fiber-spectrograph estimate from the DESI pipeline as a more transparent line.}
    \label{fig:exCalVector}
\end{figure}

\begin{figure}[tb]
    \centering
    \includegraphics[width=\linewidth]{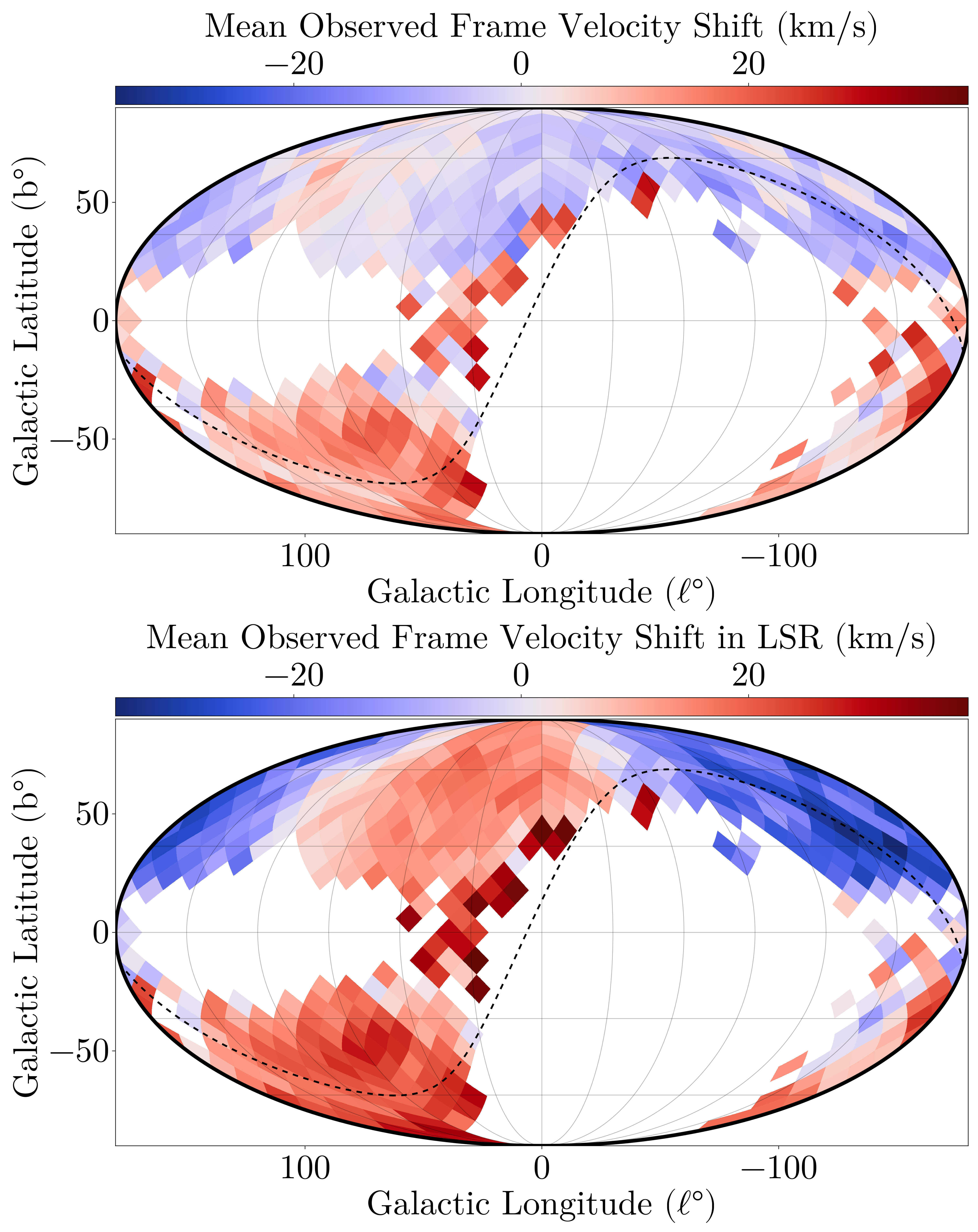}
    \caption{Expected velocity signatures from species in the Earth's atmosphere in the barycentric velocity frame (top) and local standard of rest velocity frame (bottom). Pixels without any data are shown in white. Projections are Mollweide with a dashed line showing the ecliptic plane.}
    \label{fig:baryCorLSR}
\end{figure}

\begin{figure*}[htb]
    \centering
    \includegraphics[width=\linewidth]{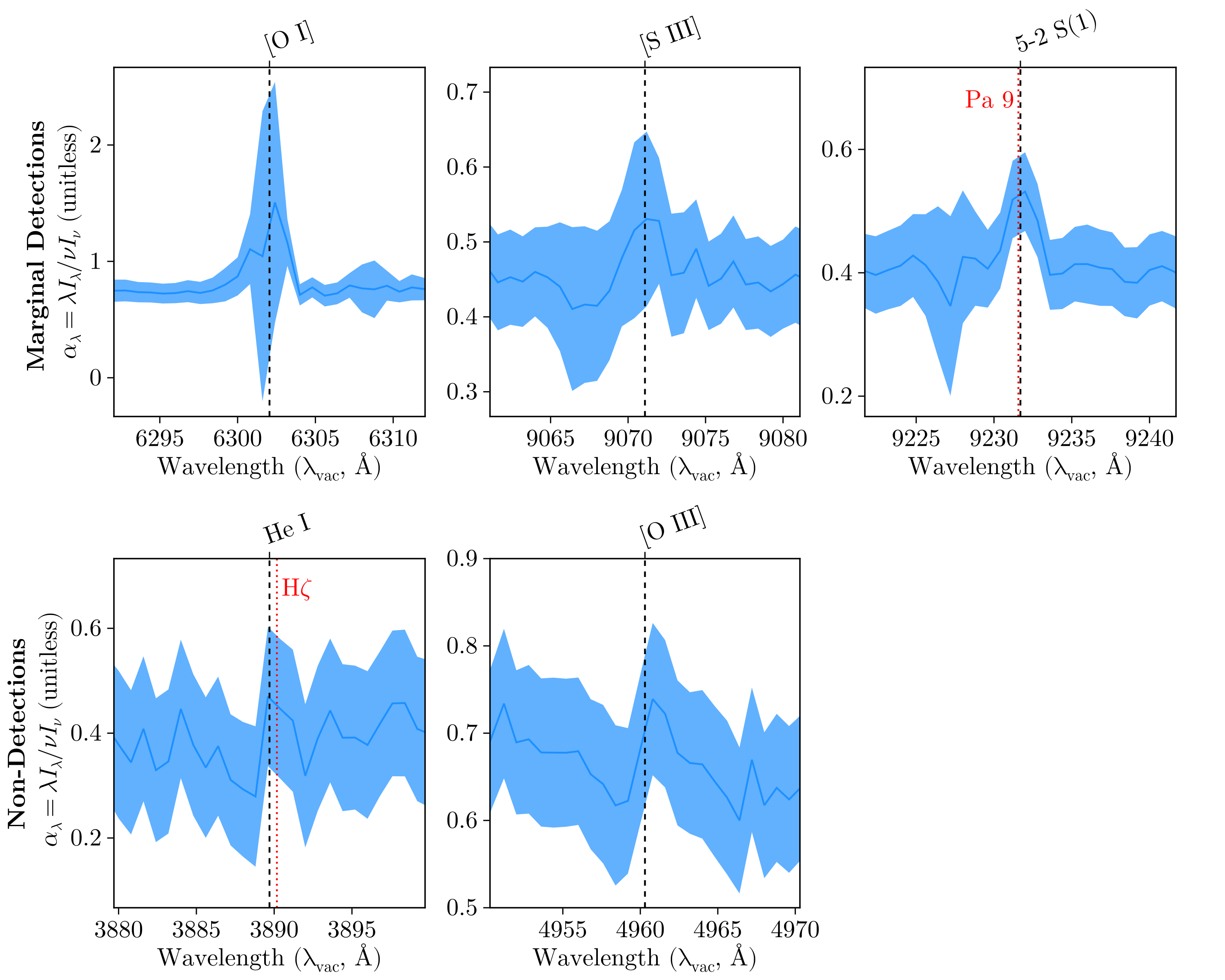}
    \caption{DGL correlation spectrum centered on lines with non-detections (bottom row) or marginal detections (top row). Central value of the correlation spectrum is shown in solid line, with shaded bands representing $3\sigma$ uncertainties under the diagonal approximation to the full covariance matrix.}
    \label{fig:marginalDetect}
\end{figure*}

We also compared the resulting DGL correlation spectrum when starting from \texttt{sframe} files, which are sky subtracted, but not flux calibrated. In the blue/red arms, the \texttt{sframe} and \texttt{cframe} approaches yielded nearly identical spectra other than a small, global multiplicative difference, \texttt{sframe} being $\sim10\%$ larger than the \texttt{cframe} DGL correlation spectrum. That is, both show the same arm-to-arm offsets. In contrast, the ratio of the \texttt{frame} and \texttt{cframe} shows significant structure resembling differences in the stellar absorption lines. While we do not attempt to fully run down the source of these differences, the fact that these arm-to-arm offsets appear between approaches using the \texttt{frame} and \texttt{sframe} files suggests that they are associated with the sky removal step in the DESI pipeline. This step includes fitting a linear spatial model for variation in the sky spectrum, which may absorb spatial variations in the sky spectra from the DGL. It is possible details related to sky continuum subtraction in the SDSS/BOSS spectra contribute to the differences in the absolute calibration factor $C$ between \citetalias{Brandt_2012_ApJ}/\citetalias{Chellew_2022_ApJ} and this work (Section \ref{sec:systematics}).

In the main text, we chose to limit our DGL measurement to leveraging correlations in sky fibers part of a single petal-exposure. In the bottom row of Figure \ref{fig:calCompare}, we show the DGL spectrum that results from instead using all sky fibers (across multiple petals) in a given exposure. Clearly this results in a higher signal-to-noise spectrum, but this choice introduces inter-arm offsets, even when starting from the \texttt{frame} files. Future work may be able to mitigate these systematics by learning multiplicative (and possibly wavelength dependent) offsets between different spectrographs and even leverage inter-spectrograph variation to reject artifacts. However, in this work, we chose the simpler path of limiting our analysis to using correlations within petal-exposures.  

\section{Example Calibration} \label{sec:calEx}

In Figure \ref{fig:exCalVector}, we show examples of the \texttt{fiberflatexp} and \texttt{fluxcalib} vectors from the DESI pipeline as well as the mean sky spectrum for comparison. The first would be a constant value of 1 for an ideal spectrograph with an achromatic response; the second is a basis spectrum capturing as much fiber-to-fiber variability as possible, including telluric absorption from molecules in the Earth's atmosphere. Figure \ref{fig:exCalVector} illustrates the location of large $>10\%$ corrections to the observed spectrum caused by the dichroics and other optics, which may not be perfectly removed, as well as the variability and uncertainty in the telluric correction, especially in the z-arm. We do not know the origin of the negative spikes in the mean sky flux panel (bottom row), but one seems close to the artifact noted at 6555.0~\r{A} in Figure \ref{fig:emission_lines}. 

\section{Velocity Signature of Telluric Features} \label{sec:baryCor}

In Figure \ref{fig:baryCorLSR}, we show the velocity pattern on the sky one would expect for a spectral feature of telluric origin. That is, if we were latching onto a species in the Earth's atmosphere (e.g.,~Na or [\ion{O}{1}]), Figure \ref{fig:baryCorLSR} shows the apparent velocity that we would report. Because DESI reports its spectra re-interpolated onto a standard wavelength axis after barycentric correction, we compute the negative of the velocity associated with the redshift applied to the observed wavelength frame during calibration. Figure \ref{fig:baryCorLSR} uses exactly the same cuts to include only the petal-exposures used through the main text, and reports a simple mean per HEALPix pixel. In the top row, we show this velocity in the ``barycentric'' frame, where there is a strong north-south divide, with predominantly negative velocities in the north and positive velocities in the south. The mean velocity also tends to zero in the north at the boundary of the first and second quadrants. 

In the bottom panel of Figure \ref{fig:baryCorLSR}, we apply the LSR velocity correction, which simply adds the dipole from anomalous solar motion to the top plot. The LSR kinematics of the tracers we measure in the DGL do \emph{not} resemble the pattern in the bottom row of Figure \ref{fig:baryCorLSR}, which suggests they are \emph{not} of primarily telluric origin.

\section{Marginal Detections} \label{sec:margDetect}

In the top row of Figure \ref{fig:marginalDetect} we present three emission lines where we deem the detections more complicated because of confusion and thus do not present them in the main paper. In the top left panel we show [\ion{O}{1}] $\lambda6302$. Measuring this line would mean we had lines from 3 ionization states [\ion{O}{1}], [\ion{O}{2}], and [\ion{O}{3}] of oxygen. However, the same [\ion{O}{1}] line is also a strong telluric, auroral line.  This results in a very high background at zero velocity offset from the terrestrial reference frame, and prevents any confident measurements of [\ion{O}{1}] $\lambda6302$ in the DGL. 

The top middle panel of Figure \ref{fig:marginalDetect} shows [\ion{S}{3}] $\lambda9071$.  The peak emission is small relative to the correlation spectrum uncertainties and it is surrounded by baseline oscillations that resemble telluric absorption features that are strong at these near-infrared wavelengths. Detecting [\ion{S}{3}] $\lambda9071$ would be valuable as its sum with [\ion{S}{3}] $\lambda9534$ and ratio against [\ion{S}{2}] $\lambda6718$ plus [\ion{S}{2}] $\lambda6733$ are sensitive to ionization ($U$) while remaining fairly insensitive to pressure and metallicity variations. This ratio is often used to distinguish between photo-ionization and shock-ionization processes.

The top right panel is centered on the 5-2 S(1) transition of $\text{H}_2$, which is discussed at the end of Section \ref{sec:h2lines}. This line is almost perfectly coincident with Paschen 9, and we argue that this line is likely a blend of the two, with 5-2 S(1) the larger contributor by a factor of $\sim$1.6.

In the bottom row of Figure \ref{fig:marginalDetect} we present two lines we deem non-detections. The bottom left panel shows \ion{He}{1} $\lambda3890$, which we might expect to detect given that we measured \ion{He}{1} $\lambda 5877$. The bottom right panel shows [\ion{O}{3}] $\lambda4960$, which we might expect to detect given that we measured [\ion{O}{3}] $\lambda5008$. Neither of these lines were detected in \citetalias{Brandt_2012_ApJ} and \citetalias{Chellew_2022_ApJ}. 

\clearpage
\global\suppressAffiliationsfalse
\allauthors
\bibliography{DGL}{}
\bibliographystyle{aasjournalv7}

\end{document}